\documentclass[12pt, oneside]{article}
\usepackage[left=1in,right=1in,top=1in,bottom=1in]{geometry}
\usepackage{graphicx}

\usepackage{amssymb,amsmath,bm,accents}
\usepackage{csquotes}
\usepackage{setspace}
\usepackage[colorlinks=true,linkcolor=blue,citecolor=blue]{hyperref}
\usepackage{natbib}
\usepackage{subcaption}
\usepackage{multirow}
\usepackage{tabularx}
\usepackage{adjustbox}
\usepackage[justification=centering]{caption}
\usepackage{fancyhdr}

\usepackage{multibib}
\newcites{App}{References for Appendices}
\newlength{\bibitemsep}
\newlength{\bibparskip}
\let\oldthebibliography\thebibliography
\renewcommand\thebibliography[1]{%
 \oldthebibliography{#1}%
 \setlength{\parskip}{\bibitemsep}%
 \setlength{\itemsep}{\bibparskip}%
}

\expandafter\def\expandafter\quote\expandafter{\quote\setstretch{0.85}}

\renewcommand{\small}{\fontsize{10pt}{11pt}\selectfont}

\title{Money, Time, and Grant Design}
\author{
Kyle Myers\thanks{Harvard Business School \& Laboratory for Innovation Science at Harvard, \href{mailto:kmyers@hbs.edu}{kmyers@hbs.edu}.} \quad Wei Yang Tham\thanks{Harvard Business School \& Laboratory for Innovation Science at Harvard, \href{mailto:wtham@hbs.edu}{wtham@hbs.edu}. \newline
This project received financial support from The Alfred P. Sloan Foundation and the Harvard Business School. This project would not have been possible without the excellent work of Rachel Mural, Nina Cohodes, and Yilun Xu as well as input from Marie Thursby, Jerry Thursby, and Karim Lakhani. We have received helpful comments from Amitabh Chandra, Matt Clancy, Ryan Hill, Jeffrey Furman, Fabian Gaessler, Joshua Graff Zivin, Benjamin Jones, Megan MacGarvie, Gustavo Manso, Ramana Nanda, Sam Ransbotham, Henry Sauermann, Jesse Shapiro, Carolyn Stein, Michael Tushman, John Walsh, and seminar participants at the BSE Summer Forum, DRUID, JOMT, NBER Productivity Seminar, NBER Summer Institute, The Ohio State University, Northwestern University, WEFI, and REER.
}
}

\begin{document}

\maketitle
\begin{abstract}\normalsize
The design of research grants has been hypothesized to be a useful tool for influencing researchers and their science. We test this by conducting two thought experiments in a nationally representative survey of academic researchers. First, we offer participants a hypothetical grant with randomized attributes and ask how the grant would influence their research strategy. Longer grants increase researchers’ willingness to take risks, but only among tenured professors, which suggests that job security and grant duration are complements. Both longer and larger grants reduce researchers' focus on speed, which suggests a significant amount of racing in science is in pursuit of resources. But along these and other strategic dimensions, the effect of grant design is small. Second, we identify researchers' indifference between the two grant design parameters and find they are very unwilling to trade off the amount of funding a grant provides in order to extend the duration of the grant --- money is much more valuable than time. Heterogeneity in this preference can be explained with a straightforward model of researchers' utility. Overall, our results suggest that the design of research grants is more relevant to selection effects on the composition of researchers pursuing funding, as opposed to having large treatment effects on the strategies of researchers that receive funding.
\end{abstract}
\thispagestyle{empty}

\clearpage
\setcounter{page}{1}
\onehalfspacing

\section{Introduction}
Firms, governments, and philanthropies that fund science must decide not only how much to invest in researchers, but how to structure those investments. These funders often invest in science invest in science in the form of grants, which award researchers a fixed amount of money for a fixed amount of time.\footnote{Research grants are ``\emph{upfront payments for the delivery of incompletely specified and non-contractable R\&D output}'' (\citealt{azoulay2021scientific}; pg. 120); see \cite{azoulay2021scientific} for a detailed review. Grants are explicitly used in academic science, and often implicitly in for-profit science when firms pre-commit some amount of time and capital to an uncertain R\&D project (\citealt{kerr2015financing}).} Many funders express an interest in using these parameters of money and time to incentivize more socially valuable science. For example, the prestigious Howard Hughes Medical Institute recently extended their grants from five to seven years, claiming that those ``\emph{years of stable support allows [researchers] to take more risk and achieve more transformative advances.}''\footnote{See the following links for more on this and other funding institutions proclaiming the importance of grant design: \href{https://www.hhmi.org/news/hhmi-investigator-program-opens-national-competition}{Howard Hughes Medical Institute}, \href{https://nexus.od.nih.gov/all/2014/07/17/formula-for-innovation-people-ideas-time/}{National Institutes of Health}, \href{https://wellcome.org/news/wellcomes-approach-research}{Wellcome Trust}.}

This argument --- the design of grants can influence researchers' strategies --- has roots in incentive-based theories of innovation and venture financing.\footnote{For example, \cite{holmstrom1989agency} and \cite{manso2011motivating} have potential implications for the optimal structure of investments in research. We discuss other related work below.} Different grant designs might induce researchers to, for example, acquire  different inputs, plan over different time horizons, or diversify their resources across different types of projects. However, the practical relevance and specific implications of any particular theory remains unclear given the backdrop of institutions and incentives that researchers operate amongst (\citealt{dasgupta1994toward,stephan1996economics}) and the rarity of natural experiments in this setting.

Understanding the usefulness of grant design as a policy lever is further complicated by the fact that researchers usually choose what grants to pursue. Thus, funders must understand researchers' preferences and incorporate this knowledge of demand when choosing which grants to supply. For example, in 2015 the National Institutes of Health (NIH) introduced a new grant structure that offered more time, but less money, in hopes of encouraging researchers ``\emph{to pursue new research directions as opportunities arise}'' and ``\emph{avoid abrupt termination of laboratory support}.'' Ideally, the NIH would have known researchers' preferences over money and time when they designed this program, but this information is hard to elicit and quantify; the market for science is largely devoid of the prices and variation in attributes that can be used to identify specifics of demand.\footnote{
See \href{https://www.nigms.nih.gov/Research/mechanisms/MIRA/Pages/Answers-to-Frequently-Asked-Questions-About-MIRA-PAR-19-367.aspx}{here} for more on this particular grant, referred to as the ``MIRA'' mechanism, and \href{https://genestogenomes.org/mixed-feelings-about-mira}{here} for more on researchers' mixed responses to the money-time tradeoff that MIRA required of participants.} So while the potential for distortions in researchers' strategies has long been appreciated (\citealt{nelson1959simple,arrow1962economic}), we still have a limited empirical understanding of researchers’ preferences over grant attributes and how these attributes could be used to influence the direction of science; what are the treatment and selection effects induced by research grant designs?

In this paper, we identify researchers' preferences and beliefs over grant designs using a nationally representative sample of research-active professors spanning all major fields of science across all major academic research institutions in the US (\citealt{myers2023new}). We focus on two specific empirical questions: (1) how would researchers change their strategies if they were awarded grants of different funding amount (dollars available) and duration (years those dollars are available); and (2) what are researchers’ preferences over grant designs; how willing are they to trade off money and time? Of course, all researchers want more money and more time, but we are able to identify the \emph{relative} value researchers place on these two attributes and their usefulness in managing the rate and direction of innovation. Funders have attempted to solicit this sort of preference before (\citealt{nsf2002survey,royal2004survey}), but not in a systematic way tied to an economic model as we do here. 

Our first empirical exercise studies how effective grant design could be in influencing researchers' strategies. We ask researchers what strategic changes they would undertake if they received a hypothetical grant, the amount and duration of which is randomized. This hypothetical grant is posed as a sudden influx of funding, and we provide specific details to eliminate idiosyncratic features of grants that might contaminate researchers' responses.\footnote{For example, we emphasize that the funding may be used for any research project and the funding amount and duration are fixed and non-negotiable in order to rule out \href{https://www.ninds.nih.gov/funding/manage-your-award/during-award/no-cost-extension}{no-cost extensions}.} This sort of unanticipated grant receipt is rare in practice. However, it allows us to isolate researchers' responses to grant designs without any selection or competitive effects driven by researchers' endogenous choices (which would have occurred if we instead posed the thought experiment about funding competitions).

Researchers respond by choosing from a menu of five subjective options, which are described in discipline-agnostic language, as to how the randomized grant would lead them to change some strategic aspects of their work.\footnote{Specifically, the five options are: ``Pursue riskier projects'', ``Increase speed'', ``Pursue projects less related to your current work'', ``Increase the size of ongoing projects'', and ``Increase accuracy or reliability''. We do not take a stance on the relative social value of any strategy. Instead our goal is to inform a policymaker or manager who already has preferences over these strategies. ``Strategy'' is a nebulous concept in the context of basic science and our formulations are not clearly mutually exclusive. But they are informed by extensive interviews with researchers across all major fields in the sample. Furthermore, the summary statistics and results of the experiment are consistent with researchers treating these options as relatively exclusive.} Two options are based on the ``risk-versus-speed'' trade-off at the core of principal-agent models that seek to balance risk-taking against shirking  with the use of long-term incentives (e.g., \citealt{holmstrom1989agency,manso2011motivating,lerner2007innovation,aghion2013innovation,tian2014tolerance,nanda2017innovation}). Two options are based on the ``explore-versus-exploit'' trade-off vis-à-vis the role of financial constraints in shaping innovators' willingness to pursue novel and uncertain opportunities (e.g., \citealt{froot1993risk,hall2002financing,brown2009financing,hottenrott2012innovative,krieger2022missing}). Lastly, one option focuses on researchers' willingness to engage in activities that improve the accuracy or reliability of their work, which is motivated by the growing attention placed on so-called replication crises that may be arising because of constraints on researchers time horizons or budgets (e.g., \citealt{pashler2012replicability,smaldino2016natural,kasy2021forking}).\footnote{We also ask researchers how they would spend the grant funds across a menu of input categories to investigate how certain inputs are connected to certain strategies.}

Through a combination of standard regressions and machine-learning methods that search for heterogeneous treatment effects, we find two main results in this experiment. First, we find that longer grants increase researchers' willingness to pursue riskier projects, but only amongst tenured professors. On the other hand, we find that all researchers are more likely to decrease the speed of their work when receiving longer grants. This result is consistent with theories such as \cite{holmstrom1989agency} and \cite{manso2011motivating}, where long-term incentives (here, in the form of longer periods of guaranteed access to funding) can induce risk-taking. However, it suggests that grant duration is only useful as an incentive for risk-taking amongst researchers who have long-term job security and/or a sufficient stock of reputation and resources with which they are willing to gamble. Furthermore, these findings emphasize the potential trade-offs involved with the provision of long-term incentives (i.e., \citealt{nanda2017innovation}).

We also find a connection between grant funding amounts and researchers' explore-versus-exploit decisions. Researchers receiving larger grants are more likely to choose the exploit-oriented option, while those receiving smaller grants are more likely to chose the explore-oriented option. At a surface level, this result stands in contrast to firms' behaviors when they experience cash windfalls, where larger windfalls induce more exploration (\citealt{krieger2022missing}). We propose two alternative ways of rationalizing this result, which has novel implications for grant designs and echoes \citeauthor{wu2019large}'s (\citeyear{wu2019large}) finding that small research teams (who presumably have less funding) are responsible for a disproportionate share of scientific ``disruptions''.\footnote{This result is also in line with theories and some empirical evidence from the strategic management and social-psychology literatures that resource constraints can foster creativity or more entrepreneurial thinking (e.g., \citealt{starr1990resource,moreau2005designing,hoegl2008financial,scopelliti2014financial}).}

We find no connection between grant designs and researchers' willingness to focus more on the accuracy or reliability of their work. To the extent this sort of effort is underincentivized by existing structures of science, it does not appear that marginal changes to grant designs would be an effective countermeasure.

In all cases, the magnitudes we identify, when scaled to match the variation we observe in real grant designs, indicate that the effects are small. Realistically-sized changes to grant designs (e.g., in the range of 15--30\%) correspond to only 1--2 percentage point changes in the probability of researchers choosing certain strategies.  Notably, our investigations into the potential for heterogeneous treatment effects rarely yields any statistically significant evidence of heterogeneity. Together, this suggests that the practical influence of grant design on researchers' strategies may be quite limited.

Our second empirical exercise is focused on understanding researchers' demand for different types of grants. We use another thought experiment to estimate researchers' willingness to trade off funding amount and funding duration. We assume that each researcher $i$’s indirect utility from a grant of amount $A$ and duration $D$ is of the form $v_i(A,D) = \alpha_i  A^\gamma  D^{1-\gamma}$, where the key parameter is $\gamma\in[0,1]$. A smaller $\gamma$ would indicate a higher willingness to trade off funding amount for duration and reflect a belief that time is an important constraint on researchers' ability to leverage funding.

We estimate that the average researcher is willing to trade off approximately \$50,000 from their grant to extend it by 1 year ($\gamma\approx0.8$). This also implies that a 1\% increase in grant size is valued nearly four times more than a 1\% increase in grant duration --- relatively speaking, grant size is much more important to researchers than duration. These estimates provide funders with a more concrete sense of researchers' demand for different grant types and suggest that researchers do not view the duration of any single grant as an important constraint on their research pursuits given their preferences, incentives, and expected access to future funding sources.\footnote{Of course, this doesn't imply that researchers' preferences are perfectly aligned with the social optimum.}

We also test the extent to which researchers have heterogeneous preferences for money and time. We motivate these tests with a simple model which predicts that we should see the strongest preference for money, relative to time, amongst researchers who are (1) more capital-intensive, (2) can more easily access funding through other sources, (3) are less risk-averse, (4) have higher discount rates, and (5) receive more direct utility from research grant funding (e.g., in the form of salary buyouts). When we split the sample along each of these dimensions using direct or proxy measures, we find results consistent with our predictions. Most importantly, these results show that manipulating grant design to change strategies will also induce selection effects that will generate changes in the composition of researchers being funded.
 
The most closely related paper to ours is \cite{azoulay2011incentives}, which provides evidence that grant design may be instrumental. They compare the publication output of premiere biomedical scientists who receive research grants awarded by either the Howard Hughes Medical Institute (HHMI) or the NIH. In short, the HHMI grants are characterized by being larger, longer, and generally having fewer constraints. \cite{azoulay2011incentives} find that HHMI awardees produce more high-impact papers \emph{and} more low-impact papers, and are more likely to change the direction of their science. Another closely related study is \cite{wang2018funding}, who investigate output differences for researchers funded under grant structures that vary in their competitiveness. However, in both cases, understanding the generalizability of the results and attributing the difference in outcomes to the specific aspects of the grants is difficult. Our work builds on theirs by clearly separating the selection and treatment effects of grant design, while also exploring the full population of academic science.

Other relevant literature includes evaluations of researchers' preferences over job attributes (e.g., \citealt{stern2004scientists,roach2010taste,sauermann2014not}), life-cycle effects (e.g., \citealt{levin1991research}), risk-taking in science (e.g., \citealt{mandler2017benefits,franzoni2022uncertainty,veugelers2022funding,clancy2023biases}), how the novelty of an idea shapes researchers' beliefs (e.g., \citealt{boudreau2016looking,wang2017bias}), and the role of speed and priority in science (e.g., \citealt{merton1957priorities,hagstrom1974competition,smaldino2016natural,hill2019scooped}). Few studies have been able to study researchers' demand for research funding specifically (e.g., \citealt{myers2020elasticity}), although a number of studies have made progress in understanding how researchers react to supply shocks (e.g., \citealt{tham2023science, cheng2023effect}) and how certain institutions can facilitate the diffusion of research inputs (e.g., \citealt{furman2011climbing,furman2012growing,agrawal2016understanding,murray2016mice,teodoridis2018understanding}).

Our approach follows a long line of prior work that has used surveys of researchers to uncover their otherwise unobservable features, choices, preferences, or beliefs (e.g., \citealt{levin1991research,fox2001careers,stern2004scientists,sauermann2010makes,curty2017attitudes,levecque2017work,shortlidge2018trade,cohen2020not,philipps2022research}). Ideally, we would pursue our questions with a field experiment or variation induced by a natural experiment. However, a field experiment for our specific questions would be prohibitively costly --- the median observed grant at institutions in our population is \$300,000 and 3 years long. And, as we show in the next section, the distribution of observed grant designs is quite concentrated as funders are far from fully exploring the space of feasible grant designs. Thus, useful natural experiments with sufficient statistical power are virtually non-existent. So while our thought experiments involve hypothetical scenarios, we are able to explore grant designs that are plausible but rarely seen in practice. While some of our measures are subjective, they provide a unique, ex-ante observation of researchers' beliefs in contrast to the typical approach in this literature, which relies on censored, ex-post observations (e.g., transformations of publication output).\footnote{See \cite{stantcheva2022run} for a full discussion of the trade-offs of survey-based empirical analyses.}

Before detailing the experiments, we pay considerable attention to the respondents' attention and the potential for non-response bias. As a test of attention, we compare researchers' self-reported annual salaries to those we observe on publicly available websites (only for researchers at institutions where such data is made public). We find strong agreement between the two, which suggests mostly truthful reporting along this dimension.

Our approach to non-response involves two parts. First, we use data obtained for both respondents and non-respondents on the amount of research funding flowing to their institutions (sourced from the National Science Foundation) and, for a subset of the population, their publication, and grant flows. In many dimensions, we do estimate that respondents are significantly different statistically speaking; however, these differences are all very small economically speaking, typically close to 5\%. 

Our second approach to potential non-response bias makes use of the randomized participation incentives and reminders in the survey. Both the incentives and reminders had a significant influence on researchers' willingness to complete the survey, which provides us with the variation necessary to implement the sample selection correction method of \cite{heckman1979sample}.

The paper proceeds as follows: Section \ref{sec_motivate} provides motivating summary statistics about the current distribution of research grants; Section \ref{sec_survey} details the survey, including the population, sampling process, the survey instrument, and our approaches to testing for, and mitigating the potential of, non-response bias; Section \ref{sec_te} describes the first thought experiment, which tests how grant designs influence researchers' strategies; Section \ref{sec_pref} describes the second thought experiment, which solicits researchers' preferences over grant attributes; lastly, Section \ref{sec_discuss} discusses our results in the context of related work and motivates future studies on policy tools for managing the rate and direction of innovation.

\section{Distribution of current grant designs}\label{sec_motivate}
Historically, the distribution of grants flowing to universities has been difficult to observe. Collecting and connecting data from the dozens of large government agencies and hundreds of smaller organizations is no small task, and surveying researchers about their full history of grants is not reasonable. Thankfully, the ``Dimensions'' data set provides a new look at what types of grants researchers are receiving by aggregating across a large number of funders who publicly report data (\citealt{dimension2018data}).\footnote{As of this paper, Dimensions sources grant data from 233 different US-based funders.} Because it relies on publicly reported grants, the Dimensions data is limited in its coverage of research investments at universities by businesses, so here we can focus only on non-business grants.\footnote{Recent federal survey evidence suggests that business-originated research funding accounts for roughly 5\% of all annual research funding at institutions of higher education (\citealt{nsf2023herd}).}

\begin{figure}[htbp]\centering
\caption{Distribution of research grant designs}\label{fig_dimension_heatmap}
\hspace{12mm}\includegraphics[width=0.8\textwidth, trim=0mm 10mm 0mm 12.5mm, clip]{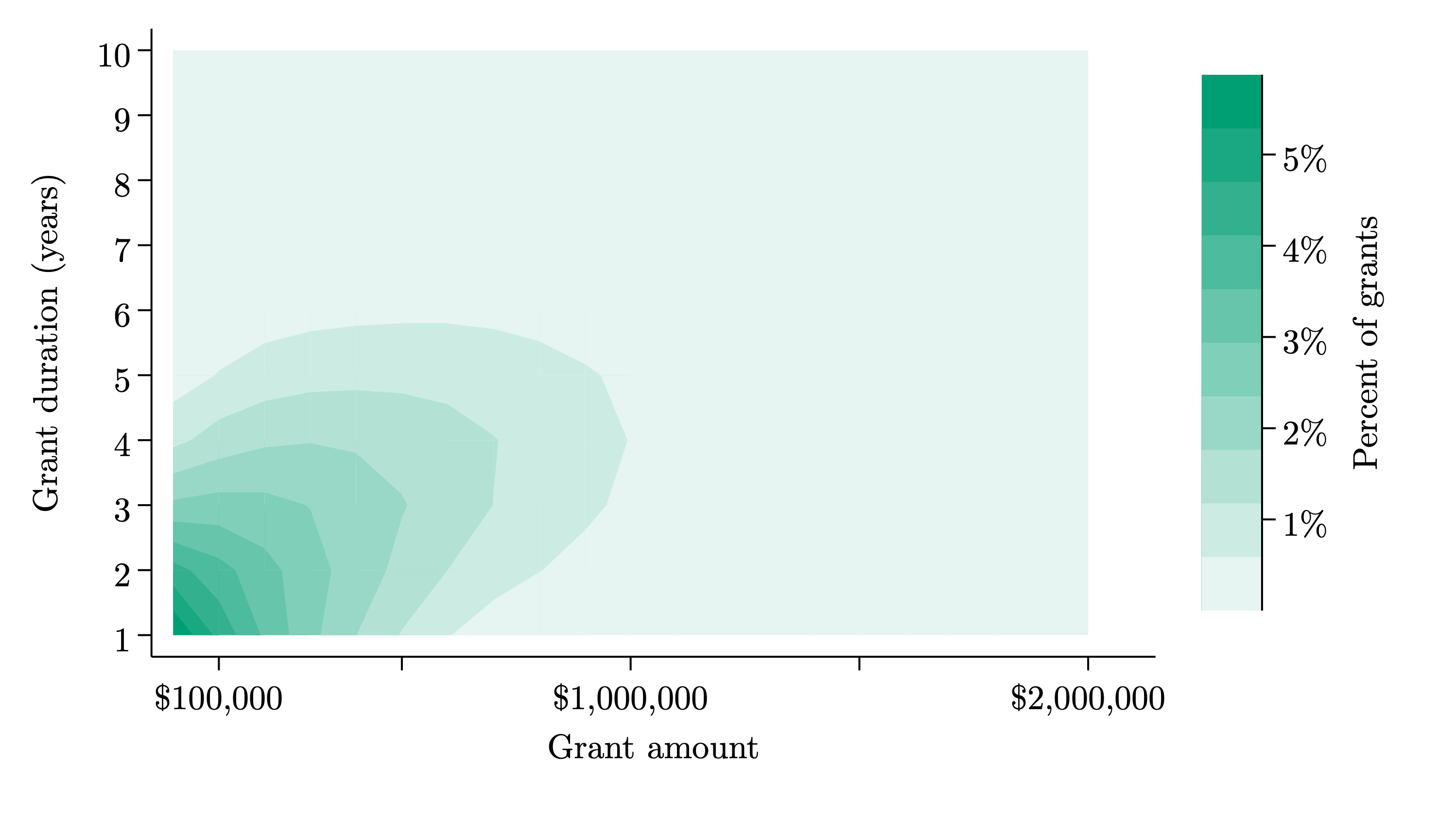}
\begin{quote} \footnotesize
\emph{Note}: \input{figtab/stub_funderdata_ingraph.tex} Data sourced from the Dimensions database (\citealt{dimension2018data}).
\end{quote}
\end{figure}

Figure \ref{fig_dimension_heatmap} provide a new look at the distribution of grant designs across U.S. academic research institutions. It shows the joint distribution of the size and duration of research grants. For consistency, we restrict our attention to grants awarded to the institutions from which we sourced our survey population, which we detail in the next section. In short, this includes the 150 largest institutions of higher education in the US per the annual amount of R\&D funding flowing to the institutions.We start with all 311,477 grants in the Dimensions data from 2000--2012 awarded to any researcher at all institutions in our population. We then restrict our attention to grants from US-based funders (295,004) with funding amount data (259,358). Of these, 222,041 are within the support of our thought experiments: 10 years long or less, and \$2 million or less.
\hspace{-2mm}\footnote{We focus on 2000--2012 to avoid censoring since duration is measured ex-post in this data.}

A few patterns emerge. Grants longer than 6 years are exceptionally rare. Even small, long grants that might resemble some form of insurance are rare; less than 1.5\% of grants are longer than 6 years and smaller than \$1 million. Another clear pattern is the positive correlation between the size and duration of research grants. On average, a grant that is one year longer will award about \$500,000 more in total.\footnote{This also holds on a per-year basis; an additional year is associated with \$40,000 more per year.}

Overall, it is clear that the full space of reasonable grant designs has not yet been explored much. The patterns illustrated here may reflect some optimal staging of investments by funders looking to learn information about risky projects (\citealt{dixit1994investment}). However, it may also reflect a missing market for certain grants, possibly driven by the fiscal costs of long-term outlays or rules and regulations that restrict the duration of financial arrangements between funders and grantees.\footnote{For example, see \href{https://grants.nih.gov/grants/policy/nihgps/html5/section_5/5.3_funding.htm}{here} for the NIH's policy that limits grant ``segments'' to a maximum of 5 years.}

Below, we use our thought experiments to understand what might happen to the type of science and composition of researchers if funders start to change this distribution. Also, we use this observed data to make some novel inferences about funders' preferences over grant designs and compare those to our estimates of researchers' preferences.

\section{Survey, population, and sample statistics}\label{sec_survey}
\subsection{Population and sampling}
We make use of the National Survey of Academic Researchers, which is detailed in \cite{myers2023new}. We provide a brief overview of the survey methodology here. The target population is US professors who conduct research at major institutions of higher education. This was formalized by selecting approximately 150 of the largest institutions in the US per their total R\&D funding reported in the National Science Foundation's 2019 Higher Education R\&D (HERD) survey (\citealt{nsf2023herd}) and manually collecting professors' information from public websites. Recruitment e-mails were sent to a random 50\% of the population from October 2022 to March 2023.

The population consisted of 264,036 unique e-mails. We e-mailed a total of 131,672 individuals and 4,388 (3.33\%) completed the survey.\footnote{This response rate is more than twice what has been obtained from sourcing academic researcher contacts from the corresponding author data contained within the publication record (e.g., \citealt{myers2020unequal}).} We then restrict the sample to the 4,175 individuals (95.1\% of respondents) who reported being a professor and spending a non-zero amount of time on research. Our final sample consists of professors from engineering, math, and related fields (703), humanities and related fields (786; e.g., history, linguistics), medical sciences (1,165; e.g., schools of medicine or public health), natural sciences (655; e.g., biology, chemistry, physics), and social sciences (866; e.g., economics, psychology, sociology).
We use these four aggregate groupings of fields throughout our discussion and empirical analyses given the small sample sizes within narrower field definitions. Table \ref{tab_survsumstat} reports some key summary statistics for our sample. The summary statistics related to the thought experiments are reported in the sections below, and Appendix \ref{sec_app_addstat_allxsumstat} contains a larger table with all covariates in the survey used in any of our analyses.

\begin{table}[ht]\centering \small
\caption{Summary statistics --- survey sample}\label{tab_survsumstat}
\-\\
{
\def\sym#1{\ifmmode^{#1}\else\(^{#1}\)\fi}
\begin{tabular}{l*{1}{rrrrr}}
\hline\hline
                    &       count&        mean&          sd\\
\hline
\underline{\emph{Broad field} \{0,1\}}&            &            &            \\
Engineering, math, \& related&       4,175&        0.17&        0.37\\
Humanities \& related&       4,175&        0.19&        0.39\\
Medical \& health sciences&       4,175&        0.28&        0.45\\
Natural sciences    &       4,175&        0.16&        0.36\\
[0.5em] \underline{\emph{Research}}&            &            &            \\
Guaranteed funding  &       4,175&  430,699.40&1,085,486.21\\
Fundraising expectations&       4,175&  549,238.32&1,055,926.36\\
[0.5em] \underline{\emph{Professional}}&            &            &            \\
Has tenure \{0,1\}  &       4,175&        0.59&        0.49\\
Total salary        &       4,175&  156,820.36&   93,741.46\\
Soft-money salary   &       4,175&   25,336.53&   45,407.18\\
Total work hrs./week&       4,175&       49.88&       13.06\\
Share of time spent on research [0,1]&       4,175&        0.39&        0.20\\
Share of time spent on fundraising [0,1]&       4,175&        0.09&        0.11\\
[0.5em] \underline{\emph{Socio-demographic}}&            &            &            \\
Age                 &       4,084&       49.31&       12.52\\
Female \{0,1\}      &       3,992&        0.41&        0.49\\
Non-white race/ethnicity \{0,1\}&       4,110&        0.24&        0.43\\
Num. dependents in household&       4,104&        0.98&        1.13\\
U.S. immigrant \{0,1\}&       4,088&        0.27&        0.44\\
\hline\hline
\end{tabular}
}

\begin{quote} \footnotesize
\emph{Note}: ``Guaranteed'' research funding includes funds from prior awards and any other guaranteed sources over the next 5 years; ``Fundraising expectations'' are researchers' beliefs about how much funds they will obtain from non-guaranteed sources over the next 5 years. All variables are continuous and bounded below at zero unless otherwise specified. Responses to socio-demographic questions were not mandatory, hence the differences in the count of observations.
\end{quote}
\end{table}

It is important to demonstrate that survey participants have some demand for funding for their research since the experiments are irrelevant to any researcher with no demand for funding (e.g., their only input is their own effort). To illustrate this, Appendix Figure \ref{bigfig_histfund_agg} plots the distribution of expected research funding over the five years from the time of survey forward for each of the four broad scientific fields. This is the sum of guaranteed funding the researcher has from either prior awards or guaranteed future funding flows plus their expected fundraising over the same period. The vast majority ($>$75\%) of researchers in each field have non-zero funding expectations. Furthermore, each distribution shows significant support over the range studied in the experiments below; research grants are important across all of science.

\subsection{Addressing potential survey biases}

\subsubsection*{Attention and representativeness}
Ideally, our respondents would report all answers truthfully and these responses would reflect the full population in terms of their preferences over, and responses to, different grant designs. We can never formally test this, but we can take some steps to investigate the possibility of inattention and non-response bias and, in the case of non-response bias, possibly account for it. 

As a test of respondents' attention and their willingness to report truthfully, \cite{myers2023new} compares respondents' self-reported salaries to their publicly-reported salaries for the subset of researchers at institutions that make such data public. The two show a close degree of alignment with a correlation of about 0.75, and the difference between the self- and publicly-reported salary is less than 30\% for roughly three-quarters of observations.\footnote{Discrepancies between self- and publicly-reported salaries can be due to a combination of recall error, inattention, or the time lag between the publicly-reported salary and when the survey was taken.} This suggests the vast majority of respondents are responding truthfully along this dimension. 

\cite{myers2023new} also reports two comparisons of the population and sample professors using observable data on institutional funding levels and professor-level grant funding and publication output. In both cases, the respondent sample is quite similar to the population. The average difference in funding amounts between the institutions of respondents and non-respondents is generally in the range of 4--6\%, and there are no statistically significant differences in the grant funding or publication output metrics between the two groups. Some of these exercises are replicated in Appendix \ref{sec_app_addstat_popsampcompare}Overall, our respondent sample appears very similar to the population along many observable dimensions.

\subsubsection*{Sample selection correction}

Despite the overlap in observable data, our respondents may still differ on other unobservable dimensions. If these unobservable differences are related to our focal parameters of interest (e.g., researchers preferences over grant designs), this may generate a non-response bias. To account for this possibility, we implement the sample selection correction approach first developed by \cite{heckman1979sample}. This approach makes use of excluded variables that cause entry into the sample (i.e., completion of our survey), but are plausibly orthogonal to the parameters of interest.

We generated the variation necessary for this approach with randomized recruitment strategies. During the recruitment process, we randomly allocated each e-mail to one of four different incentive arms and one of three different reminder arms. The four incentive arms were: (1) no incentive, (2) entry into a lottery to win a gift card, (3) the ability to vote for a set of charities to receive a donation, and (4) both the second and third incentives. The three reminder arms involved no, one, or two reminders, respectively.

Appendix \ref{sec_app_addstat_incentives} reports the effect of the incentives and reminders. We find the stronger incentives lead to a small increase in completion rates, while the reminders are much more powerful. Both induce a significant increase in completion rates; the completion rate for researchers that receive no incentives and no reminders is roughly 1.9\% whereas the completion rate for those who receive both incentives and both reminders is roughly 4.5\%.

Using the randomized variation from these incentives, combined with the information on each professor's rank, institution, and inferred field of study, we then estimated individuals' propensity to complete the survey conditional on being sampled. From these predicted probabilities, we constructed the inverse mills ratio and include this variable in all regressions. Assuming that that researchers responsiveness to either participation instrument is orthogonal to a focal parameters, this will control for any unobservable differences (in a focal parameter) between our sample and the population (\citealt{heckman1979sample}).\footnote{Obviously, this approach does sacrifice some potential sample size. But using the full set of reminders and incentives in all recruitments would have eliminated our ability to implement any correction for potential non-response bias.}

\section{Grant design and research strategy}\label{sec_te}

\subsection{Theoretical framework}
How might grant design affect researchers' strategies? To begin, it is useful to map the two focal grant design parameters, money and time, to the two levers that much of the literature on managing innovation has focused, financial slack and long-term incentives. 

In many contexts, financial slack is often operationalized as free cash flow. Here, in the setting of academic research grants, the clear analogue is the amount of funding. Larger grants can enable the acquisition of more scientific inputs because funding for one's research is quite costly to obtain. Academic researchers spend an average of about 15\% of their time fundraising for their work (\citealt{myers2020unequal}), and success rates at major funding agencies are on the scale of 10--20\% (\citealt{devrieze2017new}).

Long-term incentives are often operationalized as the duration of employment contracts, the time-horizon of stock options, or the use of ``golden parachutes''. Here, the analogue is the duration that grant funds are made available. Longer grants can change researchers' expectations and uncertainty about their future funding, which may have important consequences. For example, \cite{tham2023science} and \cite{cheng2023effect} show how interruptions to researchers' funding expectations can disrupt their spending flows and their ability to retain skilled labor in their laboratories.

In what follows, we outline the strategic dimensions along which these parameters of grant design may be instrumental. We don't take a stance on the social value of any of these dimensions or the degree to which the private and social value may diverge. Rather, our goal with the following experiment is to understand the extent to which policymakers or managers might incentivize particular strategies with particular grant designs.

\textbf{Long-term incentives, risk-taking, and speed.} Principal-agent and multi-tasking models of innovation predict that long-term reward structures can induce risk-taking (i.e., increasing the expected variance of outcomes) at the potential cost of shirking or other forms of negative selection (\citealt{holmstrom1989agency,manso2011motivating,hellmann2011incentives,nanda2017innovation}). This prediction has been confirmed in laboratory experiments (\citealt{ederer2013pay}) and observed in natural experiments amongst for-profit firms in R\&D-intensive sectors (\citealt{lerner2007innovation,aghion2013innovation,tian2014tolerance}). The results of \cite{azoulay2011incentives} are consistent with this prediction where HHMI-funded scientists (who have longer guaranteed funding arrangements) produce more very-high and very-low cited publications, which is evidence of a riskier strategy. The extent to which longer grants might induce risk-taking compared to reducing the speed with which researchers work is unclear. 

\textbf{Financial slack, exploration, and exploitation.} Individuals and organizations that are financially constrained and face costly external finance are expected to invest less in novel opportunities that are less related to their prior pursuits (\citealt{hall2002financing,brown2009financing,hottenrott2012innovative}). A recent, clear example of the corollary is \cite{krieger2022missing} who show that when pharmaceutical firms experience cash windfalls they are more likely to explore novel molecular therapies. \cite{azoulay2011incentives} is also consistent with this prediction, finding that the HHMI-funded scientists (who receive larger funding amounts) are more likely to change the direction of their research. But the extent to which this could be attributed to their funding levels per se is unclear.

\textbf{Constraints and replicability.} Models of racing in science suggest that the priority reward structure has the potential to incentivize low-quality pursuits of high-value research questions (\citealt{bryan2017direction}). This sort of competitive pressure appears to be relevant in practice, especially when researchers are financially- or time-constrained, leading to replication concerns, biased statistical reporting, and low quality experimental designs (\citealt{pashler2012replicability,smaldino2016natural,andrews2019identification,hill2021race,bryan2022r}). There continue to be numerous calls for new institutions and incentives to address these issues,\footnote{For example, see \href{https://www.thelancet.com/journals/lancet/article/PIIS0140-6736(15)60696-1/fulltext}{this}, \href{https://www.newscientist.com/article/mg25433810-400-the-replication-crisis-has-spread-through-science-can-it-be-fixed/}{this}, or \href{https://www.vox.com/future-perfect/23489211/replication-crisis-project-meta-science-psychology}{this} commentary and coverage.} but whether grants with certain designs may be able to counteract this pressure remain unclear.

There is certainly heterogeneity across professors in terms of the relevance of grants for their work, as well as the institutions and incentives that may modulate the effect of grant designs. In this vein, the thought experiment is designed to be as discipline-agnostic as possible, and our analyses make full use of modern machine learning techniques to search for heterogeneous treatment effects.

\subsection{Experimental design and summary statistics}
We ask researchers how they would change their research strategies if they received some hypothetical grant, the size and duration of which is randomized. The value and duration of this hypothetical grant are randomized over \{\$100,000, \$250,000, \$500,000, \$1,000,000, \$2,000,000\} and \{2, 3, 4, 5, 6, 7, 8, 9, 10\} years, respectively.

If grant attributes can be used to modulate the scientific production function, we should see researchers systematically changing their strategies when receiving grants with different designs. To capture this, we present researchers with five possibilities of how they might change their strategies: ``Pursue riskier projects'', ``Increase speed'', ``Pursue projects less related to your current work'', ``Increase the size of ongoing projects'', and ``Increase accuracy or reliability''. We ask them to select the two most important changes the grant would enable them to make in their research, requiring two choices to be made.

The first two options (``Pursue riskier projects'', ``Increase speed'') are motivated by the potential relationship between risk-taking (i.e., increasing the expected variance of outcomes) and pace-of-work underlying the aforementioned principal-agent models of innovation. The next options (``Pursue projects less related to your current work'', ``Increase the size of ongoing projects'') are motivated by the potential explore-versus-exploit decisions underlying the aforementioned models of financial constraints. The last option (``Increase accuracy or reliability'') is motivated by the fact that, while problems with reliability and replication in science have become increasingly publicized, the extent to which policy levers like grant designs might be able to address this problem remains unclear.\footnote{We lack any ground truth with which to validate the descriptions of these strategies; however, their phrasing was informed by numerous pilot interviews with researchers across a variety of fields.}

Validated, ex-ante measures of strategy in the context of science do not yet exist, but each of these options are clearly motivated by prior studies of researchers. The obvious limitation here is that these are subjective descriptions of a subset of all possible strategies researchers could undertake. Still, the traditional approach of making inferences from researchers' output (e.g., publications) have their own set of drawbacks,\footnote{Those measures are ex-post realizations and may suffer from a truncation bias when failed projects do not produce observable output. For example, see \cite{franzoni2022uncertainty} for a discussion of the difficulties of quantifying risk-taking in science.} and so we view this approach as a new, complementary one.

Table \ref{tab_stratsumstat} reports the summary statistics for the randomized and response variables in this thought experiment. 

\begin{table}\centering \small
\caption{Summary statistics -- strategy thought experiment}\label{tab_stratsumstat}
\-\\
{
\def\sym#1{\ifmmode^{#1}\else\(^{#1}\)\fi}
\begin{tabular}{l*{1}{rrrrr}}
\hline\hline
                    &       count&        mean&          sd&         min&         max\\
\hline
\underline{\emph{Randomized elements}}&            &            &            &            &            \\
Grant duration (years)&       4,175&        6.06&        2.56&           2&          10\\
Grant amount (\$)   &       4,175&  778,239.52&  684,075.23&     100,000&   2,000,000\\
\\ \underline{\emph{Strategic changes \{0,1\}}}&            &            &            &            &            \\
Increase speed      &       4,175&        0.36&        0.48&           0&           1\\
Pursue riskier projects&       4,175&        0.53&        0.50&           0&           1\\
Pursue new directions&       4,175&        0.33&        0.47&           0&           1\\
Increase size of ongoing projects&       4,175&        0.61&        0.49&           0&           1\\
Increase accuracy   &       4,175&        0.17&        0.38&           0&           1\\
\hline\hline
\end{tabular}
}

\begin{quote} \footnotesize
\emph{Note}: Summary statistics of the two randomized elements of the research strategy thought experiment, and the respondent's answers to the question of ``What are the two most important changes this grant would enable you to make in your research?''.
\end{quote}
\end{table}

\subsection{Empirical approach}
To explore the effect of grant design on researchers' probability of choosing a particular strategy, we estimate five regressions of the form:
\begin{equation}\label{eq_Sreg}
Y^k_i = h^k(A_{i}, D_{i}, \mathbf{X}_i) \,\,,
\end{equation}
where the $Y$ is an indicator that equals one if strategy $k$ is chosen by the researcher in response to the randomized grant, the funding amount and duration are $A$ and $D$, and $\mathbf{X}$ is a vector of covariates solicited elsewhere in the survey (see Appendix \ref{sec_app_addstat} for a full listing of these covariates and their summary statistics). We explore a range of functional forms of $h$. In the simplest scenario, we test only the main effects of the randomized grant amount and duration and estimate five separate OLS regressions using each of the five indicators as the dependent variable. In the appendix, we also report results from OLS regressions after using Lasso to select controls from the rest of the survey data to possibly improve our precision as well as estimates from a single discrete choice model that jointly estimates how researchers' propensity to choose a particular strategy depends on grant attributes.

Because we force researchers to choose two of the strategic options, our analyses here are all focused on \emph{relative} shifts in the mix of strategies that researchers pursue after receiving a grant. That is, we are not estimating how researchers' total scientific output (e.g., publications) depends on the grant structure. Rather, we assume that researchers are always undertaking the same fixed ``amount of strategy'', and their decision is only about the mix of strategies to employ --- it is purely a directional measure.

For clarity, consider our OLS regression specification:
\begin{equation}\label{eq_SregOLS}
Y^k_i = \alpha^k + \beta^k_A A_{i} + \beta^k_D D_{i} + \epsilon^k_i \,\,,
\end{equation}
which we estimate for each of the five strategies indexed by $k$. Written this way, our coefficient estimates ($\beta^k_A,\beta^k_D$) will sum to zero across the five models (i.e., $\sum_k \beta^k_A=0$); again, we are only estimating how grant design leads to substitution across these options. Positive values of our coefficients indicate that, as grants get larger or longer, researchers increasingly prioritize some strategy because, with those additional resources, that strategy has become marginally more useful to the researcher. Conversely, negative values indicate which strategies researchers substitute away from as grants get larger or longer. Our statistical tests of the null are therefore tests of whether or not researchers systematically substitute to or away from particular strategies.

We are particularly interested in the possibility of heterogeneous responses given the idiosyncrasies of each researchers' situation. To do this, we use the causal forests algorithm developed by \cite{wager2018estimation} to estimate conditional average partial effects (CAPEs) of the grant attributes. This approach yields informative distributions of effects, but do not directly reveal the features of researchers associated with larger or smaller CAPEs. To understand the heterogeneity in a more low-dimensional manner, we then use the Best Linear Projection methodology to test how the CAPEs vary with a select number of covariates (\citealt{chernozhukov2018generic}).\footnote{All of these analyses are performed in \texttt{R} using the \texttt{grf} package (\citealt{tibshirani2022grf}).}

\subsection{Results}
\begin{figure}[htbp]\centering \small
\caption{Heterogeneous effects of grant designs on researchers' strategies}\label{fig_capes_xp1}
\subfloat[Faster]{\label{fig_capes_speed_xp1}\includegraphics[width=0.475\textwidth, trim=0mm 0mm 0mm 0mm, clip]{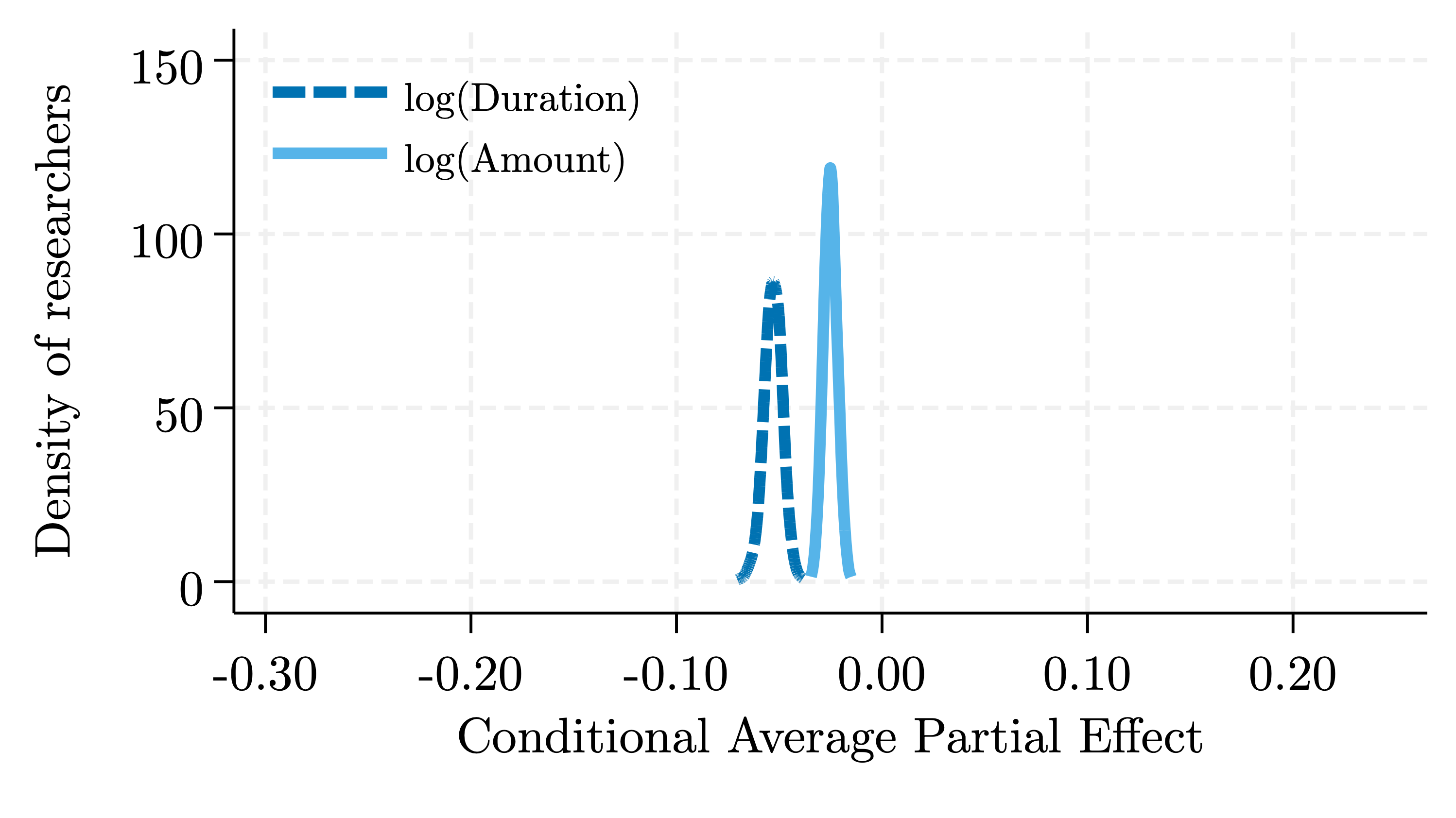}}
\subfloat[More risk]{\label{fig_capes_riski_xp1}\includegraphics[width=0.475\textwidth, trim=0mm 0mm 0mm 0mm, clip]{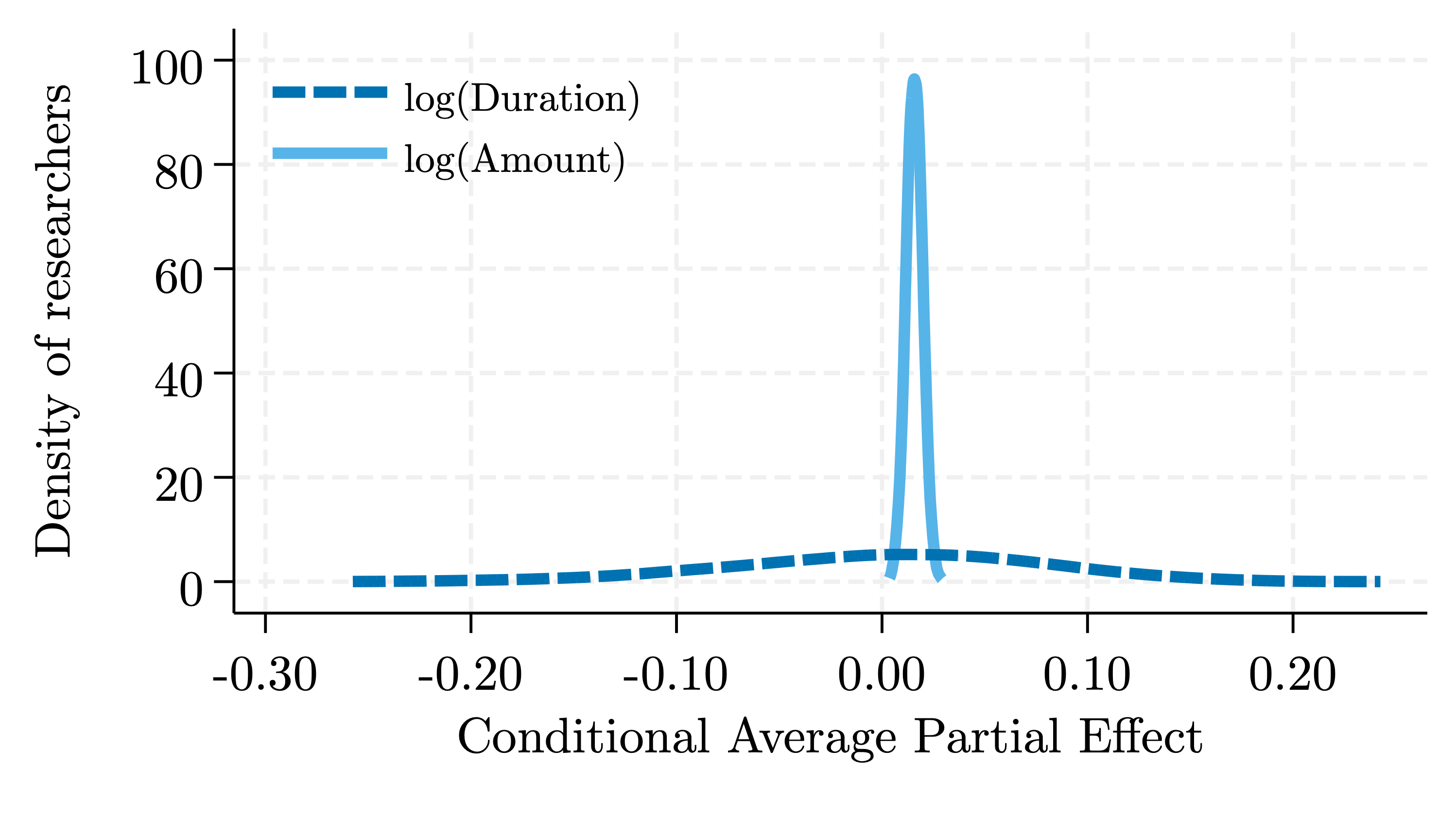}}\-\\\-\\
\subfloat[New directions]{\label{fig_capes_newdi_xp1}\includegraphics[width=0.475\textwidth, trim=0mm 0mm 0mm 0mm, clip]{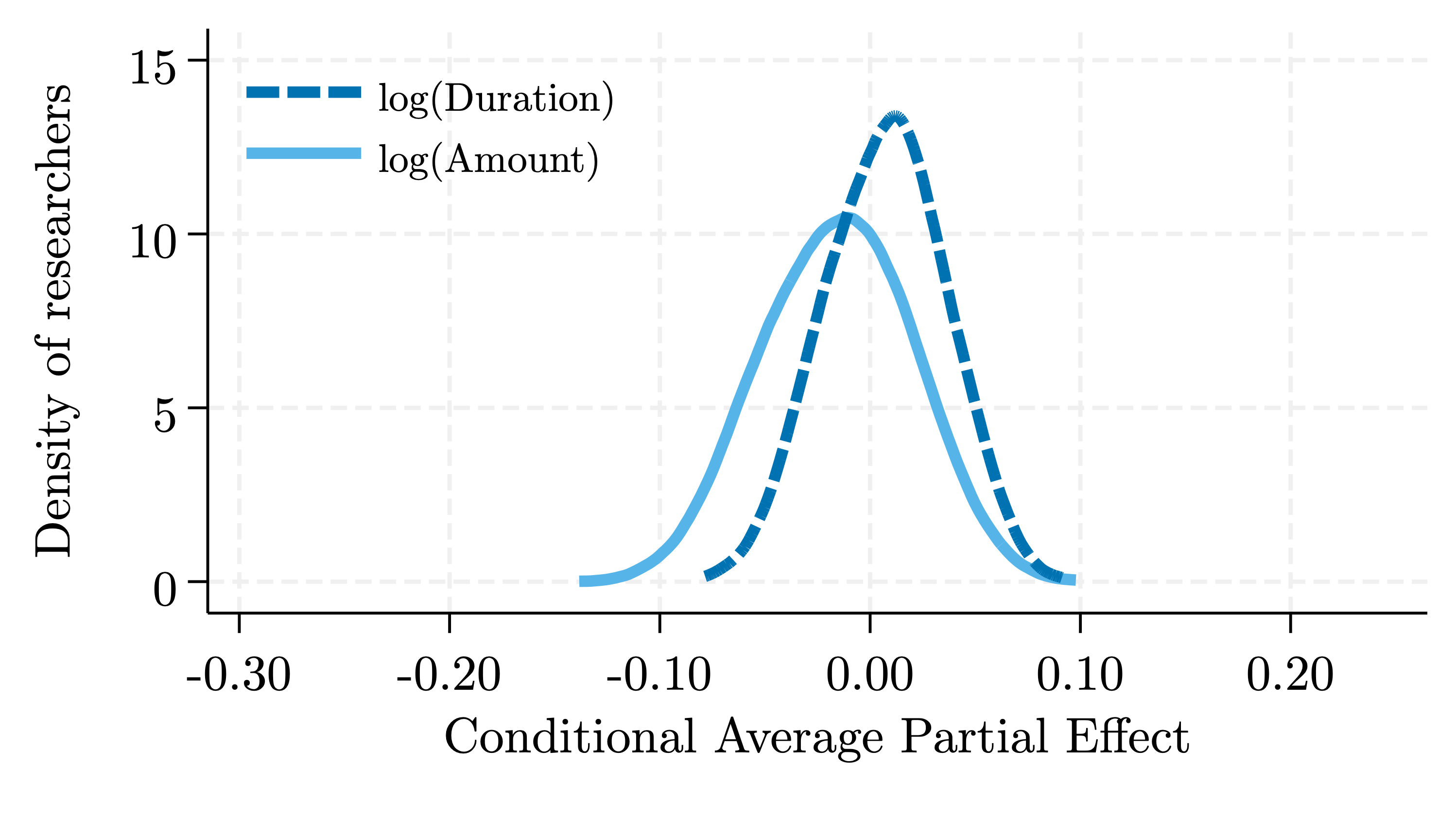}}
\subfloat[Larger ongoing projects]{\label{fig_capes_sizeo_xp1}\includegraphics[width=0.475\textwidth, trim=0mm 0mm 0mm 0mm, clip]{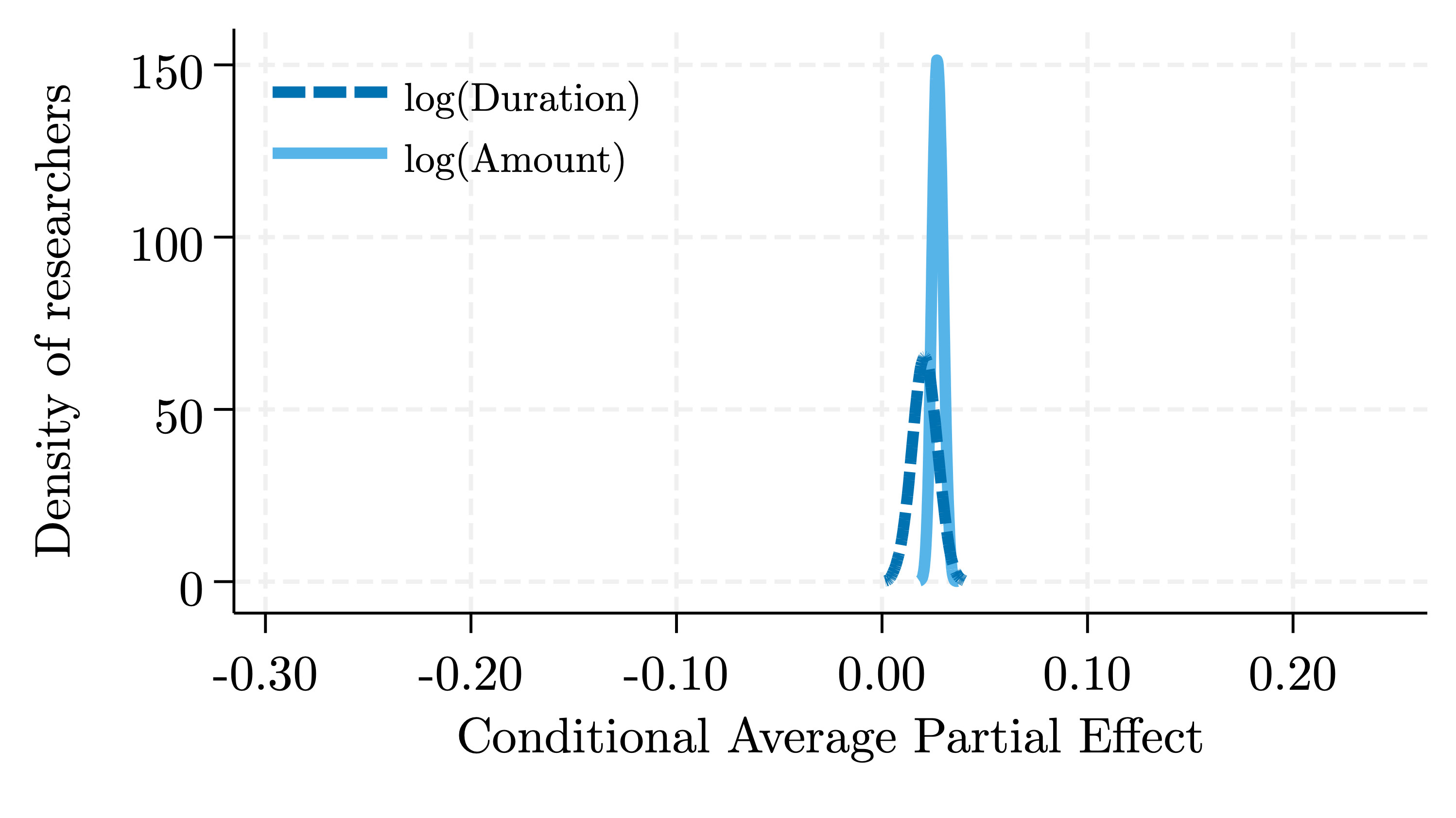}}\-\\\-\\
\subfloat[More accurate]{\label{fig_capes_accur_xp1}\includegraphics[width=0.475\textwidth, trim=0mm 0mm 0mm 0mm, clip]{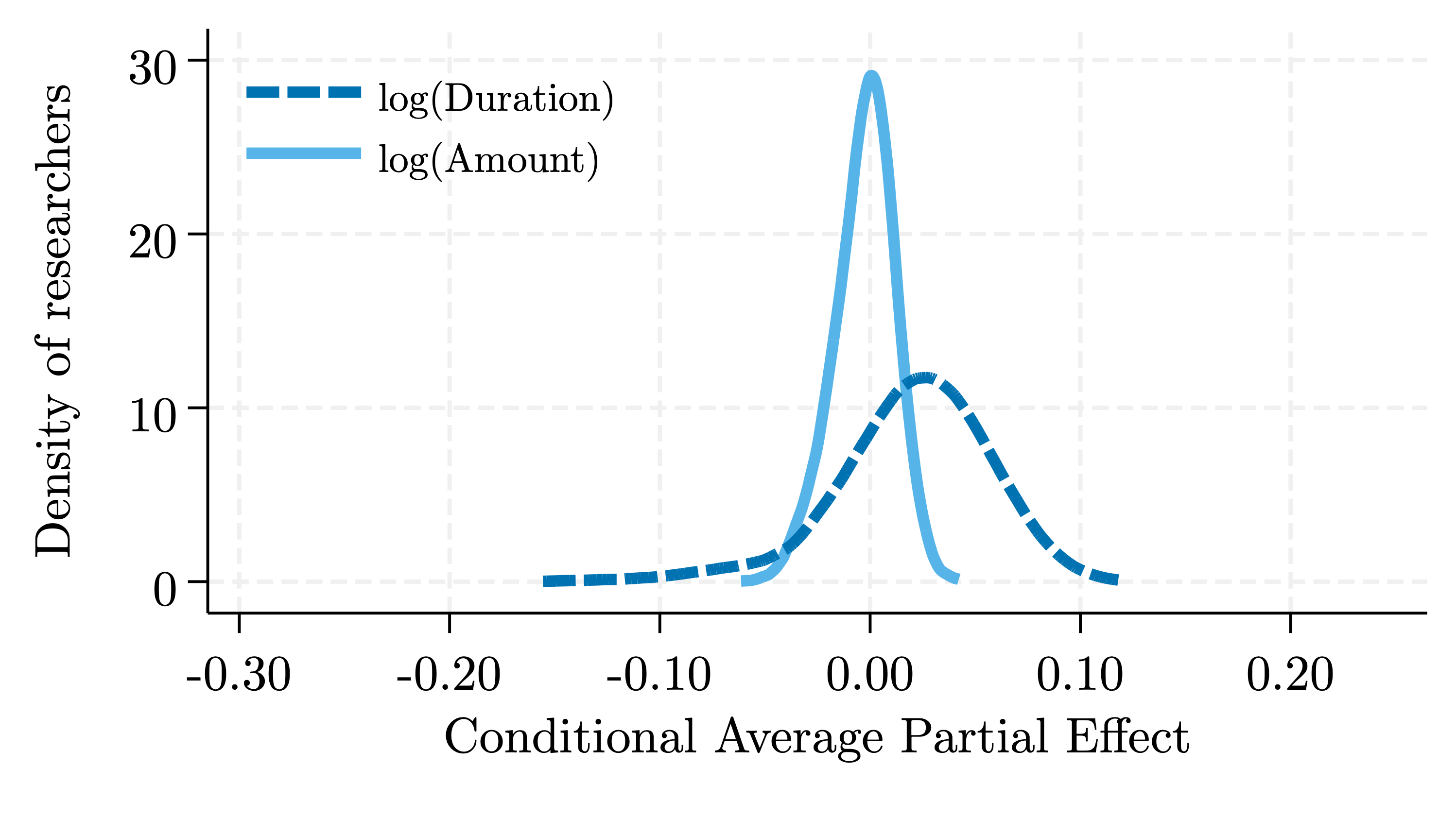}}
\begin{quote} \footnotesize
\emph{Note}: Each panel shows the distribution of the Conditional Average Partial Effects (CAPEs), which are estimated separately for the two randomized grant attributes using the causal forest method of \cite{wager2018estimation}; the CAPE is defined as $E[cov[W, Y | X] / var[W | X]]$, where $W=\{\text{log(duration), log(size)}\}$, $Y$ is the dependent variable, and $X$ are the covariates. The only two distributions that show statistically significant heterogeneity are (1) the effect of grant duration on risk-taking ($p\input{./figtab/stub_capeomnibustesthet_riski_len_xp1.tex}$) in Panel (b), and (2) the effect of grant funding amount on undertaking new directions ($p\input{./figtab/stub_capeomnibustesthet_newdi_siz_xp1.tex}$) in Panel (c). 
\end{quote}
\end{figure}

Table \ref{tab_grantstrat_all} reports our simplest version of the analyses where we conduct five separate OLS regressions of strategic choice indicators on the grant attributes.\footnote{Since the independent variables are identical in each regression, estimating each OLS regression separately is identical to joint estimation via , for example, seemingly-unrelated-regression approach.} We report conventional robust standard errors as well as the family-wise $p$-values, which adjust for the fact that we are testing ten hypotheses (five outcome variables and two independent variables).\footnote{We make use of \citeauthor{jones2019workplace}'s (\citeyear{jones2019workplace}) implementation of the free step-down procedure of \cite{westfall1993resampling} as codified in their Stata package \texttt{wyoung}.} Appendix \ref{sec_app_strat} reports the results from a logit choice model that jointly estimates researchers' propensities to choose each strategy. It more closely reflects the structure of the choice facing the respondent, but it yields very similar results. Likewise, the post-Lasso OLS regressions reported there, where covariates are included to potentially improve precision, also show very similar results. Figure \ref{fig_capes_xp1} reports the results from using causal forests to estimate the Conditional Average Partial Effect (CAPE) distributions.\footnote{The means of each CAPE reported in Figure \ref{fig_capes_xp1} are approximately equal to the mean effects reported in Table \ref{sec_app_strat}.}

Three patterns emerge, which we investigate and discuss further in the sub-sections below. First, we find that larger and longer grants lead researchers to prioritize speed less, but that such grants can also increase researchers' willingness to take risks. Second, as grant size increases, researchers shift their strategic focus \emph{towards} increasing the size of ongoing projects and \emph{away} from pursuing work in new directions. In other words, smaller grants appear to promote exploration and larger grants appear to promote exploitation. Third, grant designs appear to have no effect on researchers' focus on the accuracy of their science.

\begin{table}[htbp]\centering \small
\caption{Effect of grant designs on researchers' strategies}\label{tab_grantstrat_all}
\-\\
{
\def\sym#1{\ifmmode^{#1}\else\(^{#1}\)\fi}
\begin{tabular}{l*{9}{c}}
 \hline\hline
& & & & & & &Larger & & \\
& &  &More & &New & &ongoing & &More \\
&Faster & &risk & &directions & &projects &  &accurate \\
                    &\multicolumn{1}{c}{(1)}         &            &\multicolumn{1}{c}{(2)}         &            &\multicolumn{1}{c}{(3)}         &            &\multicolumn{1}{c}{(4)}         &            &\multicolumn{1}{c}{(5)}         \\
\hline \\
log(Duration)       &      --0.056\sym{***}&            &       0.012         &            &       0.008         &            &       0.018         &            &       0.018         \\
                    &     (0.015)         &            &     (0.016)         &            &     (0.015)         &            &     (0.015)         &            &     (0.011)         \\
[0.5em]
log(Amount)         &      --0.025\sym{***}&            &       0.017\sym{**} &            &      --0.016\sym{**} &            &       0.026\sym{***}&            &      --0.003         \\
                    &     (0.007)         &            &     (0.007)         &            &     (0.007)         &            &     (0.007)         &            &     (0.006)         \\
[0.5em]
\hline
\underline{Family--wise $ p$--values} \\ Duration&    $ <$0.01         &            &        0.83         &            &        0.85         &            &        0.61         &            &        0.49         \\
Amount              &    $ <$0.01         &            &        0.12         &            &        0.12         &            &    $ <$0.01         &            &        0.85         \\
\hline dep. var. mean&        0.36         &            &        0.53         &            &        0.33         &            &        0.61         &            &        0.17         \\
$ N$ obs.           &       4,175         &            &       4,175         &            &       4,175         &            &       4,175         &            &       4,175         \\
\hline\hline
\end{tabular}
}

\begin{quote} \footnotesize
\emph{Note}: Shows results from OLS regressions where the dependent variables are indicators that equal one if the strategy listed at the top of the column was chosen as one of the ``two most important changes'' that researchers would make in response to receiving a grant of a given (randomized) size and duration. Robust standard errors shown in parentheses; stars indicate significance levels using the unadjusted $p$-values: $^{*}$ \(p<0.1\), $^{**}$ \(p<0.05\), $^{***}$ \(p<0.01\). The family-wise $p$-values adjust for the ten hypotheses tested and are based on 10,000 bootstraps of the free step-down procedure of \cite{westfall1993resampling}.
\end{quote}
\end{table}

\subsubsection*{Speed versus risk}
The first column of Table \ref{tab_grantstrat_all}, clearly shows that larger and longer grants lead to an decreased emphasis on speed (equivalently, smaller and shorter grants lead to an increased emphasis on speed). This result is intuitive, but by no means guaranteed given the wide array of time horizons over which researchers are likely optimizing (e.g., due to the heterogeneity in tenure status, contract lengths, ages, etc.). The socially optimal pace of science is far from clear given that there are good reasons both that sciences moves ``too fast'' (e.g., as in the racing models of \citealt{loury1979market}, which has some empirical support from \citealt{bryan2022r}), and ``too slow'' (e.g., as is emphasized by \citealt{nanda2017innovation} who study how the provision of long-term incentives affects the extensive margin of innovation). Still, it does suggest that grant design is a tool for manipulating researchers' speed.

The large body of work showing how long-term incentives can induce risk-taking in firms (\citealt{lerner2007innovation,aghion2013innovation,tian2014tolerance}) and the analogous results of \cite{azoulay2011incentives} suggest that we should find some increased willingness to take risks when receiving longer grants. But we do not see an average effect of this sort (see Column 2 of Table \ref{tab_grantstrat_all}).

However, the CAPE distribution in Figure \ref{fig_capes_xp1} Panel (b) clearly shows a large degree of heterogeneity in researchers' responsiveness to grant duration with respect to risk-taking. This particular relationship is one of only two CAPE distributions that represent statistically significant degrees of heterogeneity. To better understand this heterogeneity, Appendix \ref{sec_app_strat_blp} reports the Best Linear Predictors of the CAPEs. The best predictor of researchers' responsiveness to grant duration with respect to risk-taking is their tenure status --- the effect of a log-point increase in grant duration on researchers' willingness to take risks is roughly 10 percentage points larger for tenured professors compared to nontenured professors (compared to statistically insignificant mean effect of 0.01). This suggests that grant duration may be a useful incentive for risk-taking as the aforementioned theories would suggest, but only amongst researchers who have the complementary job security, reputation, or resources associated with tenure. Conversely, nontenured researchers may be facing performance evaluations on timescales that render marginal changes in grant duration irrelevant.

This result is consistent with \cite{azoulay2011incentives}, in that their sample consists entirely of elite researchers who likely already have, or are clearly on a path, to a tenured or de-facto tenured positions but are still on the younger end of the age spectrum. Our findings help guide the generalizability of those results and identify the conditions under which researchers might be incentivized to take risks with longer grants.

\subsubsection*{Exploration versus exploitation}
Prior work focusing primarily on science in for-profit firms would suggest that an important mechanism for shifting researchers' decisions along the explore-exploit axis would be the size of the grant. Specifically, empirical studies have shown that when firms' access to capital increases or they receive a windfall of cash, they undertake more exploratory research trajectories in new directions (\citealt{krieger2022missing}). We find something close to the opposite. 

In our setting, larger grants lead researchers to increase the size of ongoing projects at the expense of moving in new directions --- that is, as funding amounts increase, researchers shift from using those funds to explore to using them to exploit. Equivalently, as grants get smaller, researchers are more likely to use those smaller amounts of funds to explore new directions.

There are many ways this could be rationalized. 
One rationalization would be that starting new projects involves high fixed costs due to the scarcity of specialized inputs. That is, if there are certain inputs that researchers need specifically for the purposes of starting new projects (e.g., a new postdoc to join their laboratory) and these inputs are scarce enough, researchers may be forced to commit larger grants to ongoing projects. The net returns to large investments in new directions may be small because the costs are higher (e.g., it is worth spending a small amount of money to explore a new direction, but scaling this new-direction project quickly becomes expensive because of the scarce inputs required). We return to this idea below when we examine how grant designs influence researchers' choices of inputs to acquire with the grant.

\subsubsection*{Accuracy and reliability}
The least popular strategic choice overall is the \emph{increase accuracy or reliability} option, and there are no average or heterogeneous effects to suggest that grant design can shift researchers' choices along this dimension. The importance of preventing and detecting false positives and negatives in science has come to be appreciated (e.g., \citealt{pashler2012replicability,andrews2019identification}). Our results suggest the reason researchers may be underinvesting in efforts related to accuracy or reliability is \emph{not} because they lack the resources to do so, but is because those efforts are underincentivized in the market for science.

\subsection{Grants and input allocations}
In the preceding analyses, we find that grant design can affect researchers' strategies (i.e., grant size influencing the choice of new directions versus larger ongoing projects). How much of this responsiveness is due to the grants allowing researchers to obtain new inputs for their work? Or, more generally, how much might grant design influence researchers' input choices?

A full investigation into such questions is beyond the scope of this paper, but we did solicit a second set of outcomes in this thought experiment that allow us to estimate how grants with different designs are used. In the survey, after respondents were asked to report how their strategies would change in response to receiving the hypothetical grant, we asked them to report how they would spend those funds across five different input categories: (1) their own salary; (2) their own training; (3) travel; (4) equipment, data, or supplies; and (5) salaries or wages for other new or current researchers, students, or staff.

Pilot testing with researchers indicated that soliciting precise allocations across these categories (e.g., percentage of dollars spent) was an excessive burden and would likely involve a tremendous amount of uncertainty since most researchers do not have specific knowledge about input costs. As a lower-burden alternative, we simply asked respondents to indicate if they would spend ``None'', ``Some'', or ``A lot'' of the hypothetical grant funds on each category. Pilot testing indicated that researchers were much more comfortable with this format.

To conduct empirical analyses of the responses, we code the three possible responses to values of 0,1, and 2. Then, we convert these values into shares by dividing each response by the sum of the responses across all categories (e.g., if a researcher reports ``Some'' for all categories, then all categories will receive a 0.2 share of the funding from the hypothetical grant). As with our analyses of researchers' strategic responses, we are less concerned with the precise magnitudes of the effects and instead our focus is on testing whether we can reject the null in particular directions. 

\begin{table}[ht] \centering \small
\caption{Summary statistics --- input allocations}\label{tab_inputsumstat}
\-\\
{
\def\sym#1{\ifmmode^{#1}\else\(^{#1}\)\fi}
\begin{tabular}{l*{1}{rrrrr}}
\hline\hline
                    &       count&        mean&          sd\\
\hline
\underline{\emph{Randomized elements}}&            &            &            \\
Grant duration (years)&       4,175&        6.06&        2.56\\
Grant amount (\$)   &       4,175&  778,239.52&  684,075.23\\
\\ \underline{\emph{Input allocation share [0,1]}}&            &            &            \\
Own salary          &       4,175&        0.18&        0.13\\
Own training        &       4,175&        0.08&        0.10\\
Travel              &       4,175&        0.18&        0.10\\
Equipment or supplies&       4,175&        0.21&        0.13\\
Others' salaries    &       4,175&        0.34&        0.14\\
\hline\hline
\end{tabular}
}

\begin{quote} \footnotesize
\emph{Note}: Summary statistics of the two randomized elements of the research strategy thought experiment (duplicating the first two rows of Table \ref{tab_stratsumstat}) and respondents' allocation of funding amounts across input categories. Input allocation responses are standardized such that input shares sum to one.
\end{quote}
\end{table}

Table \ref{tab_inputsumstat} shows the summary statistics for the share of funding that researchers allocate to each of the five inputs. For simplicity, we report the results from five separate OLS regressions using the input shares for each category as the dependent variables in Table \ref{tab_grantinput_all}.\footnote{Unreported results from joint estimation of all choices via a fractional response regressions with a logit model for the conditional mean yield very similar results.} Overall, grant duration has no significant effect on input allocations. But we do find grant size to have some clear effects.  

First, we see that larger grants lead researchers to allocate a greater share of funding towards both their own salary and their own training. By construction, this increase must come with a decrease elsewhere, and it appears that almost all of the decrease is occurring in the ``Others' salaries'' category. To be clear, this does not mean that researchers are claiming to reduce their total spending on this category. Instead, they are indicating that, as grants get larger in size, a smaller share of the additional funding will be spent on others' salaries or wages.

\begin{table}[ht]\centering \small
\caption{Effect of grant designs on researchers' input allocations}\label{tab_grantinput_all}
\-\\
{
\def\sym#1{\ifmmode^{#1}\else\(^{#1}\)\fi}
\begin{tabular}{l*{9}{c}}
 \hline\hline
& & & & & & & & & \\
 &Own & &Own & & & &Equipment & &Others' \\
 &salary & &training & &Travel & &or supplies & &salaries \\
                    &\multicolumn{1}{c}{(1)}         &            &\multicolumn{1}{c}{(2)}         &            &\multicolumn{1}{c}{(3)}         &            &\multicolumn{1}{c}{(4)}         &            &\multicolumn{1}{c}{(5)}         \\
\hline \\
log(Duration)       &      --0.007\sym{*}  &            &       0.000         &            &       0.002         &            &      --0.001         &            &       0.005         \\
                    &     (0.004)         &            &     (0.003)         &            &     (0.003)         &            &     (0.004)         &            &     (0.004)         \\
[0.5em]
log(Amount)         &       0.008\sym{***}&            &       0.008\sym{***}&            &      --0.001         &            &      --0.003         &            &      --0.012\sym{***}\\
                    &     (0.002)         &            &     (0.001)         &            &     (0.002)         &            &     (0.002)         &            &     (0.002)         \\
[0.5em]
\hline
\underline{Family--wise $ p$--values} \\ Duration&        0.37         &            &        0.97         &            &        0.93         &            &        0.97         &            &        0.68         \\
Amount              &    $ <$0.01         &            &    $ <$0.01         &            &        0.97         &            &        0.51         &            &    $ <$0.01         \\
\hline dep. var. mean&        0.18         &            &        0.08         &            &        0.18         &            &        0.21         &            &        0.34         \\
$ N$ obs.           &       4,175         &            &       4,175         &            &       4,175         &            &       4,175         &            &       4,175         \\
\hline\hline
\end{tabular}
}

\begin{quote} \footnotesize
\emph{Note}: Shows results from OLS regressions where the dependent variables are the share of the (randomized) grant that would be allocated to each input category. Robust standard errors shown in parentheses; stars indicate significance levels using the unadjusted $p$-values: $^{*}$ \(p<0.1\), $^{**}$ \(p<0.05\), $^{***}$ \(p<0.01\). The family-wise $p$-values adjust for the ten hypotheses tested and are based on 10,000 bootstraps of the free step-down procedure of \cite{westfall1993resampling}.
\end{quote}
\end{table}

We cannot test whether this shift in allocations is due to supply constraints (e.g., researchers cannot attract additional labor to work with them), or whether this is due to researchers' demand for labor shrinking as grant size increases (e.g., their own time is the main constraint that the increased funding allows them to address). 

The fact that we find larger grants increase researchers' spending shares on themselves (i.e., via their salary and training) and decrease spending shares on others (e.g., students, staff scientists) while simultaneously increasing the probability of increasing the size of ongoing projects while decreasing their likelihood of pursuing new research directions is quite interesting. This is consistent with researchers' teams, their students and staff, being a key part of helping them explore new research directions. 

We lack the variation or data to dig further into this question to disentangle how much these results are driven by the supply- or demand-side, and we cannot formally connect specific inputs to specific strategies given the variation in our data.\footnote{It is tempting to devise a two-stage analyses where we test how grant design influences input choices, and in turn, how these input choices effect strategic choices. However, our randomized grant designs are not valid instrumental variables because grant design may influence other features of the researchers' production functions besides these specific input choices (i.e., we cannot guarantee that the exclusion restriction holds).} Still, recent work has begun to highlight how the wages and employment outcomes of researchers' teams can be influenced by funding shocks (\citealt{cheng2023effect}), and continued work on how inputs and strategies are connected within the scientific production function seems especially promising.

\section{Researchers' preferences}\label{sec_pref}

\subsection{Empirical model}\label{sec_ADmodel}
The prior analyses focused on the identifying the treatment effect(s) of grant design conditional on receiving the grant. Here, our focus is on researchers' preferences over grant designs and how, when given freedom to choose which grants to pursue, these preferences may lead to selection effects that influence the composition of researchers who pursue particular grants.

We assume that each grant $g$ is defined by a pair of parameters ($A_g$,$D_g$), which describe the total dollar amount $A$ of the grant and the duration $D$ of the grant, in terms of how many years the researcher has access to the funds. Researcher $i$'s indirect utility from a grant is:
\begin{equation}\label{eq_ADutility}
v_i(A_{g}, D_{g}) = \alpha_i A^\gamma_{g} D^{1-\gamma}_{g} \,\,,
\end{equation}
where $\alpha_i$ is a researcher-specific taste shifter and $\gamma$ is a common parameter that describes scientists' relative preference for funding amount versus duration. 

To motivate the analyses and arrive at our main regression equation, consider a thought experiment where each researcher $i$ observes a pair of grants $g=(1,2)$ where ($A_{i1}, D_{i1}$) and ($D_{i2}$) are randomly specified by the experimenter. The respondent is asked to report the value $A_{i2}$ that makes them indifferent between the two grants such that $v(A_{i1}, D_{i1})=v(A_{i2}, D_{i2})$. We assume that researchers' responses ($\widetilde{A}_{i2}$) are the truth plus i.i.d. noise, such that $\widetilde{A}_{i2}=A_{i2} + \epsilon_i$. This allows us to write:
\begin{align}\label{eq_ADutiltoreg}
\begin{split}
\gamma \log(A_{i1}) + (1-\gamma) \log(D_{i1}) & = \gamma \log(\widetilde{A}_{i2} - \epsilon_i) + (1-\gamma) \log(D_{i2}) \,\,.
\end{split}
\end{align}
We obtain our main estimating equation by solving Equation \ref{eq_ADutiltoreg} for $\widetilde{A}_{i2}$:
\begin{equation}\label{eq_ADreg}
\widetilde{A}_{i2} = \frac{A_{i1} }{\exp\left(\frac{1-\gamma}{\gamma}\log\left(\frac{D_{i2}}{D_{i1}}\right)\right)} + \epsilon_i \,\,,
\end{equation}
which we estimate via non-linear least squares.\footnote{Note that the design of this grant is designed to be uncorrelated with the design of the hypothetical grant in the first thought experiment. Amongst the pairwise correlations of the two grant attributes in the first thought experiment and the three attributes in this thought experiment, the absolute value of any correlation is never larger than 0.03.}

Equation \ref{eq_ADutility} is a reduced form representation of a more complex model of researchers' production and consumption decisions detailed in Appendix \ref{sec_app_prefs}. Comparative statics derived from stylized simplifications of that model illustrate how the $\gamma$ parameter should be related to five key features: (1) the capital intensity of a researchers' work (i.e., their returns to scale with respect to funding), where more capital intensive researchers should have a larger $\gamma$; (2) the ease with which the researcher can acquire additional funding through other means (i.e., their fundraising productivity), where researchers more capable at obtaining funding should have a larger $\gamma$; (3) researchers' risk aversion, where more risk-averse researchers should have a smaller $\gamma$; (4) the discount rate, where researchers that discount future periods more should have a larger $\gamma$; and (5) researchers' direct utility from grant funding (e.g., in the form of salary buyout), which should be increase in $\gamma$. These are the dimensions along which we may see heterogeneity across researchers that would signal the importance of selection effects induced by grant designs.

\subsection{Thought experiment}
To generate the data necessary to estimate Equation \ref{eq_ADreg}, we present respondents with a hypothetical scenario where an anonymous donor would like to award the researcher with a single research grant. The donor presents the researcher two options: Grant \#1 and Grant \#2, where the first is set to be shorter than the second. The amount of Grant \#1 is shown, but the amount of Grant \#2 is not. The respondent is then asked to state the funding amount of Grant \#2 that would make them indifferent between both grants.\footnote{Text in the survey also indicates to the respondent that all funding associated with the grant would be awarded immediately, but only could be used within the time-frame specified. Furthermore, no so-called ``no-cost extensions'' of the grant would be considered and all unused funds would expire after the duration of the grant.}

All features shown are randomized. The amount of Grant \#1 is a uniform random draw from \{\$100,000, \$250,000, \$500,000, \$1,000,000, \$2,000,000\}. The duration of Grants \#1 and \#2 are uniform random draws from \{2, 3, 4, 5\} and \{6, 7, 8, 9, 10\} years, respectively. Table \ref{tab_indifsumstat} reports the summary statistics for the four key variables related to this thought experiment. The randomized elements have the expected statistics (e.g., the mean of the variables are approximately equal the mean of the distributions from which they're drawn).

\begin{table}\centering \small
\caption{Summary statistics -- grant indifference thought experiment}\label{tab_indifsumstat}
\-\\
{
\def\sym#1{\ifmmode^{#1}\else\(^{#1}\)\fi}
\begin{tabular}{l*{1}{rrrrr}}
\hline\hline
                    &       count&        mean&          sd&         min&         max\\
\hline
\underline{\emph{Randommized elements}}&            &            &            &            &            \\
Grant \#1: Duration (years) &       4,175&         3.5&         1.1&           2&           5\\
Grant \#1: Amount (\$)&       4,175&   781,904.2&   690,313.1&     100,000&   2,000,000\\
Grant \#2: Duration (years)&       4,175&         8.0&         1.4&           6&          10\\
\\ \underline{\emph{Researcher's response at indifference}}&            &            &            &            &            \\
Grant \#2: Amount (\$)&       4,175&   607,129.1&   584,626.5&      10,000&   2,000,000\\
\hline\hline
\end{tabular}
}

\begin{quote} \footnotesize
\emph{Note}: Summary statistics of the randomized elements of, and response to, the grant preferences thought experiment.
\end{quote}
\end{table}

Table \ref{tab_indifsumstat} also starts to illustrate the degree to which researchers are willing to trade off funding amount and duration. With Grant \#2 being 4.5 years longer than Grant \#1 on average, researchers report that they would be indifferent between both grants if Grant \#2 is roughly \$165,000 smaller. This suggests researchers would be willing to trade approximately \$35,000 per year gained on a grant. Below, we report the marginal rates of substitution based on our estimation of the formal model of researchers' utility.

\subsection{Results}
Table \ref{tab_indif1} reports our estimates of $\gamma$ obtained by estimating Equation \ref{eq_ADreg} via non-linear least squares. In the full sample (Col. 1), we estimate $\gamma$ to be 0.77, which implies a marginal rate of substitution of approximately \$57,000 per year at the means of the Grant \#1 distribution. Relatively speaking, dollars appear to be three and a half times as important as duration in the context of grant design.

\begin{table}[htbp] \centering \small
\caption{Researchers' willingness to trade off grant funding amount and duration}\label{tab_indif1}
\-\\
{
\def\sym#1{\ifmmode^{#1}\else\(^{#1}\)\fi}
\begin{tabular}{l*{4}{c}}
\hline\hline
            &\multicolumn{1}{c}{(1)}         &\multicolumn{1}{c}{(2)}         &\multicolumn{1}{c}{(3)}         &\multicolumn{1}{c}{(4)}         \\
\hline
            &                     &                     &                     &                     \\
$\gamma$    &       0.767\sym{***}&       0.779\sym{***}&       0.817\sym{***}&       0.848\sym{***}\\
            &   (0.00698)         &   (0.00676)         &   (0.00592)         &   (0.00533)         \\
[1em]
\hline
\\ Marginal rate&      57,375         &      53,869         &      42,862         &      34,420         \\
\hspace{3mm} of substitution&     [7.3\%]         &     [6.9\%]         &     [5.5\%]         &     [4.4\%]         \\
[1em] \hline Excl. above ptile.&                     &  99$ ^{th}$         &  95$ ^{th}$         &  90$ ^{th}$         \\
$ N$ obs.   &       4,175         &       4,134         &       3,978         &       3,777         \\
\hline\hline
\end{tabular}
}

\begin{quote} \footnotesize
\emph{Note}: Reports results from estimating Eq. \ref{eq_ADreg} using alternative samples; larger $\gamma$ values reflect a stronger taste for funding amount compared to duration; marginal rate of substitution (\$ per year) is reported at the means and is shown in brackets as a percentage of average funding amount; the ``Excl. above ptile.'' samples drop responses where $(A_{i1} -\widetilde{A}_{i2})/A_{i1}$ is at or above the reported percentile; robust standard errors in parentheses; $^{*}$ \(p<0.1\), $^{**}$ \(p<0.05\), $^{***}$ \(p<0.01\).
\end{quote}
\end{table}

Columns (2--4) of Table \ref{tab_indif1} drop small portions of the sample that appear to be those most willing to trade size for duration in order to test the sensitivity of our results to outliers. Excluding the 1--10 percent of the most responsive researchers yields $\gamma$ estimates in the range of 0.78--0.85, which imply marginal rates of substitution of approximately \$53,000-35,000 per additional year of grant duration. These are small amounts relative to the size of the grants, roughly 4-7\%.

Appendix \ref{sec_app_prefs_sampselect} includes an alternative version of Table \ref{tab_indif1} that reports the coefficient on the sample selection correction control (i.e., the standardized Inverse Mills Ratio). In all cases the control is not significant at conventional levels, and the magnitude is small.\footnote{We find that a one standard deviation increase in the control is associated with a 1\% increase in the respondent's answer to the thought experiment compared to the sample mean.} This suggests that our sample is not comprised of individuals with especially high or low $\gamma$ values compared to the population. 

\begin{figure} \centering
\caption{Grant design indifference curves}\label{fig_indif1}
\includegraphics[width=0.85\textwidth, trim=5mm 20mm 0mm 10mm, clip]{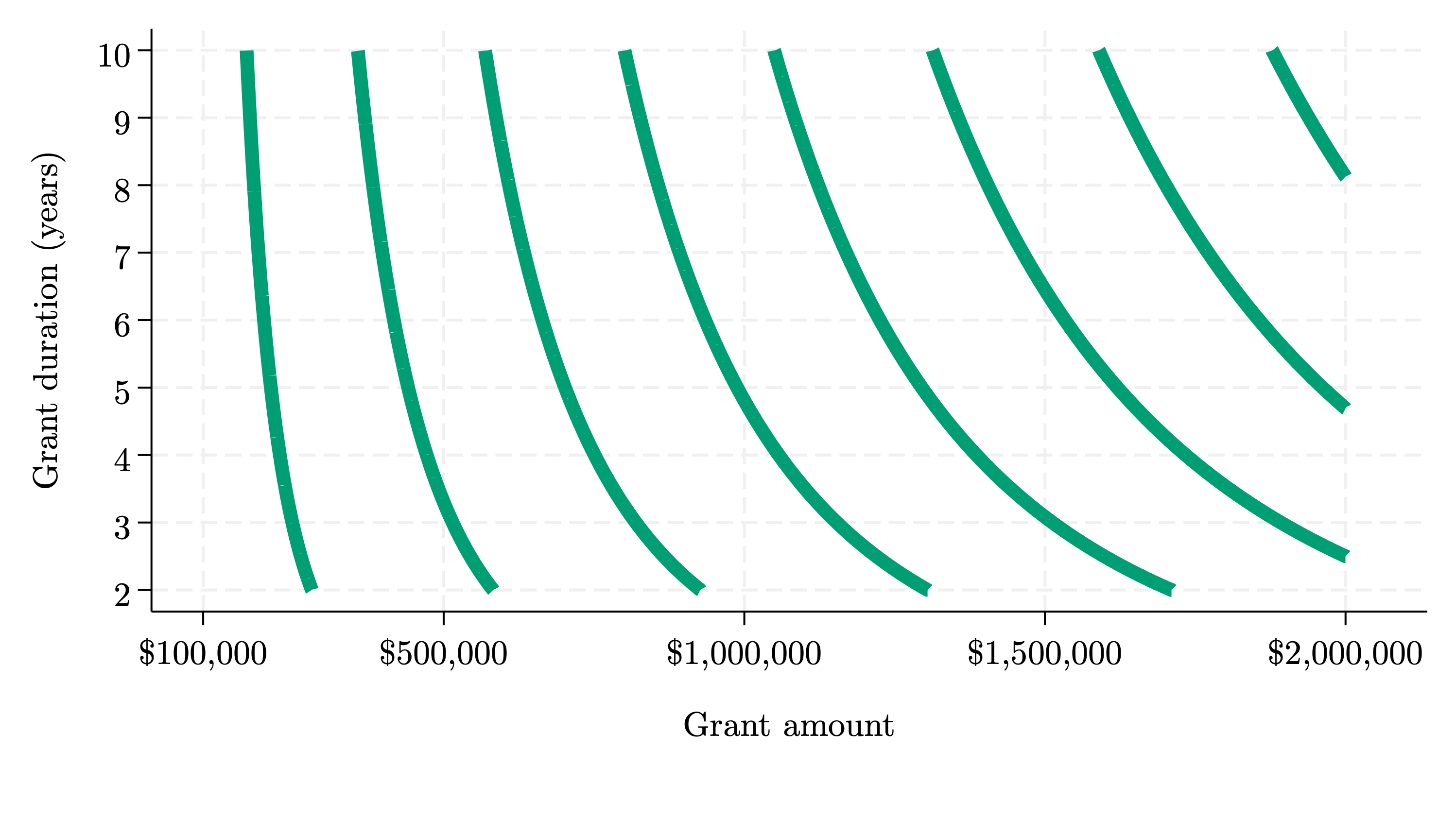}
\begin{quote} \footnotesize
\emph{Note}: Plots indifference curves over the support of the variation in the duration and size of the hypothetical research grants considered by respondents; based on the preferred $\gamma$ estimate (Table \ref{tab_indif1}, Col. 1).
\end{quote}
\end{figure}

To visualize the magnitudes implied by our estimates, Figure \ref{fig_indif1} plots the indifference curves over the support of the experimental variation. The plot illustrates how, for grants less than roughly \$500,000, researchers are relatively unwilling to sacrifice any of those funds to extend the duration of the grant. Once grant sizes extend beyond \$1,000,000, this tradeoff becomes more meaningful.

Appendix \ref{sec_app_prefs_noncrs} reports results from an alternative specification of the researchers' utility function that does not assume the constant-returns-to-scale assumption of Equation \ref{eq_ADutility}. This is done by leveraging another component of the survey, which we detail further in the Appendix. In this alternative specification, we can estimate a separate parameter for both size and duration preferences. Still, we find that the relative value researchers place on funding amount versus duration is very similar to the estimates obtained using our main specification shown above.

\subsubsection*{Heterogeneous preferences}

As discussed previously, we expect the researchers who have the strongest preference for money versus time (larger $\gamma$) to be those that discount the future at higher rates, are capital-intensive, receive more direct utility from grants, are less risk averse, and can more easily access research funding. 

Appendix \ref{sec_app_prefs_results} contains the results from regressions where we split the sample along these five dimensions, using either direct or proxy measures, and estimate separate $\gamma$ parameters within sub-samples. In each case, we find results that align with the predictions. Appendix \ref{sec_app_prefs_results} also reports on more tests of heterogeneity in $\gamma$ across other dimensions including field of study and tenure status. Notably, the variation across these dimensions is often smaller than what we observe across the five dimensions motivated by our intuitive theory.

We can obtain some sense of the aggregate amount of heterogeneity in the sample. To do so, we construct a summary metric that is the sum of these five variables (after standardizing each variable), which we construct in a way that larger values should correspond to a stronger preference for money over time. We then split the sample into five equal groups using the quintiles of this summary metric and estimate a $\gamma$ for each sub-sample.

The results of this exercise are shown in Table \ref{tab_indifhet_combined}, which indicates that the range of $\gamma$ in our sample spans (at least) from 0.7 to 0.8. This corresponds to roughly a 40\% difference in the marginal rate of substitution across the full sample -- there are some researchers who value money over time nearly twice as much as others, and our stylized theory of scientific production provides some guidance as to which researchers will have stronger preferences for larger or longer grants.

\begin{table}[ht] \centering \small
\caption{Heterogeneity in the amount-duration tradeoff}\label{tab_indifhet_combined}
\-\\
{
\def\sym#1{\ifmmode^{#1}\else\(^{#1}\)\fi}
\begin{tabular}{l*{5}{c}}
\hline\hline
            &\multicolumn{1}{c}{(1)}         &\multicolumn{1}{c}{(2)}         &\multicolumn{1}{c}{(3)}         &\multicolumn{1}{c}{(4)}         &\multicolumn{1}{c}{(5)}         \\
\hline
            &                     &                     &                     &                     &                     \\
$\gamma$    &       0.725\sym{***}&       0.750\sym{***}&       0.751\sym{***}&       0.801\sym{***}&       0.814\sym{***}\\
            &    (0.0146)         &    (0.0161)         &    (0.0170)         &    (0.0152)         &    (0.0142)         \\
[1em]
\hline
\\ m.r.s.   &      70,980         &      62,846         &      62,473         &      47,162         &      43,719         \\
\hspace{3mm} &     [9.1\%]         &     [8.0\%]         &     [8.0\%]         &     [6.0\%]         &     [5.6\%]         \\
[1em] \hline 
$ N$ obs.   &         835         &         835         &         835         &         835         &         835         \\
Sub-sample&           1         &           2         &           3         &           4         &           5         \\
& low & \multicolumn{3}{c}{$\leftarrow$ age $\rightarrow$} & high \\
& low & \multicolumn{3}{c}{$\leftarrow$ capital intensity $\rightarrow$} & high \\
& low & \multicolumn{3}{c}{$\leftarrow$ soft money share $\rightarrow$} & high \\
& low & \multicolumn{3}{c}{$\leftarrow$ risk tolerance $\rightarrow$} & high \\
& low & \multicolumn{3}{c}{$\leftarrow$ fundraising productivity $\rightarrow$} & high \\
\hline\hline
\end{tabular}
}

\begin{quote} \footnotesize
\emph{Note}: Larger $\gamma$ values reflect a stronger taste for grant size compared to duration; (m.r.s.) marginal rate of substitution (\$ per year) is reported at the means and is shown as a percentage of average grant size in brackets; robust standard errors in parentheses; $^{*}$ \(p<0.1\), $^{**}$ \(p<0.05\), $^{***}$ \(p<0.01\). The ``sub-sample'' row refers to the quintiles of the heterogeneity metric, which is the sum of the five standardized variables hypothesized to influence researchers' preferences as illustrated at the bottom of the table.
\end{quote}
\end{table}

\subsection{Funders' preferences}
How well are funders' preferences for grant designs aligned with researchers'? Having funders undertake a thought experiment such as the one described above is not feasible. However, we can get an initial sense of funders' preferences from the following two exercises: (1) a case study of a unique grant program at the NIH that undertook the tradeoff of grant size and duration, and (2) an analysis of the realized grant distribution (as shown in Figure \ref{fig_dimension_heatmap}).

\subsubsection*{Case study: The ``Maximizing Investigators' Research Award''}
As noted in our introduction, in 2015 the National Institute of General Medical Sciences at the NIH introduced the Maximizing Investigators’ Research Awards (MIRA) program to provide grants that were longer than the traditional grants that dominate their budget, but included limitations on the total amount of funding researchers could receive. The program was described as providing ``award levels somewhat lower than current support in return for increased award length, stability, flexibility, and reduced administrative burden'' compared to the traditional research project grant mechanisms.\footnote{See \href{https://www.nigms.nih.gov/Research/mechanisms/MIRA/Pages/Answers-to-Frequently-Asked-Questions-About-MIRA-PAR-19-367.aspx}{this FAQ} by the NIH.}

The complex way in which the program was implemented limits our ability to conduct any formal analyses. But summary statistics reported by the NIH and survey data from a reputable third party suggest that the program offered researchers funding for approximately 30\% longer and reduced their total funding level over that period by 10--20\%.\footnote{The funding duration changes are based on estimates reported by the NIH \href{https://loop.nigms.nih.gov/2022/02/mira-renewals-award-rates-and-budget-changes/}{here}. The funding level changes are based on estimates reported by the Genetics Society of America \href{https://genestogenomes.org/mixed-feelings-about-mira}{here}.} These magnitudes imply a $\gamma^{\text{MIRA}}$ of approximately 0.6--0.75, which is about 15\% smaller than the same parameter for researchers and implies the NIH is more willing to trade off size for duration.

The MIRA program is still in its early stages, but preliminary feedback indicated a mixed response with some researchers claiming they were ``happy to trade less funding for more stability of funding and more flexibility to pursue new research directions,’’ while others noted that the reductions in the grant size ``might be too severe for labs to be able to continue doing their science at the current level, and that people might have to be laid off.’’\footnote{See \href{https://genestogenomes.org/mixed-feelings-about-mira/}{this commentary} by the Genetics Society of America for more.} These quotes are consistent with the relative difference between the $\gamma^{\text{MIRA}}$ implied by the program design and the $\gamma$ we estimate for our sample.

\subsubsection*{Analysis of realized grant design distribution}
We can make some inferences about funders' preferences more generally based on the observed distribution of grant designs shown in Figure \ref{fig_dimension_heatmap}. Appendix \ref{sec_app_funderprefs} details an exercise that treats all funders as a multi-product monopolist facing inelastic demand from researchers. In this case, just as changes in a monopolists prices reflects changes in marginal costs, the variation in the number of grants of different designs reflects the ``aggregate funder's'' relative preference for awarding grants of different amount and duration. 

This revealed preference approach yields an estimate of $\gamma^{\text{funder}} \approx 0.4$. Intuitively, we should expect funders to be rather willing to trade off grant amount for duration since there are much tighter constraints on their annual budgets compared to the constraints on their time horizons. More importantly, this exercise suggests that funders are currently providing researchers with a much more ``long-short'' set of grant opportunities than the researchers would prefer. Perhaps this is why most of the effects of grant designs on researchers' strategies seen in Section \ref{sec_te} (i.e., Table \ref{tab_grantstrat_all}) are driven by variation in funding amounts as opposed to duration. Whether the effects of grant design would be different in alternative equilibria (e.g., in other populations such as academic institutions in Europe, which have very different structures, or developing countries, which have very different levels of inputs) is an interesting open question.

\section{Discussion}\label{sec_discuss}

Understanding how grant design can be used to manage science requires understanding the incentives and institutions that transform inputs and outputs into objects that researchers value (e.g., job security, salary, prestige). Practical examples of such factors include researchers' taste for science (\citealt{stern2004scientists,roach2010taste}), the tenure process and output measurement schemes (\citealt{macleod2020does}), intellectual property regimes (\citealt{hvide2018university}), the nature of competition within fields (\citealt{hill2019scooped}), gate-keeping (\citealt{azoulay2019does}), and other social factors more generally (\citealt{shapin1995here}). Our analyses provide new insights about grant designs given the incentives and institutions surrounding researchers. We investigate both the treatment and selection effects that can be induced by grants with different designs.

Grants appear to affect researchers' strategies in some intuitive and some counter-intuitive ways. As predicted by prior theories, longer grants increase researchers' plans to take risks in their work, but only if the researcher has the job security of tenure. This finding aligns with the work of \cite{azoulay2011incentives}, but provides new evidence as to some important boundary conditions. Furthermore, these longer grants reduce \emph{all} researchers' focus on speed. More work remains to be done on the connection between these two strategies and its implications for optimal grant design.\footnote{See \cite{nanda2017innovation} for a discussion of how alternative innovation policies can influence the types of projects funders are willing to support, and the importance of understanding how this selection process influences the extensive margin.} 

In contrast to conventional theory, we find that larger grants promote more ``exploit'' strategies and smaller grants promote more ``explore'' strategies. This may be partly driven by the subjective nature of our definitions of strategy. But it may also signal some unique features of the scientific production function that have not yet received much attention. We can rationalize this result either with a model of uncertain production by risk-averse researchers, or with a scarcity of inputs that are especially productive when starting new projects. Further work to estimate the supply of, demand for, and productivity of specific scientific inputs (e.g., graduate student labor, data, equipment) is certainly an important avenue for future work and is increasingly becoming possible using linked administrative data (e.g., \citealt{lane2015new,chang2019federal}).

When scaled to reflect the degree of grant design variation we see in practice, these treatment effects are small. In the example of the NIH's use of MIRA grants versus traditional R01 grants (where the MIRA grant is roughly 15\% smaller and 30\% longer than the R01 grant), our estimates imply that receiving one or the other would only change researchers' probabilities of choosing certain strategies by about 1 percentage point. But these (small) treatment effects are not the full story.

Since the vast majority of grants require researchers to self-select their grant pursuits, the usefulness of grant designs depends both on these treatment effects as well as the presence of any selection effects --- grant designs may change the number and composition of researchers who pursue them. To this point, our second thought experiment sheds new light on researchers' preferences over grant design. Our results indicate that while there is some willingness by researchers to tradeoff grant size (total grant \$) for duration (years grant \$ are available), they value relative changes in size nearly four times as much as changes in duration. The $\gamma$ parameter we estimate subsumes several structural parameters, and our heterogeneity analyses indicate there is some significant variation in preferences across researchers. Disentangling this heterogeneity further to better understand selection effects would require a much richer model and more complicated experiments. For example, we use age as a proxy for researchers discount rates, but it would be important to disentangle researchers' actual discount rates versus dynamic incentives they face (e.g., due to career concerns) over their life-cycle. Overall, our results from this second thought experiment indicate that the selection effects of grant design and how they might influence the composition of researchers in an area are likely more important than the treatment effect of how those grant designs might influence \emph{how} researchers conduct their science.

A key limitation of this paper is that we rely on non-incentivized stated preferences from a small sample of researchers. As noted by \cite{stantcheva2022run}, this allows us to ``create our own variation'' and measure variables that are difficult or impossible to observe in existing datasets (e.g. bibliometric data). Nonetheless, this comes at the cost of having to assume that these stated preferences are informative about actual behavior and that our sample is representative of the population. Our investigations into respondents' attention and accuracy (using salary comparisons) and their representativeness (using data comparisons and sample selection correction techniques) suggest that these issues may not loom large in this specific survey. Our results indicate that grants can be instrumental, and that the large costs of field experiments with stronger guarantees of representativeness may be worth the investment to better understand how grant design can be used to manage science.

More generally, it is still clear that grants are a key tool for public investments in science (\citealt{azoulay2018public}), can have important effects on the research workforce (\citealt{cheng2023effect}), and, if large enough, can be used to influence researchers' scientific directions (\citealt{myers2020elasticity}) even if they may not have large effects on \emph{how} researchers pursue those new directions. Recent evidence has highlighted the interaction of public and private funding of science (\citealt{babina2023cutting}), which introduces more dimensions of funding arrangements (e.g., intellectual property, control rights) that may play an important role in influencing researchers' decisions.

\clearpage
\singlespacing
\bibliographystyle{apalike}
\bibliography{bibliography.bib}

\appendix
\onehalfspacing

\clearpage
\fancypagestyle{mystyle}{%
    \fancyhead{}
    \fancyhead[C]{Appendices}\fancyfoot{}
}%
\thispagestyle{mystyle}
\setlength{\headheight}{14.49998pt}
\addtolength{\topmargin}{-2.49998pt}.
\section{Additional summary statistics}\label{sec_app_addstat}
\setcounter{figure}{0}
\renewcommand{\thefigure}{A\arabic{figure}}
\setcounter{table}{0}
\renewcommand{\thetable}{A\arabic{table}}
\setcounter{equation}{0}
\renewcommand{\theequation}{A\arabic{equation}}

\subsection{Full summary statistics for covariates}\label{sec_app_addstat_allxsumstat}
\begin{figure}[htbp] \centering
\caption{Expected research funding by broad field}\label{bigfig_histfund_agg}
\subfloat[Engineering, math, \& related]{\label{fig_histfund_agg1}\includegraphics[width=0.33\textwidth, trim=0mm 0mm 0mm 0mm, clip]{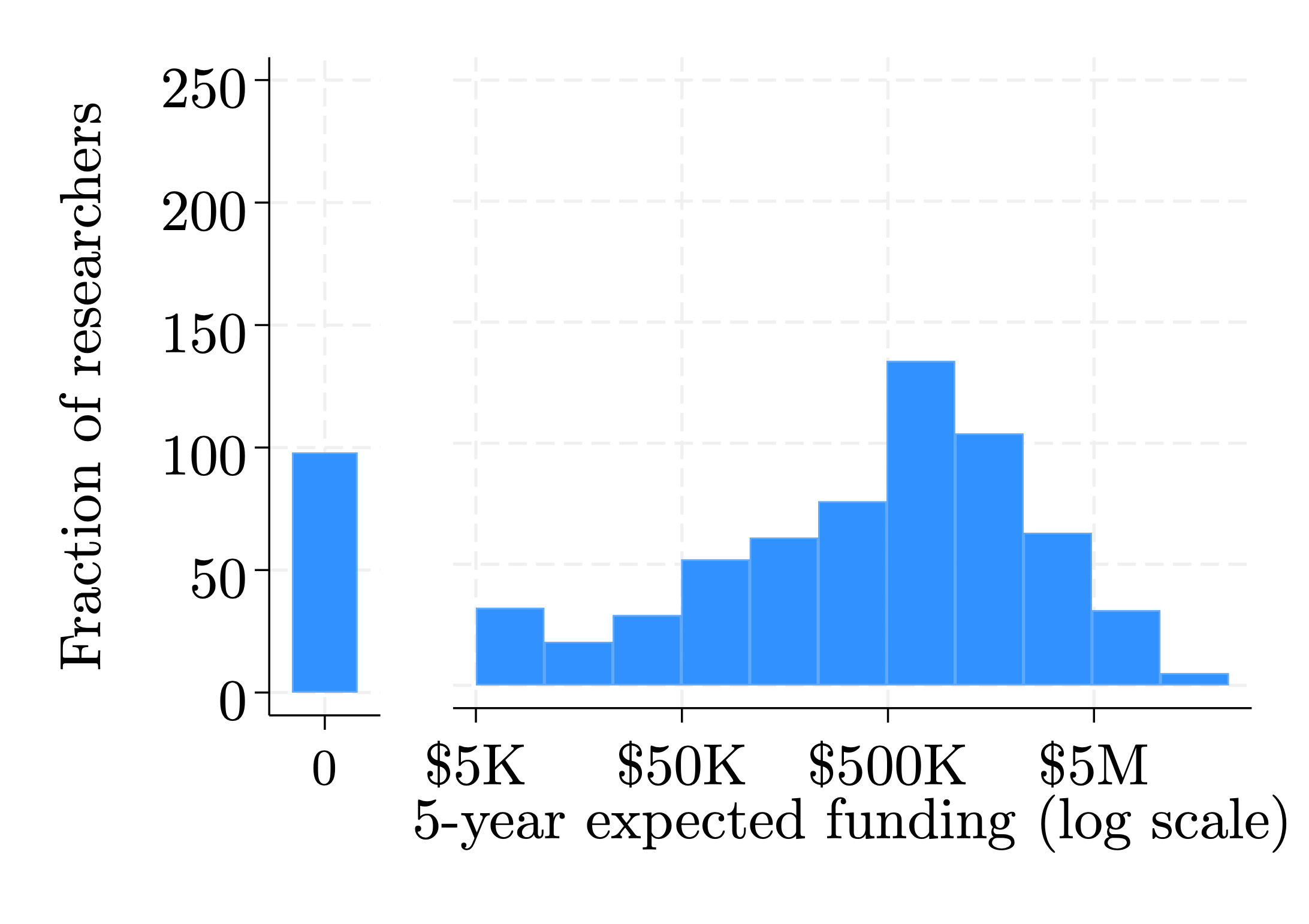}}
\subfloat[Humanities \& related]{\label{fig_histfund_agg2}\includegraphics[width=0.33\textwidth, trim=0mm 0mm 0mm 0mm, clip]{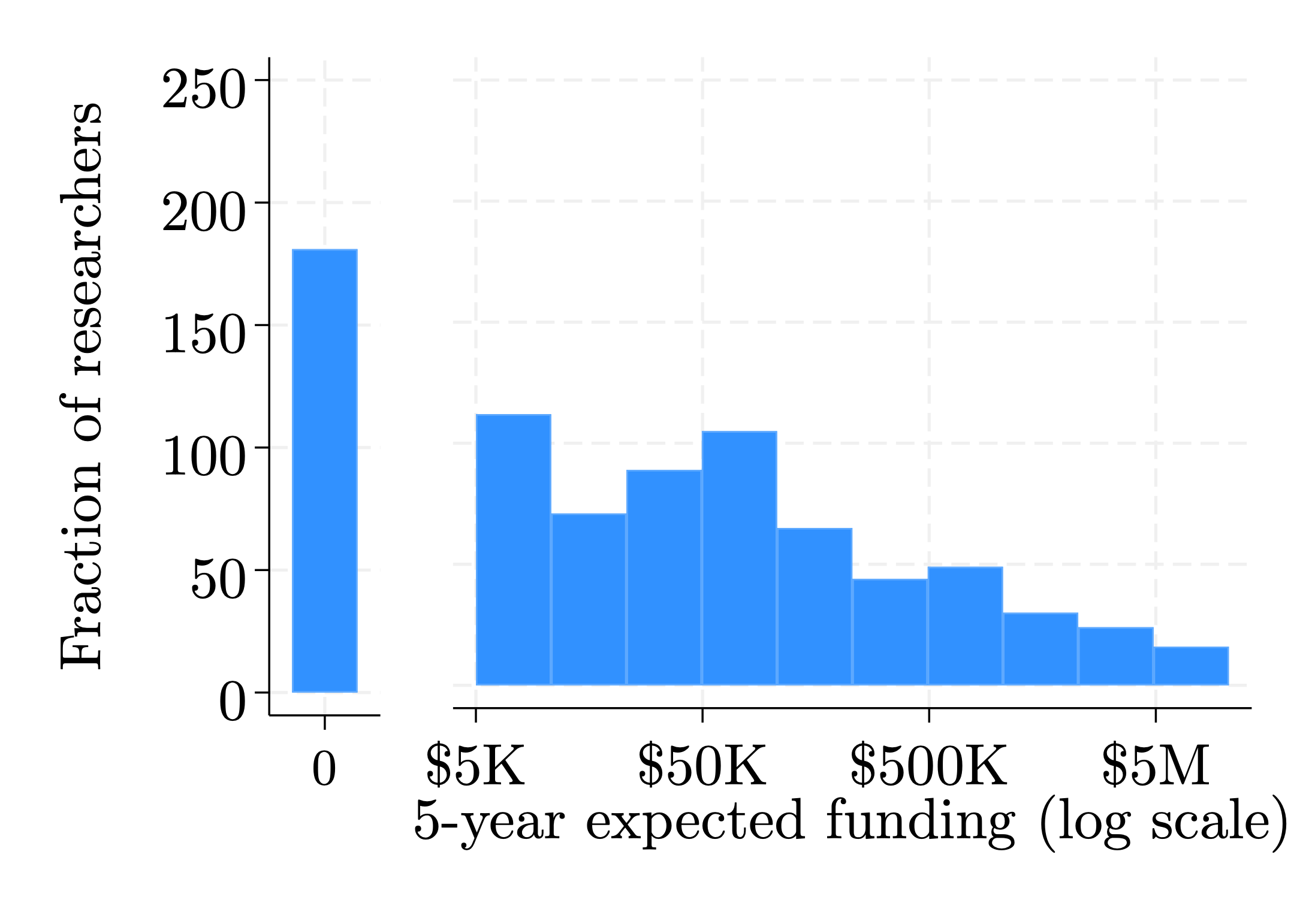}}
\subfloat[Medical \& health sciences]{\label{fig_histfund_agg3}\includegraphics[width=0.33\textwidth, trim=0mm 0mm 0mm 0mm, clip]{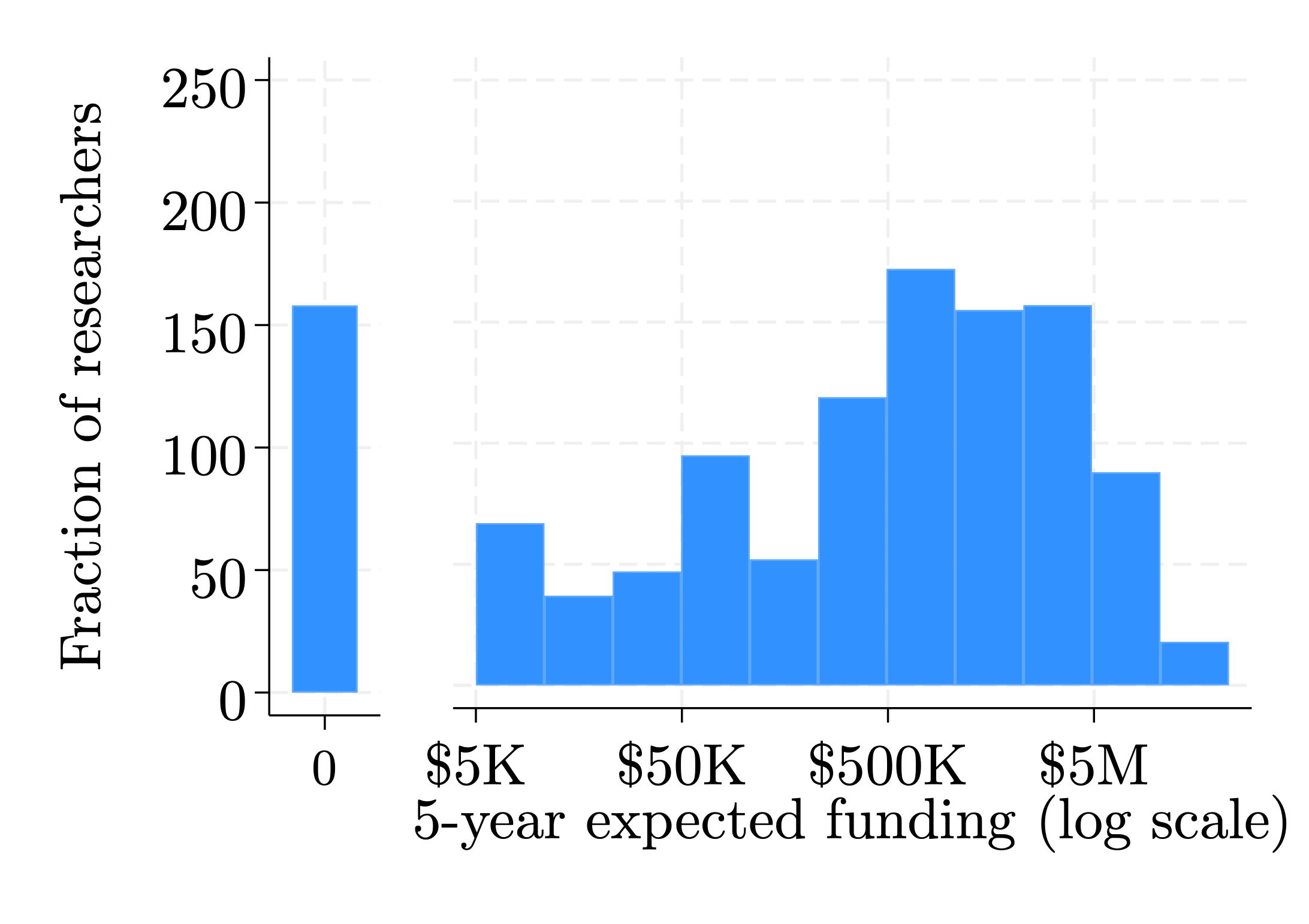}}\-\\\-\\
\subfloat[Natural sciences]{\label{fig_histfund_agg4}\includegraphics[width=0.33\textwidth, trim=0mm 0mm 0mm 0mm, clip]{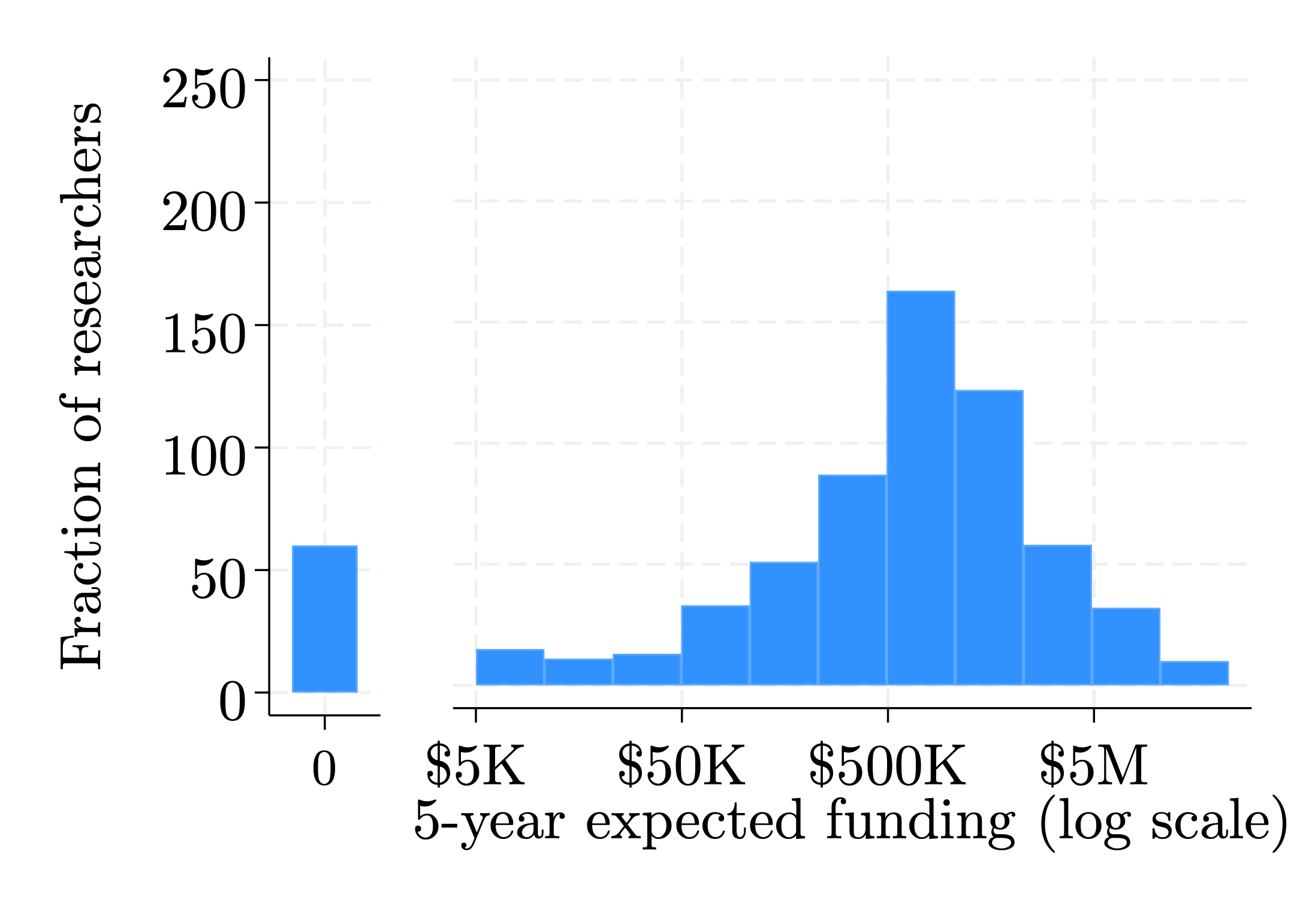}}
\subfloat[Social sciences]{\label{fig_histfund_agg5}\includegraphics[width=0.33\textwidth, trim=0mm 0mm 0mm 0mm, clip]{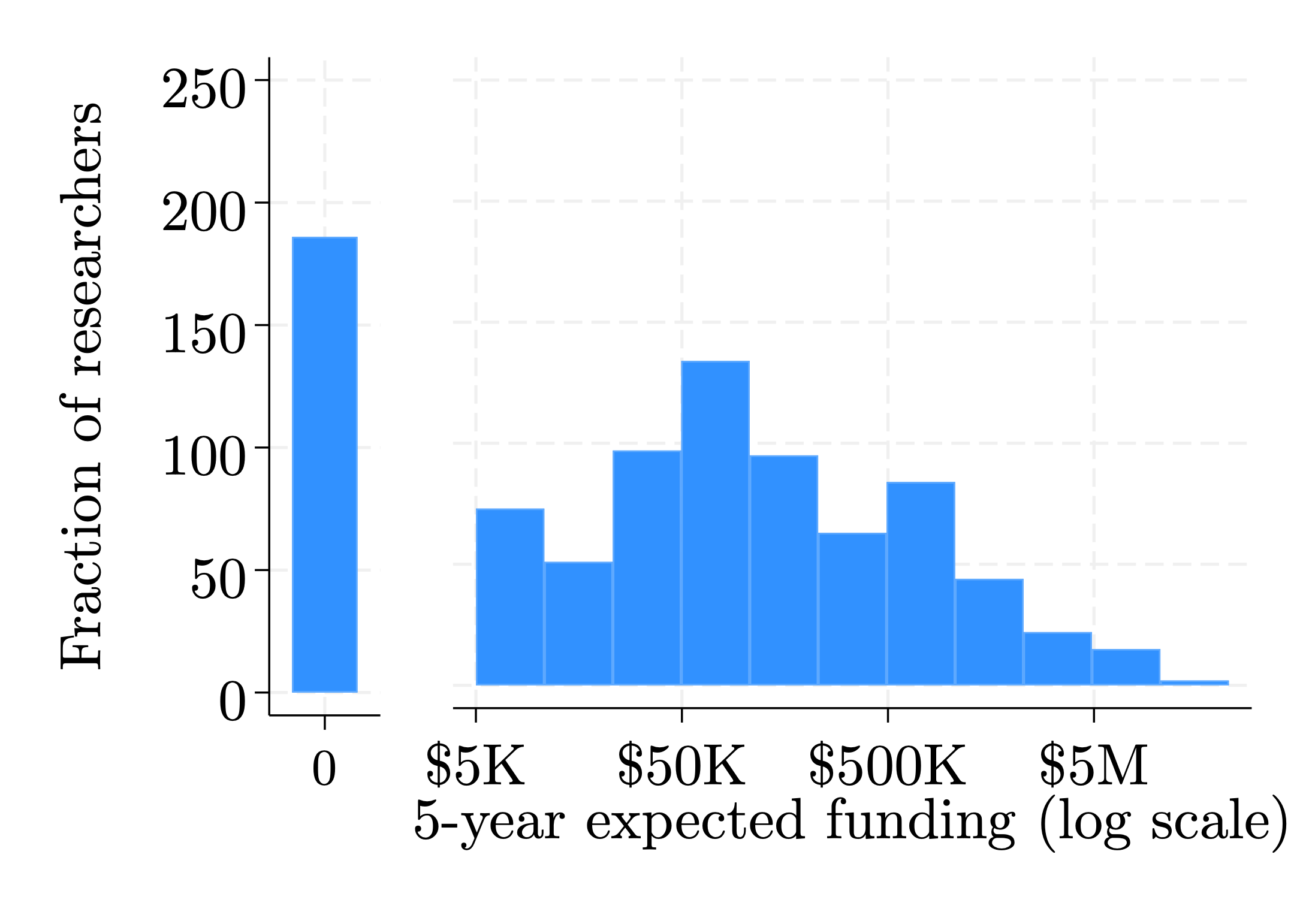}}
\begin{quote} \footnotesize
\emph{Note}: Shows the within-field distributions of researchers' total expected research funding over the next five years, which is the sum of existing funding and expected fundraising amounts over the same period.
\end{quote}
\end{figure}

Tables \ref{tab_allxsumstat1}-\ref{tab_allxsumstat2} report the mean and standard deviation for all variables ever used in the analyses. If the count of observations is less than the full sample (4,175), it is because respondents were not required to answer the question and some chose not to --- this only is relevant for the socio-demographic questions. Figure \ref{bigfig_histfund_agg} plots the distribution of expected research funding across the five broad fields to illustrate the pervasiveness of demand for research grants.

\begin{table}[ht] \centering \small
\caption{Full summary statistics for covariates}\label{tab_allxsumstat1}
\-\\
{
\def\sym#1{\ifmmode^{#1}\else\(^{#1}\)\fi}
\begin{tabular}{l*{1}{rrrrr}}
\hline\hline
                    &       count&        mean&          sd\\
\hline
\textbf{Professional } \\ \\ \underline{\emph{Rank} \{0,1\}} \\ Adjunct, clinical, teaching, other&       4,175&        0.06&        0.23\\
Assistant           &       4,175&        0.25&        0.44\\
Associate           &       4,175&        0.25&        0.43\\
Emeritus or retired &       4,175&        0.03&        0.17\\
Full                &       4,175&        0.41&        0.49\\
\\ \underline{\emph{Tenure} \{0,1\}} \\ Not on tenure track, not applicable&       4,175&        0.20&        0.40\\
On tenure track, pre-tenure&       4,175&        0.22&        0.41\\
Tenured             &       4,175&        0.59&        0.49\\
\\ \underline{\emph{Salary}} \\ Base&       4,175&  120,649.10&   79,692.66\\
Grant-sponsored or  &       4,175&   25,336.53&   45,407.18\\
Supplementary       &       4,175&    5,142.51&   16,566.07\\
Clinical            &       4,175&    5,692.22&   37,375.73\\
Outside activities  &       4,175&    7,798.80&   29,703.94\\
Guaranteed research funding&       4,175&  430,699.40&1,085,486.21\\
Research fundraising expectations&       4,175&  549,238.32&1,055,926.36\\
Total work hrs./week&       4,175&       49.88&       13.06\\
\\ \underline{\emph{Share of work time} [0,1]} \\ Research&       4,175&        0.39&        0.20\\
Fundraising         &       4,175&        0.09&        0.11\\
Teaching            &       4,175&        0.27&        0.17\\
Administration      &       4,175&        0.15&        0.14\\
Clinical            &       4,175&        0.04&        0.14\\
Other               &       4,175&        0.06&        0.09\\
\\ \underline{\emph{Aggregate field} [0,1]} \\ Humanities&       4,175&        0.17&        0.37\\
Medical \& health   &       4,175&        0.19&        0.39\\
Natural             &       4,175&        0.28&        0.45\\
Social \& mathematical&       4,175&        0.16&        0.36\\
\hline\hline
\end{tabular}
}

\end{table}

\begin{table} \centering \small
\caption{Full summary statistics for covariates, continued}\label{tab_allxsumstat2}
\-\\
{
\def\sym#1{\ifmmode^{#1}\else\(^{#1}\)\fi}
\begin{tabular}{l*{1}{rrrrr}}
\hline\hline
                    &       count&        mean&          sd\\
\hline
\textbf{Socio-demographic} \\ \\ Age&       4,084&       49.31&       12.52\\
Female \{0,1\}      &       3,992&        0.41&        0.49\\
Married or in partnership \{0,1\}&       4,044&        0.84&        0.36\\
U.S. immigrant \{0,1\}&       4,088&        0.27&        0.44\\
3rd generation in U.S. or less \{0,1\}&       2,848&        0.44&        0.50\\
Household annual income&       4,045&  260,819.53&  215,128.41\\
Num. household dependents&       4,104&        0.98&        1.13\\
\\ \underline{\emph{Race/ethnicity} \{0,1\}} \\ Asian&       4,175&        0.12&        0.33\\
Black               &       4,175&        0.03&        0.17\\
Hispanic            &       4,175&        0.06&        0.23\\
White               &       4,175&        0.78&        0.42\\
Other               &       4,175&        0.05&        0.22\\
\\ \textbf{Risk-taking} \\ \\ Research, own belief&       4,175&        4.63&        2.35\\
Research, others' belief&       4,175&        4.56&        2.40\\
Personal life       &       4,175&        5.25&        2.12\\
\\ \textbf{Research} \\ \\ Objective: ask or answer questions [0,10]&       4,175&        4.85&        2.57\\
\\ \underline{\emph{Output} \{0,1,2\}} \\ Journal articles&       4,175&        1.87&        0.37\\
Books               &       4,175&        0.52&        0.69\\
Materials, methods, or tools&       4,175&        0.70&        0.68\\
Products            &       4,175&        0.47&        0.64\\
\\ \underline{\emph{Audience} \{0,1,2\}} \\ Academics&       4,175&        1.88&        0.36\\
Policymakers        &       4,175&        0.84&        0.69\\
Businesses and orgs.&       4,175&        0.54&        0.62\\
General public      &       4,175&        0.83&        0.62\\
\\ \textbf{Other} \\ \\ Inverse Mills Ratio&       4,175&        2.40&        0.10\\
\hline\hline
\end{tabular}
}

\end{table}

\subsection{Population and sample comparisons}\label{sec_app_addstat_popsampcompare}
\begin{figure}[htbp] \centering
\caption{Recruitment versus completion sample comparison per HERD metrics}\label{bigfig_nonresponse_herd}
\subfloat[Distributions per HERD R\&D measures]{\label{fig_nonresponse_herd}\includegraphics[width=0.95\textwidth, trim=0mm 0mm 0mm 0mm, clip]{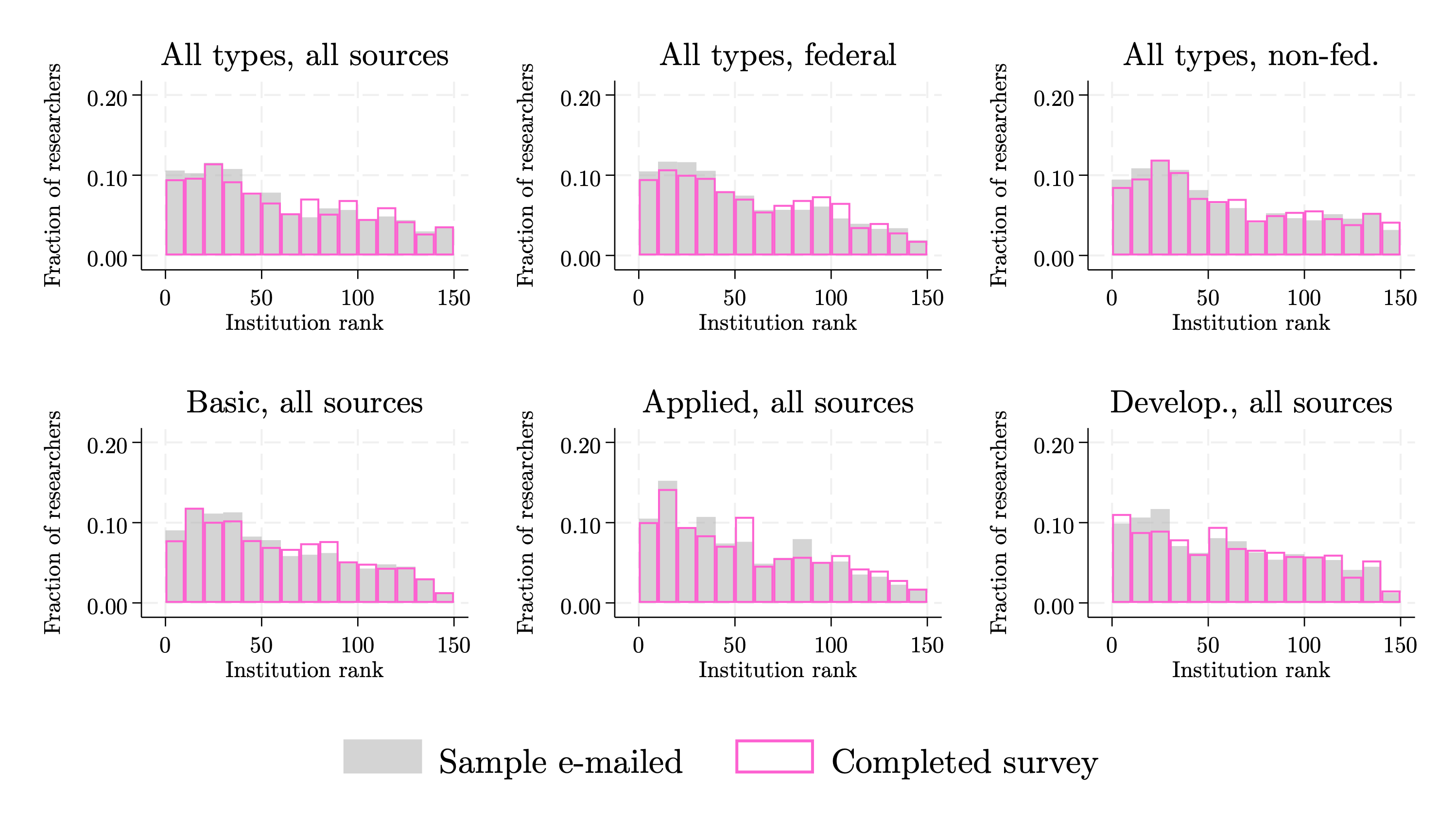}}\-\\
\subfloat[Regression estimates of mean differences]{\label{tab_nonresponse_herd}\centering \footnotesize {
\def\sym#1{\ifmmode^{#1}\else\(^{#1}\)\fi}
\begin{tabular}{l*{7}{c}}
\hline\hline
& \multicolumn{3}{c}{All sources, by type} & &\multicolumn{3}{c}{All types, by source} \\ 
& All & Federal & Non--fed. & & Basic & Applied & Develop. \\
            &\multicolumn{1}{c}{(1)}         &\multicolumn{1}{c}{(2)}         &\multicolumn{1}{c}{(3)}         &            &\multicolumn{1}{c}{(4)}         &\multicolumn{1}{c}{(5)}         &\multicolumn{1}{c}{(6)}         \\
\hline
Completed survey&      --35.01\sym{***}&      --21.30\sym{***}&      --13.71\sym{***}&            &      --24.31\sym{***}&      --7.170\sym{***}&      --3.812\sym{***}\\
            &     (6.953)         &     (4.511)         &     (3.062)         &            &     (4.867)         &     (2.492)         &     (1.328)         \\
[1em]
Constant    &       649.1\sym{***}&       351.7\sym{***}&       297.4\sym{***}&            &       413.2\sym{***}&       176.7\sym{***}&       60.46\sym{***}\\
            &     (1.308)         &     (0.863)         &     (0.578)         &            &     (0.917)         &     (0.459)         &     (0.254)         \\
[1em]
\hline
\% diff.    &      --5.4\%         &      --6.1\%         &      --4.6\%         &            &      --5.9\%         &      --4.1\%         &      --6.3\%         \\
$ N$ obs.   &     130,785         &     130,785         &     130,785         &            &     130,785         &     130,735         &     128,169         \\
\hline\hline
\end{tabular}
}
}
\begin{quote} \footnotesize
\emph{Note}: Panel (a) compares the distributions of e-mailed professors and respondents per the rank of their institution along each dimension of R\&D funding. Panel (b) reports estimates from a regression of each sampled professor's institutional R\&D funding (in 2019 \$-M) on a dummy for whether the sampled individual completed the survey; the ``\% diff.'' row reports the mean difference in the measure as a percentage of the non-respondent average (i.e., it is the ratio of the two coefficients); robust standard errors in parentheses; $^{*}$ \(p<0.1\), $^{**}$ \(p<0.05\), $^{***}$ \(p<0.01\). All institutional data is from the 2019 NSF HERD (\citealt{nsf2023herd}).
\end{quote}
\end{figure}

Here we replicate the two tests of representativeness for the survey as also done in \cite{myers2023new}. First, \ref{bigfig_nonresponse_herd} shows this comparison based on data from the National Science Foundation's HERD survey (\citealt{nsf2023herd}), which reports institutional-level data on the total amount of funding flows into all of the institutions in our population (recall, our population was constructed using the HERD, which is why this data is available for the full population).

Figure \ref{bigfig_nonresponse_herd} reports the distributions and regression tests for mean differences in six metrics of institution-level research funding that compare respondents to the full set of researchers invited to participate (``Sample e-mailed''). The distributions overlap to a large degree (see Panel a). We do estimate statistically significant differences in means (see Panel b), but the magnitudes of these differences are all in the range of approximately 4--6\%. 

\begin{figure}[htbp] \centering
\caption{Recruitment versus completion sample comparison per Dimensions metrics}\label{bigfig_nonresponse_dim}
\subfloat[Distributions per Dimensions publications and grants]{\label{fig_nonresponse_dim}\includegraphics[width=0.8\textwidth, trim=0mm 0mm 0mm 0mm, clip]{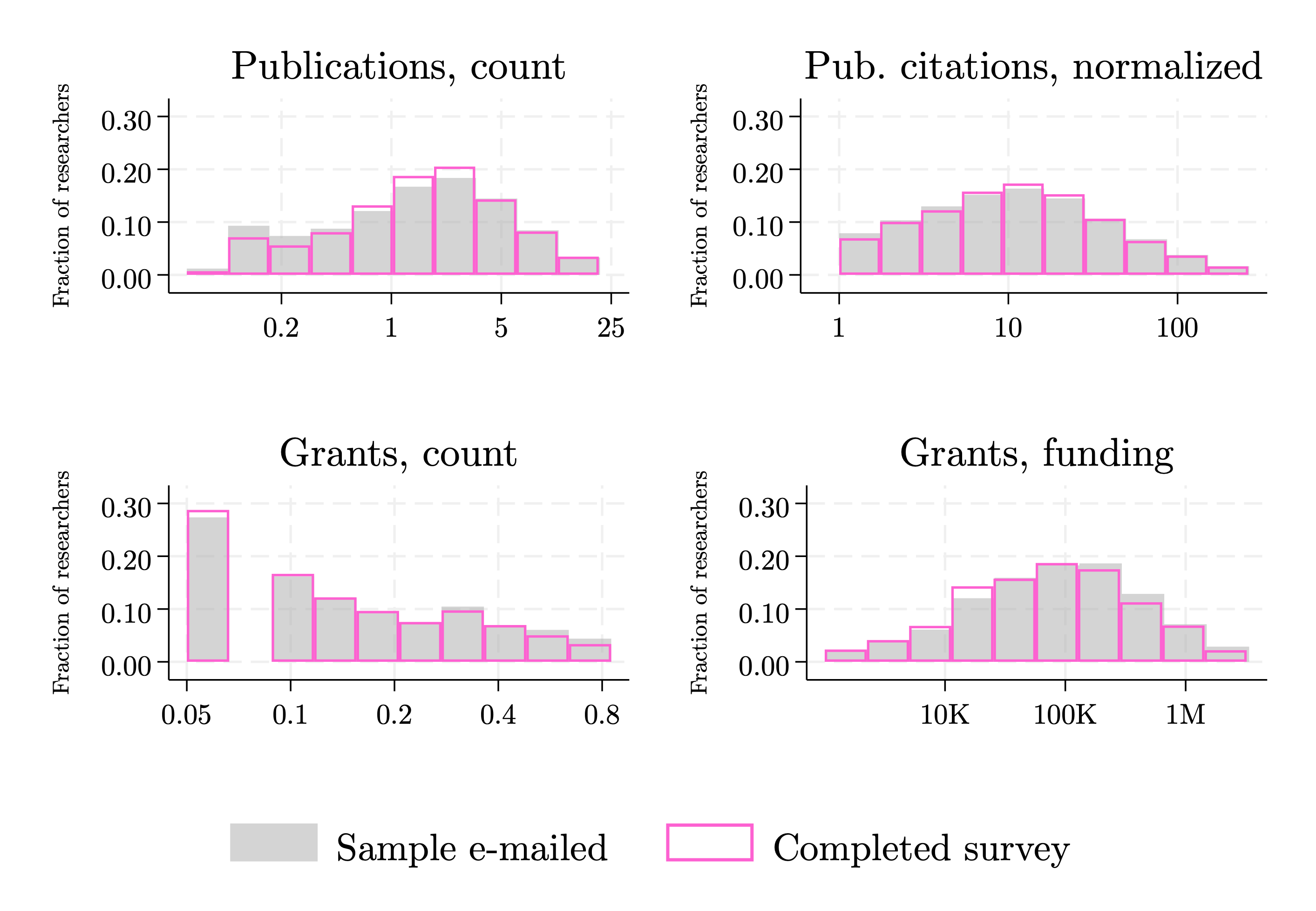}}\-\\
\subfloat[Regression estimates of mean differences]{\label{tab_nonresponse_dim}\centering \footnotesize {
\def\sym#1{\ifmmode^{#1}\else\(^{#1}\)\fi}
\begin{tabular}{l*{4}{c}}
\hline\hline
& Pub., & Pub. cites, & Grant, & Grant, \\
& count & normalized & count & total \$ \\
            &\multicolumn{1}{c}{(1)}         &\multicolumn{1}{c}{(2)}         &\multicolumn{1}{c}{(3)}         &\multicolumn{1}{c}{(4)}         \\
\hline
Completed survey&      --0.111         &      --0.575         &     0.00878\sym{**} &     --5568.8         \\
            &    (0.0688)         &     (0.807)         &   (0.00343)         &    (8519.4)         \\
[1em]
Constant    &       2.906\sym{***}&       20.59\sym{***}&       0.117\sym{***}&    121272.4\sym{***}\\
            &    (0.0158)         &     (0.185)         &  (0.000715)         &    (2093.0)         \\
[1em]
\hline
\% diff.    &      --3.8\%         &      --2.8\%         &       7.5\%         &      --4.6\%         \\
$ N$ obs.   &      87,000         &      87,000         &      87,000         &      87,000         \\
\hline\hline
\end{tabular}
}
}
\begin{quote} \footnotesize
\emph{Note}: Panel (a) compares the distributions of e-mailed professors and respondents per each dimension of individual-level publication output and grant funding per year (2003--2022). Panel (b) reports estimates from a regression of each sampled professor's publication or grant metric on a dummy for whether the sampled individual completed the survey; the ``\% diff.'' row reports the mean difference in the measure as a percentage of the non-respondent average (i.e., it is the ratio of the two coefficients); robust standard errors in parentheses; $^{*}$ \(p<0.1\), $^{**}$ \(p<0.05\), $^{***}$ \(p<0.01\). All publication and grant data is from the Dimensions database (\citealt{dimension2018data}).
\end{quote}
\end{figure}

Figure \ref{bigfig_nonresponse_dim} reports the results of a similar exercise, instead using individual-level data on researchers' publication output and grant receipts. This data was obtained by performing a fuzzy match of our population (i.e., using names and institution data) to the Dimensions database (\citealt{dimension2018data}), which includes disambiguated researcher-level records. We focus on researchers' publications and grants during the twenty years prior (2003-2022) and see very little differences between our respondents and the full set of individuals invited to the survey. The distributions have strong overlap over the full support (see Panel a), and the mean differences are all insignificant and/or smaller than 7.5\%.

\subsection{Randomized incentives}\label{sec_app_addstat_incentives}
In order to implement the sample selection correction methods of \cite{heckman1979sample}, we must have a source of variation in survey participation that is uncorrelated with any of the focal parameters. With such variation in hand, a control function can be constructed and included in the focal regressions to condition on any variation in respondents' answers that otherwise would have been unobservable, endogenous, and led to biased estimates. Furthermore, the statistical significance of this control function can shed light on the degree to which selection into the survey on (relevant endogenous) unobservables may have occurred.

\begin{figure}[htbp]\centering \small
\caption{Completion rate by incentive and reminder arms}
\-\\
\subfloat[Raw data]{\label{fig_responserate_byincent}{\includegraphics[width=0.6\textwidth, trim=0mm 0mm 0mm 0mm, clip]{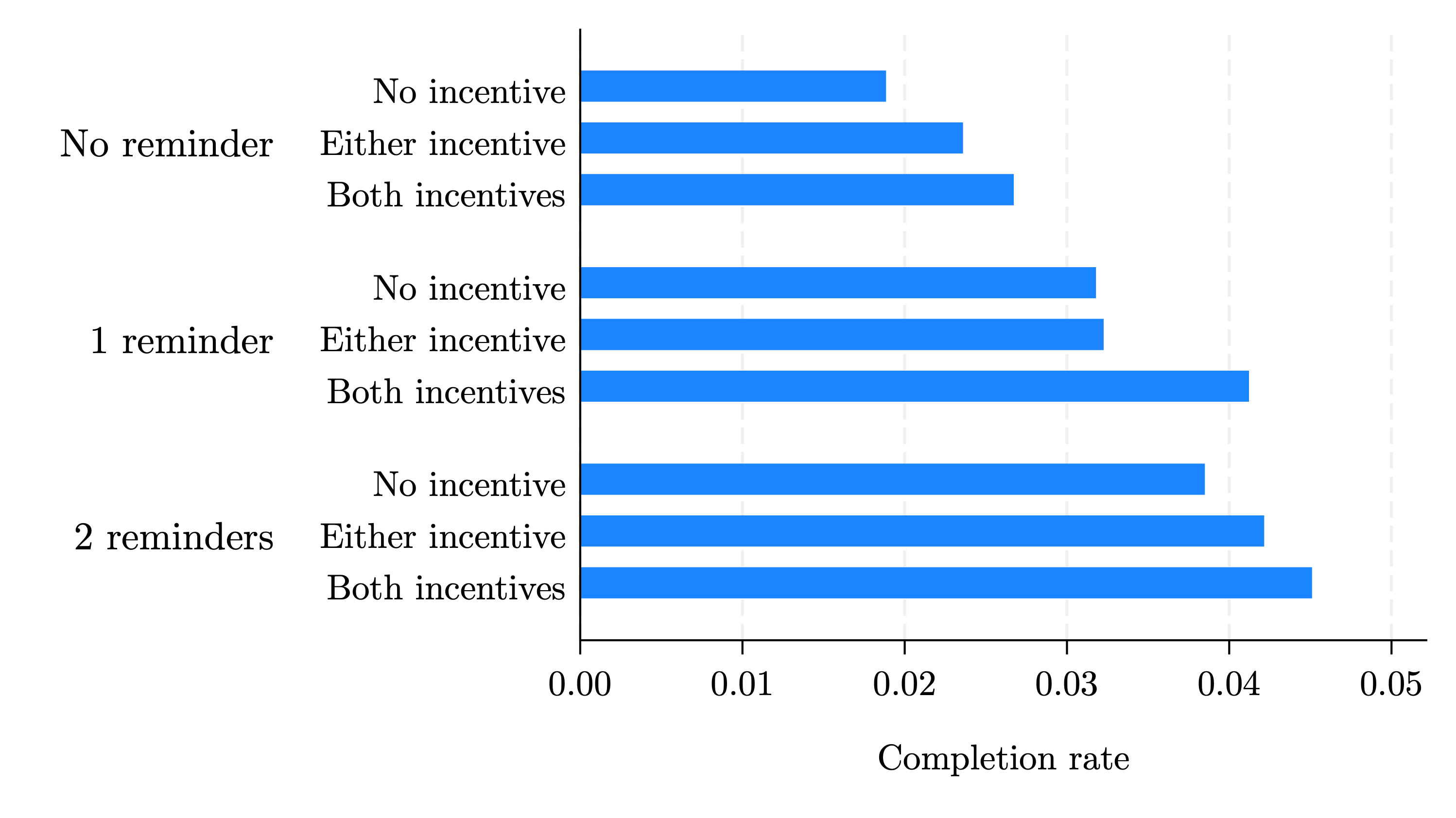}{\color{white}\hspace{0mm}}}}
\subfloat[Regression estimates]{\label{tab_responserate_byincent}{\footnotesize {
\def\sym#1{\ifmmode^{#1}\else\(^{#1}\)\fi}
\begin{tabular}{l*{1}{c}}
\hline\hline
& \\
Either incentive    &     0.00297\sym{**} \\
                    &   (0.00117)         \\
[0.5em]
Both incentives     &     0.00797\sym{***}\\
                    &   (0.00141)         \\
[0.5em]
1 reminder          &      0.0112\sym{***}\\
                    &   (0.00114)         \\
[0.5em]
2 reminders         &      0.0188\sym{***}\\
                    &   (0.00119)         \\
[0.5em]
Constant            &      0.0197\sym{***}\\
                    &   (0.00107)         \\
[0.5em]
\hline
$ N$ obs.           &     131,672         \\
\hline\hline
\end{tabular}
}
}}
\begin{quote} \footnotesize
\emph{Note}: Panel (a) shows the average completion rate by each combination of incentive and reminder arm; Panel (b) shows estimates from a linear probability model of completion as a function of incentives and reminders; robust standard errors in parentheses; $^{*}$ \(p<0.1\), $^{**}$ \(p<0.05\), $^{***}$ \(p<0.01\).
\end{quote}
\end{figure}

To generate the necessary variation in participation, we implemented randomized incentives and reminders in our e-mail strategy. Each e-mail was randomly assigned to (1) one of four incentive arms comprising a gift card lottery or a donation on the respondents' behalf to a charity of their choice, and (2) one of three reminder schedules including zero, one, or two reminders. Figure \ref{fig_responserate_byincent} shows the raw data and statistical tests of the effects of these arms (aggregating the incentive arms in Panel a, and assuming linear separability of the treatments in the regressions shown in Panel b). There are clear effects, with the ``most treated'' invites (i.e., both incentives and two reminders) increasing the probability of survey participation by roughly 2.7 percentage points to roughly 4.5\%, yielding an average response rate roughly more than twice as large as the ``least treated'' invites (i.e., no reminders and no incentives) whose response rate was 1.9\%.

\clearpage
\section{Additional results on strategic responses to grants}\label{sec_app_strat}
\setcounter{figure}{0}
\renewcommand{\thefigure}{B\arabic{figure}}
\setcounter{table}{0}
\renewcommand{\thetable}{B\arabic{table}}
\setcounter{equation}{0}
\renewcommand{\theequation}{B\arabic{equation}}

\subsection{Main results on strategic responses, reporting sample selection correction estimates}\label{sec_app_strat_sampselect}
Table \ref{tab_grantstrat_all_withimr} replicates the results from the same analyses as reported in Table \ref{tab_grantstrat_all}, but here also reporting the coefficient estimates associated with the Inverse Mills Ratio (IMR) covariate and the constant term. In all cases, we do not reject the null hypothesis of no sample selection effects with none of the IMR coefficients being significantly different from zero (\citealt{heckman1979sample}). The strong effects of the participation incentives and reminders reported in Section \ref{sec_app_addstat_incentives} suggests that the IMR control is valid in the sense that there is non-trivial variation in the IMR driven by those randomized incentives and reminders. We still cannot rule out that there are no other unobservable dimensions of heterogeneity that are both correlated with researchers' responsiveness to either the incentives or the reminders as well as our focal parameters (i.e., researchers' strategic responsiveness to grant designs). However, we have no qualitative, theoretical, or other empirical reason to suggest any specific sources of such lingering selection that we could investigate further.

\begin{table}[ht]\centering \small
\caption{Effect of grant designs on researchers' strategies, reporting sample selection coefficient}\label{tab_grantstrat_all_withimr}
\-\\
{
\def\sym#1{\ifmmode^{#1}\else\(^{#1}\)\fi}
\begin{tabular}{l*{9}{c}}
 \hline\hline
& & & & & & &Larger & & \\
& &  &More & &New & &ongoing & &More \\
&Faster & &risk & &directions & &projects &  &accurate \\
                    &\multicolumn{1}{c}{(1)}         &            &\multicolumn{1}{c}{(2)}         &            &\multicolumn{1}{c}{(3)}         &            &\multicolumn{1}{c}{(4)}         &            &\multicolumn{1}{c}{(5)}         \\
\hline \\
log(Duration)       &      --0.056\sym{***}&            &       0.012         &            &       0.008         &            &       0.018         &            &       0.018         \\
                    &     (0.015)         &            &     (0.016)         &            &     (0.015)         &            &     (0.015)         &            &     (0.011)         \\
[0.5em]
log(Amount)         &      --0.025\sym{***}&            &       0.017\sym{**} &            &      --0.016\sym{**} &            &       0.026\sym{***}&            &      --0.003         \\
                    &     (0.007)         &            &     (0.007)         &            &     (0.007)         &            &     (0.007)         &            &     (0.006)         \\
[0.5em]
IMR                 &      --0.008         &            &       0.003         &            &       0.010         &            &      --0.005         &            &       0.000         \\
                    &     (0.007)         &            &     (0.008)         &            &     (0.007)         &            &     (0.008)         &            &     (0.006)         \\
[0.5em]
Constant            &       0.776\sym{***}&            &       0.285\sym{***}&            &       0.533\sym{***}&            &       0.228\sym{**} &            &       0.178\sym{**} \\
                    &     (0.097)         &            &     (0.101)         &            &     (0.095)         &            &     (0.099)         &            &     (0.077)         \\
[0.5em]
\hline
dep. var. mean      &        0.36         &            &        0.53         &            &        0.33         &            &        0.61         &            &        0.17         \\
$ N$ obs.           &       4,175         &            &       4,175         &            &       4,175         &            &       4,175         &            &       4,175         \\
\hline\hline
\end{tabular}
}

\begin{quote} \footnotesize
\emph{Note}: Shows results from OLS regressions where the dependent variables are indicators that equal one if the strategy listed at the top of the column was chosen as one of the ``two most important changes'' that researchers would make in response to receiving a grant of a given (randomized) size and duration. The IMR row reports the coefficient on the standardized Inverse Mills Ratio. Robust standard errors shown in parentheses; $^{*}$ \(p<0.1\), $^{**}$ \(p<0.05\), $^{***}$ \(p<0.01\).
\end{quote}
\end{table}


\subsection{Joint estimation of strategic responses}\label{sec_app_strat_joint}
In the main text, we estimate separate five separate OLS models where we regress an indicator variable for whether the researcher chose each strategy on the (randomized) grant attributes they were provided with. This approach abstracts away from the fact that the researcher is effectively making a single, discrete choice decision about which pair of strategies to select.\footnote{Recall, the researcher is told they have received some grant and are asked to choose the ``top 2'' most important changes to their research the grant would enable.} To more closely model the structure of this problem, we can convert our data into researcher ($i$) by strategy ($j$) level observations and estimate a logit model of the form:
\begin{equation}\label{eq_stratjoint}
\Pr(\text{Strategy change}_{ij}=1 \,|\, D_{i}, A_{i}, X_i) = \frac{\exp(\log(D_{i})\beta^D_j + \log(A_{i})\beta^A_j + \mathbf{X}\bm{\delta})}{1+\exp(\log(D_{i})\beta^D_j + \log(A_{i})\beta^A_j + \mathbf{X}\bm{\delta})} \,\,,
\end{equation}
which allows the effect of the grant attributes to depend on the strategy being considered. Now, instead of five OLS regressions, we have a single logit regression. Table \ref{tab_grantstrat_joint} reports the results of estimating Equation \ref{eq_stratjoint} either without any covariates (Col. 1) or by using cross-fit partialing-out Lasso to select and include some covariates (Col. 2) per \cite{chernozhukov2018double}. In both cases, the results mirror those of our simpler, OLS-based approach shown in the main text.

\begin{table}[htbp] \centering \small
\caption{Logit model of strategy changes in response to grant design}\label{tab_grantstrat_joint}
\-\\
{
\def\sym#1{\ifmmode^{#1}\else\(^{#1}\)\fi}
\begin{tabular}{l*{2}{c}}
\hline\hline
                    &\multicolumn{1}{c}{(1)}         &\multicolumn{1}{c}{(2)}         \\
\hline
                    &                     &                     \\
More risk           &                     &                     \\
[0.5em]
\hspace{3mm} $\times$ log(Duration)&       0.048         &       0.050         \\
                    &     (0.062)         &     (0.063)         \\
[0.5em]
\hspace{3mm} $\times$ log(Amount)&       0.069\sym{**} &       0.069\sym{**} \\
                    &     (0.030)         &     (0.030)         \\
[0.5em]
Faster              &                     &                     \\
[0.5em]
\hspace{3mm} $\times$ log(Duration)&      --0.239\sym{***}&      --0.243\sym{***}\\
                    &     (0.065)         &     (0.065)         \\
[0.5em]
\hspace{3mm} $\times$ log(Amount)&      --0.107\sym{***}&      --0.104\sym{***}\\
                    &     (0.031)         &     (0.031)         \\
[0.5em]
New direction       &                     &                     \\
[0.5em]
\hspace{3mm} $\times$ log(Duration)&       0.034         &       0.043         \\
                    &     (0.066)         &     (0.067)         \\
[0.5em]
\hspace{3mm} $\times$ log(Amount)&      --0.073\sym{**} &      --0.079\sym{**} \\
                    &     (0.031)         &     (0.032)         \\
[0.5em]
Ongoing bigger      &                     &                     \\
[0.5em]
\hspace{3mm} $\times$ log(Duration)&       0.078         &       0.083         \\
                    &     (0.063)         &     (0.064)         \\
[0.5em]
\hspace{3mm} $\times$ log(Amount)&       0.111\sym{***}&       0.117\sym{***}\\
                    &     (0.030)         &     (0.031)         \\
[0.5em]
More accurate       &                     &                     \\
[0.5em]
\hspace{3mm} $\times$ log(Duration)&       0.125         &       0.122         \\
                    &     (0.082)         &     (0.083)         \\
[0.5em]
\hspace{3mm} $\times$ log(Amount)&      --0.019         &      --0.019         \\
                    &     (0.040)         &     (0.040)         \\
[0.5em]
\hline
$ L(\mathbf{X})$    &                     &           Y         \\
$ N i$ obs.         &       4,175         &       4,175         \\
$ N ij$ obs.        &      20,875         &      20,875         \\
\hline\hline
\end{tabular}
}

\begin{quote} \footnotesize
\emph{Note}: Reports the estimates of $\beta^D_j$ and $\beta^S_j$ for each of the strategic options (indexed by $j$) per Eq. \ref{eq_stratjoint}. Col. 1 includes no covariates and Col. 2 includes any Lasso selected covariates. Robust standard errors shown in parentheses; $^{*}$ \(p<0.1\), $^{**}$ \(p<0.05\), $^{***}$ \(p<0.01\).
\end{quote}
\end{table}


\subsection{Best Linear Predictors of heterogeneity}\label{sec_app_strat_blp}
Figure \ref{fig_capes_xp1} in the main text indicates some notable heterogeneity in researchers' responsiveness to grant designs. Those plots of the individual-level Conditional Average Partial Effects (CAPEs) are all based on the high-dimensional vector of covariates included in the causal forests (see Tables \ref{tab_allxsumstat1}-\ref{tab_allxsumstat2} for the full list of those covariates). 

To better understand the sources of that heterogeneity in a more low-dimensional way, Figure \ref{tab_capeblpunivar_xp1} reports some of the Best Linear Predictors of the CAPEs (\citealt{semenova2021debiased}). Loosely speaking, the Best Linear Predictor approach regresses an individual's CAPE on some covariate in order to test how the CAPE covaries with that observable, effectively testing whether that covariate mediates the focal causal effect. To do this, we make use of the \texttt{best\_linear\_projection} function within the \texttt{grf} package (\cite{tibshirani2022grf}). 

There is no systematic way for choosing which observables to test here, so Figure \ref{tab_capeblpunivar_xp1} reports only any observable which yields a $p$-value in a univariate regression of less than 0.01. The ``Univariate coefficient'' reported is in standard deviation units and relative to the mean CAPE (i.e., the mean treatment effect). Thus, a univariate coefficient of 0.1 for Humanities would indicate that the effect of the grant attribute on the focal strategy is 0.1 larger for researchers in that field compared to the rest of the sample.

\begin{figure}[htbp] \centering \small
\caption{Univariate Best Linear Predictors of heterogeneous treatment effects}\label{tab_capeblpunivar_xp1}
\-\\
\subfloat[Faster]{\label{tab_capeblpunivar_speed_xp1}{
\def\sym#1{\ifmmode^{#1}\else\(^{#1}\)\fi}

\newcolumntype{R}{>{\raggedleft\arraybackslash}X}
\newcolumntype{L}{>{\raggedright\arraybackslash}X}
\newcolumntype{C}{>{\centering\arraybackslash}X}

\begin{tabular}{lc}

\hline \hline
& Univar. \\ 
 & coef. \\ 
\hline \\
\underline{\emph{Size}}& \\
[0.5em]Outputs: Books&--0.021\sym{***} \\
& \\
\underline{\emph{Duration}}& \\
\emph{n/a}& \\
\\ \hline \hline 

\end{tabular}

}
}
\subfloat[More risk]{\label{tab_capeblpunivar_riski_xp1}{
\def\sym#1{\ifmmode^{#1}\else\(^{#1}\)\fi}

\newcolumntype{R}{>{\raggedleft\arraybackslash}X}
\newcolumntype{L}{>{\raggedright\arraybackslash}X}
\newcolumntype{C}{>{\centering\arraybackslash}X}

\begin{tabular}{lc}

\hline \hline
& Univar. \\ 
 & coef. \\ 
\hline \\
\underline{\emph{Size}}& \\
\emph{n/a}& \\
& \\
\underline{\emph{Duration}}& \\
[0.5em]Tenured&0.087\sym{***} \\
[0.5em]Not on tenure-track&--0.101\sym{***} \\
\\ \hline \hline 

\end{tabular}

}
}\-\\ \-\\ \-\\
\subfloat[New directions]{\label{tab_capeblpunivar_newdi_xp1}{
\def\sym#1{\ifmmode^{#1}\else\(^{#1}\)\fi}

\newcolumntype{R}{>{\raggedleft\arraybackslash}X}
\newcolumntype{L}{>{\raggedright\arraybackslash}X}
\newcolumntype{C}{>{\centering\arraybackslash}X}

\begin{tabular}{lc}

\hline \hline
& Univar. \\ 
 & coef. \\ 
\hline \\
\underline{\emph{Size}}& \\
[0.5em]Outputs: Books&0.021\sym{***} \\
& \\
\underline{\emph{Duration}}& \\
\emph{n/a}& \\
\\ \hline \hline 

\end{tabular}

}
}
\subfloat[Larger ongoing projects]{\label{tab_capeblpunivar_sizeo_xp1}{
\def\sym#1{\ifmmode^{#1}\else\(^{#1}\)\fi}

\newcolumntype{R}{>{\raggedleft\arraybackslash}X}
\newcolumntype{L}{>{\raggedright\arraybackslash}X}
\newcolumntype{C}{>{\centering\arraybackslash}X}

\begin{tabular}{lc}

\hline \hline
& Univar. \\ 
 & coef. \\ 
\hline \\
\underline{\emph{Size}}& \\
\emph{n/a}& \\
& \\
\underline{\emph{Duration}}& \\
[0.5em]Outputs: Products&0.039\sym{***} \\
\\ \hline \hline 

\end{tabular}

}
}
\subfloat[More accurate]{\label{tab_capeblpunivar_accur_xp1}{
\def\sym#1{\ifmmode^{#1}\else\(^{#1}\)\fi}

\newcolumntype{R}{>{\raggedleft\arraybackslash}X}
\newcolumntype{L}{>{\raggedright\arraybackslash}X}
\newcolumntype{C}{>{\centering\arraybackslash}X}

\begin{tabular}{lc}

\hline \hline
& Univar. \\ 
 & coef. \\ 
\hline \\
\underline{\emph{Size}}& \\
\emph{n/a}& \\
& \\
\underline{\emph{Duration}}& \\
\emph{n/a}& \\
\\ \hline \hline 

\end{tabular}

}
}
\begin{quote} \footnotesize
\emph{Note}: Reports the results from univariate Best Linear Predictors (BLP) regressions of the Conditional Average Partial Effects (CAPEs) on aggregate field indicators as well as other select covariates; only variables with univariate coefficient $p$-values of 0.01 or less are shown; ``n/a'' indicates that no covariates had $p$-values less than 0.01 in the BLP regressions. The mean effects are reported in Table \ref{tab_grantstrat_all}. Robust standard errors, which account for the fact that the CAPEs are estimated objects, shown in parentheses; $^{***}$ \(p<0.01\)
\end{quote}
\end{figure}


\subsection{Alternative estimation of heterogeneity}\label{sec_app_strat_althet}
In the main analyses, we estimate researchers' heterogeneous strategic responses to grant design using the causal forest method of \cite{wager2018estimation}. That approach is appealing because it allows for the estimation of unit-specific partial effects. However, in order to obtain a lower-dimensional understanding of what covariates are influencing the heterogeneity captured by the random forest, we must ex-post select variables of interest. 

An alternative approach is to directly focus on identifying treatment-covariate interaction terms that are ``important'' with the use of machine learning tools such as the lasso. Here, we follow an algorithm inspired by the approaches of \cite{imai2013estimating} and \cite{blackwell2022reducing} that uses two stages of lassos. The algorithm is as follows:
\begin{enumerate}
    \item Generate interaction terms between each randomized grant attributes (amount $A$ and duration $D$) and all covariates $\mathbf{X}$
    \item Perform a lasso regression of a given dependent variable on ($A,D,A\times \mathbf{X},D\times \mathbf{X},\mathbf{X}$); let $L^1_{A,D,\mathbf{X}}$ denote the subset of attribute-covariate interaction terms selected in this step, i.e., from the possible set $(A\times \mathbf{X},D\times \mathbf{X})$
    \item Perform a cross-fit partialing-out lasso linear regression of a given dependent variable on ($A,D,L^1_{A,D,\mathbf{X}}$) using $\mathbf{X}$ as the set of high-dimensional controls (\citealt{chernozhukov2015post})
\end{enumerate}
This approach relies on a separate penalty parameter to be used in the selection of interaction terms (Step 2.) and in the selection of covariates (Step 3.) as suggested by \cite{imai2013estimating}. And by using a post-selection inference method (\citealt{chernozhukov2015post}), we are attempting to avoid regularization biases as noted by \cite{blackwell2022reducing}.

Table \ref{tab_grantstrat_all_lasso_interacts_s1_i1} reports the results from this alternative approach. Overall, the results are consistent with the random forests approach. Notably, tenure status is again observed to be a clear moderator of the effect of grant duration on risk-taking. The main effect estimates also align with our other estimation approach.

\begin{table}[htbp] \centering \small
\caption{Heterogeneous strategic responses to grant attributes }\label{tab_grantstrat_all_lasso_interacts_s1_i1}
\-\\
{
\def\sym#1{\ifmmode^{#1}\else\(^{#1}\)\fi}
\begin{tabular}{l*{10}{c}}
 \hline\hline
& & &  & \multicolumn{1}{c}{Larger} &  \\
& &\multicolumn{1}{c}{More} & \multicolumn{1}{c}{New}&\multicolumn{1}{c}{ongoing}&\multicolumn{1}{c}{More} \\
 &\multicolumn{1}{c}{Faster}&\multicolumn{1}{c}{risk}&\multicolumn{1}{c}{directions}&\multicolumn{1}{c}{projects}&\multicolumn{1}{c}{accurate}\\
\hline \\
log(Duration)       &      --0.030         &      --0.052\sym{**} &       0.015         &       0.019         &       0.019\sym{*}  \\
                    &     (0.023)         &     (0.023)         &     (0.017)         &     (0.015)         &     (0.011)         \\
[0.5em]
\hspace{3mm} $\times$ Male&      --0.044         &                     &                     &                     &                     \\
                    &     (0.030)         &                     &                     &                     &                     \\
[0.5em]
\hspace{3mm} $\times$ Tenured&                     &       0.094\sym{***}&                     &                     &                     \\
                    &                     &     (0.030)         &                     &                     &                     \\
[0.5em]
\hspace{3mm} $\times$ On t.t., not tenured&                     &                     &      --0.045         &                     &                     \\
                    &                     &                     &     (0.035)         &                     &                     \\
[0.5em]
\hspace{3mm} $\times$ Audience: Policymakers&                     &                     &                     &                     &       0.015         \\
                    &                     &                     &                     &                     &     (0.011)         \\
[0.5em]
log(Amount)         &      --0.018         &       0.012         &      --0.016\sym{**} &       0.026\sym{***}&      --0.003         \\
                    &     (0.013)         &     (0.010)         &     (0.007)         &     (0.008)         &     (0.006)         \\
[0.5em]
\hspace{3mm} $\times$ Age&       0.002         &                     &                     &                     &                     \\
                    &     (0.008)         &                     &                     &                     &                     \\
[0.5em]
\hspace{3mm} $\times$ Assistant professor&       0.027         &                     &                     &                     &                     \\
                    &     (0.028)         &                     &                     &                     &                     \\
[0.5em]
\hspace{3mm} $\times$ Full professor&      --0.014         &       0.011         &                     &                     &                     \\
                    &     (0.017)         &     (0.016)         &                     &                     &                     \\
[0.5em]
\hspace{3mm} $\times$ On t.t., not tenured&      --0.015         &                     &                     &                     &                     \\
                    &     (0.029)         &                     &                     &                     &                     \\
[0.5em]
\hspace{3mm} $\times$ Engineering \& related&      --0.036\sym{**} &                     &                     &                     &                     \\
                    &     (0.018)         &                     &                     &                     &                     \\
[0.5em]
\hspace{3mm} $\times$ Total salary&                     &       0.001         &                     &                     &                     \\
                    &                     &     (0.008)         &                     &                     &                     \\
[0.5em]
\hspace{3mm} $\times$ Fundraising expect.&                     &       0.005         &                     &                     &                     \\
                    &                     &     (0.006)         &                     &                     &                     \\
[0.5em]
\hspace{3mm} $\times$ Research risk, own belief&                     &                     &      --0.012         &                     &                     \\
                    &                     &                     &     (0.007)         &                     &                     \\
[0.5em]
\hspace{3mm} $\times$ Humanities \& related&                     &                     &                     &       0.009         &                     \\
                    &                     &                     &                     &     (0.018)         &                     \\
[0.5em]
\hline
Incl. main effect   &           Y         &           Y         &           Y         &           Y         &           Y         \\
Lasso method        &     plug--in         &     plug--in         &     plug--in         &     plug--in         &     plug--in         \\
$ N X$ var. sel.    &       11/62         &       19/62         &        7/62         &        4/62         &        7/62         \\
dep. var. mean      &        0.36         &        0.53         &        0.33         &        0.61         &        0.17         \\
$ N$ obs.           &       4,175         &       4,175         &       4,175         &       4,175         &       4,175         \\
\hline\hline
\end{tabular}
}

\begin{quote} \footnotesize
\emph{Note}: Robust standard errors shown in parentheses; $^{*}$ \(p<0.1\), $^{**}$ \(p<0.05\), $^{***}$ \(p<0.01\).
\end{quote}
\end{table}

\subsection{Effects at the limits of grant design}\label{sec_app_strat_limits}
\begin{figure}[htbp] \centering
\caption{Choice probabilities at observed and limits of grant designs}\label{fig_minmaxstrat}
\includegraphics[width=0.85\textwidth, trim=0mm 15mm 0mm 10mm, clip]{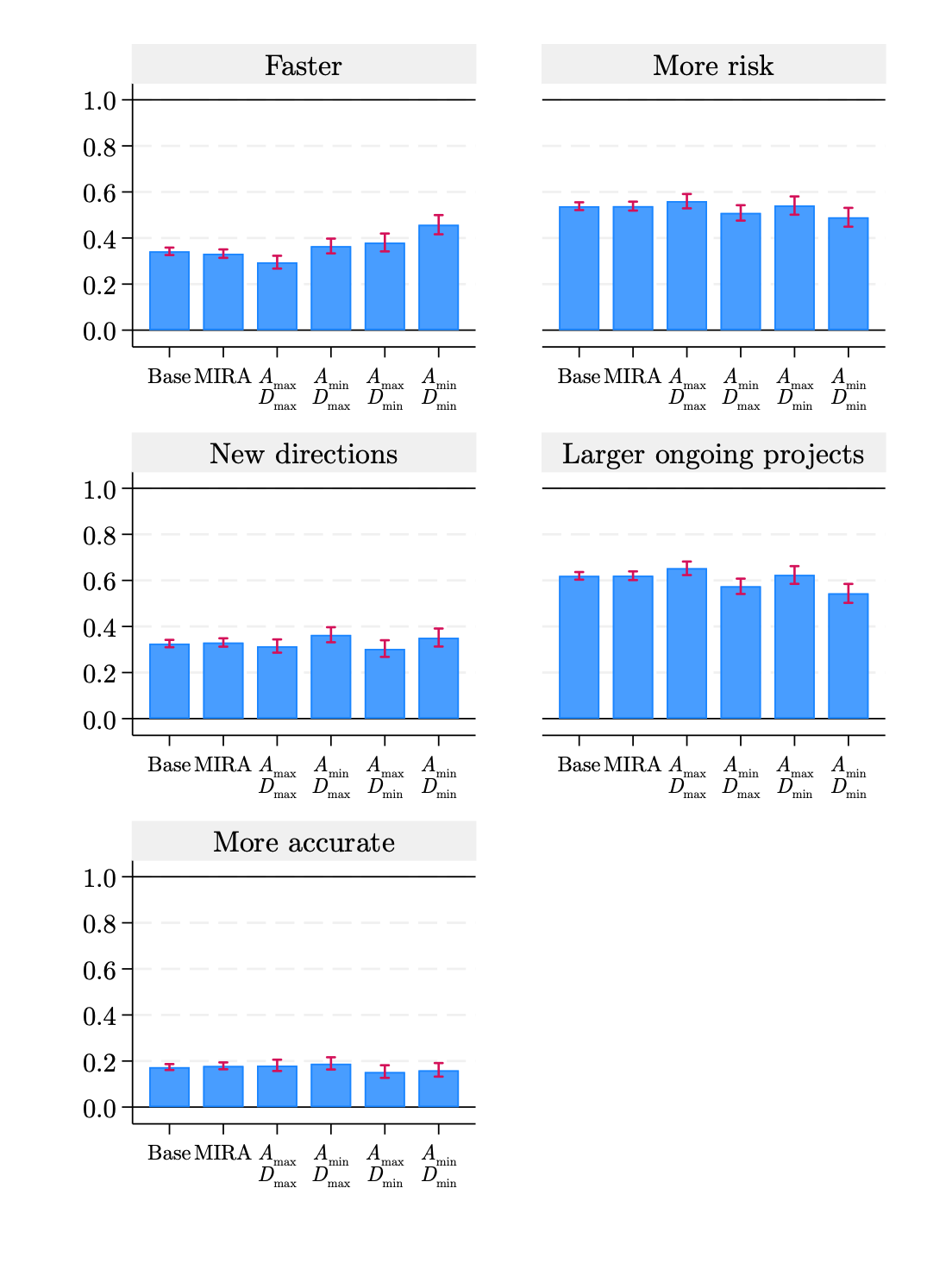}
\begin{quote} \footnotesize
\emph{Note}: Reports the predicted probabilities of choosing a strategy when receiving (1) an average grant in the thought experiment (``Base'') , (2) a grant that is 30\% longer and 15\% smaller than average, reflecting the same relative change as the MIRA grants compared to traditional R01 grants at the NIH (``MIRA''); (3-6) grants with the largest or smallest dollar amount in the thought experiment ($A_{\text{max}}=\text{\$2,000,000};  A_{\text{min}}=\text{\$100,000}$) and the longest or shortest duration in the thought experiment ($D_{\text{max}}=\text{2 years};  D_{\text{min}}=\text{10 years}$). All predicted probabilities are based on the logit models.
\end{quote}
\end{figure}

To get a sense of the limits grant designs, Figure \ref{fig_minmaxstrat} plots the predicted choice probabilities for each strategy given designs at the limits of the dollar amount and duration explored in the thought experiment. These predicted probabilities are based on the point estimates from the logistic regression model. Thus, they are ``generous'' in that we do not force any point estimates that do not reject the null (per their standard errors) to zero.

We find that a ``MIRA-sized change in grant design'' (i.e., an additional 30\% in duration and a 15\% decline in amount) leads to no more than a  1 p.p. change in the probability of any strategy being chosen. Using the full support of grant designs from the thought experiment, the point estimates and confidence intervals indicate that it takes a tremendous amount of differences in grant design (i.e., plus or minus approximately \$2 million and/or 8 years) to change choice probabilities by an amount that could reasonably be considered meaningful (i.e, approximately 10 p.p.). For comparison, \$2 million is an order of magnitude larger than the median funding expectations of professors in the sample, and 8 years is roughly 3 times the average duration of non-tenure-track professors contracts (approximately 2.5 years) and the average amount of time before a pre-tenure professor's tenure evaluation (approximately 2.5 years). Table \ref{tab_minmaxstrat} summarizes these ranges based on estimates from the full sample, as well as estimates from just the non-tenured and tenured subsamples.

\begin{table}[ht] \centering \small
\caption{Span of probabilities at observed and limits of grant designs}\label{tab_minmaxstrat}
\-\\

\newcolumntype{R}{>{\raggedleft\arraybackslash}X}
\newcolumntype{L}{>{\raggedright\arraybackslash}X}
\newcolumntype{C}{>{\centering\arraybackslash}X}

\begin{tabular}{l c c c c c c c c c}

\hline\hline
 & & & & & \multicolumn{2}{c}{Point Estimate} & & \multicolumn{2}{c}{95\% C.I.} \\
{}&{mean}&{}&{MIRA}&{}&{min}&{max}&{}&{min}&{max} \tabularnewline
\hline
\underline{Panel (a): Full sample}&&&&&&&&& \tabularnewline
Faster&0.34&&0.33&&0.29&0.46&&0.27&0.50 \tabularnewline
More risk&0.54&&0.54&&0.49&0.56&&0.45&0.59 \tabularnewline
New directions&0.33&&0.33&&0.30&0.36&&0.27&0.40 \tabularnewline
Larger ongoing projects&0.62&&0.62&&0.54&0.65&&0.50&0.68 \tabularnewline
More accurate&0.17&&0.18&&0.15&0.19&&0.13&0.22 \tabularnewline
\\ \underline{Panel (b): Non-tenured}&&&&&&&&& \tabularnewline
Faster&0.40&&0.39&&0.36&0.48&&0.32&0.54 \tabularnewline
More risk&0.47&&0.45&&0.42&0.52&&0.37&0.58 \tabularnewline
New directions&0.31&&0.32&&0.27&0.35&&0.22&0.41 \tabularnewline
Larger ongoing projects&0.63&&0.63&&0.55&0.66&&0.48&0.70 \tabularnewline
More accurate&0.20&&0.21&&0.17&0.22&&0.13&0.27 \tabularnewline
\\ \underline{Panel (c): Tenured}&&&&&&&&& \tabularnewline
Faster&0.30&&0.29&&0.25&0.44&&0.22&0.50 \tabularnewline
More risk&0.59&&0.60&&0.50&0.63&&0.44&0.67 \tabularnewline
New directions&0.34&&0.34&&0.32&0.37&&0.28&0.43 \tabularnewline
Larger ongoing projects&0.62&&0.61&&0.54&0.65&&0.49&0.69 \tabularnewline
More accurate&0.16&&0.16&&0.14&0.17&&0.11&0.20 \tabularnewline
\hline\hline 

\end{tabular}

\begin{quote} \footnotesize
\emph{Note}: Reports the predicted probabilities of choosing a strategy when receiving (Column 1) an average grant in the thought experiment (``Base''); (Column 2) a grant that is 30\% longer and 15\% smaller than average, reflecting the same relative change as the MIRA grants compared to traditional R01 grants at the NIH (``MIRA''); (Columns 3-6) report the minimum and maximum choice probabilities obtained under any grant design in the thought experiment using either the point estimates or the 95\% confidence intervals.
\end{quote}
\end{table}

\clearpage
\section{Additional results on researchers' preferences}\label{sec_app_prefs}
\setcounter{figure}{0}
\renewcommand{\thefigure}{C\arabic{figure}}
\setcounter{table}{0}
\renewcommand{\thetable}{C\arabic{table}}
\setcounter{equation}{0}
\renewcommand{\theequation}{C\arabic{equation}}

\subsection{Main results on preferences, reporting sample selection correction estimates}\label{sec_app_prefs_sampselect}

\begin{table}[htbp] \centering \small
\caption{Researchers' willingness to trade off grant funding amount and duration, reporting sample selection coefficient}\label{tab_indif1_withimr}
\-\\
{
\def\sym#1{\ifmmode^{#1}\else\(^{#1}\)\fi}
\begin{tabular}{l*{4}{c}}
\hline\hline
 &\multicolumn{1}{c}{(1)} &\multicolumn{1}{c}{(2)} &\multicolumn{1}{c}{(3)} &\multicolumn{1}{c}{(4)} \\
\hline
\\
 $\gamma$ & 0.767\sym{***}& 0.779\sym{***}& 0.817\sym{***}& 0.848\sym{***}\\
 & (0.00698) & (0.00676) & (0.00591) & (0.00533) \\
\\
 IMR & --9645.0\sym{*} & --6260.8 & --5905.1\sym{*} & --5089.1\sym{*} \\
 & (5286.4) & (4472.4) & (3391.8) & (3035.3) \\
[1em]
\hline
\\ Marginal rate& 57,353 & 53,857 & 42,848 & 34,413 \\
\hspace{3mm} of substitution& [7.3\%] & [6.9\%] & [5.5\%] & [4.4\%] \\
[1em] \hline Excl. above ptile.& & 99$ ^{th}$ & 95$ ^{th}$ & 90$ ^{th}$ \\
$ N$ obs. & 4,175 & 4,134 & 3,978 & 3,777 \\
\hline\hline
\end{tabular}
}

\begin{quote} \footnotesize
\emph{Note}: Reports results from estimating Eq. \ref{eq_ADreg} using alternative samples; larger $\gamma$ values reflect a stronger taste for funding amount compared to duration; marginal rate of substitution (\$ per year) is reported at the means and is shown in brackets as a percentage of average funding amount; the ``Excl. above ptile.'' samples drop responses where $(A_{i1} -\widetilde{A}_{i2})/A_{i1}$ is at or above the reported percentile; robust standard errors in parentheses; $^{*}$ \(p<0.1\), $^{**}$ \(p<0.05\), $^{***}$ \(p<0.01\). The IMR row reports the coefficients on the standardized Inverse Mills Ratios included in the regressions.
\end{quote}
\end{table}

\subsection{A model of researchers' production and consumption}\label{sec_app_prefs_theory}
The main empirical model of researchers' preferences over grant amount and duration (Eq. \ref{eq_ADutility}: $\alpha A^\gamma D^{1-\gamma}$) is, effectively, a reduced form representation of a more complex model of researchers' production and consumption patterns. The following subsections motivate tests of heterogeneity across researchers in terms of their preferences (i.e., $\gamma$).

First, we specify a more expansive model of researchers' utility from the receipt of a grant that has some amount $A$ and duration $D$:

\newcommand{\harpoon}{\accentset{\rightharpoonup}}
\begin{equation}\label{eq_motivate_gammahet}
\begin{aligned}
V(\harpoon{a}_i; \theta_i, A, D) = \max_{\harpoon{a}_i} & \; \sum_t^T \delta_i^t u_i(f_i(\beta_i+a_{it}),a_{it})  \\
& s.t. \\
& \sum_t a_{it} \leq A \\
& a_{it} = 0 \text{ if } t>D \;, \\
\end{aligned}
\end{equation}

where the key choice for the researcher is how to allocate the grant over periods $t=(1,2,...,T)$. The notation of the model is as follows: $\harpoon{a}=(a_1,a_2,...,a_T)$ represents the vector of choices about how to allocate the funding over time; $\theta$ is the set of parameters that govern the researcher's discount rate $\delta$, their utility function $u$, their production function $f$, and their guaranteed funding $\beta$, which is available every year; $u$ and $f$ are assumed to be concave; and all parameters and variables are implicitly unique to each researcher. The two constraints dictate how much grant funding is at the researcher's disposal and how long they have access to those funds. In the context of this model, the focal empirical parameter of $\gamma$ represents the marginal utility of an increase in grant amount ($\partial V/\partial A$) relative to the marginal utility of an increase in grant duration ($\partial V/\partial D$) after the researcher has re-optimized their allocations over time.

Researchers value grants with more funding (larger $A$) because they enable more production within a given year, and they value grants with a longer duration (larger $D$) because they enable more years of production on the steeper portion of the production function.\footnote{Anecdotal conversations with researchers indicate that they also value longer grants because of the reduced variance in their future funding streams, which is not formalized in our motivating model. Practically, this enables researchers to better plan their work over longer horizons. For example, a common remark by researchers when considering a longer grant is that it would allow them to offer their staff and students longer (formal or informal) contracts. Another (un-modeled) potential benefit of longer grants is that it provides insurance for future supply or demand shocks --- e.g., if a researcher believes there is a new technology on the horizon that would increase their own productivity, they could reserve funding for future periods.} 

In this model, researchers derive positive utility not just from the outputs of their work, but also from their inputs directly. This reflects two realities; first, grant funds can often be used for researchers' salaries; second, as sociological studies and popular critiques of science have long pointed out, researchers are not immune to ``empire-building'' tendencies, where funding levels can play a prominent role in delineating social one's status (\citealt{merton1968matthew,shapin1995here,madsen2020concentration}).\footnote{Since most grants are obtained through competitive evaluations, they also serve as signals about researchers' abilities. There is logic to this given the unobservability of researchers' effort and the fact that most grants are awarded via peer review processes --- they are signals that peers value a researchers' ideas.  Thus, they can have more direct and indirect value to each researcher depending on their position. Furthermore, grants often include overhead funding that flows to the researchers' institution, with these ``indirect costs'' accounting for a significant fraction of a grants total value (\citealt{johnston2015predictors}). However, our thought experiment abstracts away from both of these points since they are non-competitive and devoid of indirect costs.}

The model abstracts away from any of the researchers' choices about their effort or their research strategies, with the production function $f$ summarizing each researchers' ability to convert funding into output they value.

Overall, the value derived from outputs and inputs depends on the incentives and institutions that transform these things into objects that researchers value (e.g., job security, salary, prestige) and their preferences for those objects. All of these forces and preferences are summarized by $u$. Practical examples of factors shaping $u$ include researchers taste for science (\citealt{stern2004scientists,roach2010taste}), the tenure process and output measurement schemes (\citealt{macleod2020does}), intellectual property regimes (\citealt{hvide2018university}), the nature of competition within fields (\citealt{hill2019scooped}), gate-keeping (\citealt{azoulay2019does}), and social factors more generally (\citealt{shapin1995here}).

In the next sub-section, we used highly simplified versions of this model to make some predictions about heterogeneity.

\subsection{Predicting heterogeneity in grant design preferences}\label{sec_app_prefs_stylizedtheory}
Deriving comparative statics from Equation \ref{eq_motivate_gammahet} without making further assumptions is particularly challenging. But, our goal is only to motivate tests of heterogeneity in researchers' relative preferences for grant funding amount versus duration ($\gamma$). So below, we take a much simpler approach of eliminating all but one of the four sources of heterogeneity in the model ($\delta,u,f,\beta$) and then testing whether, in each stylized model, researchers' relative preferences for amount versus design is increasing or decreasing in a given parameter.\footnote{Each stylized model retains concavity in the production (or consumption) function since linear formulations eliminate any concerns for the temporal dimension and in turn eliminate the value of grant duration by construction.}

Throughout, we define $V_{A,D} \equiv V(a; \theta, A, D)$, $a^*(V_{A,D}) \equiv \arg \max_a V_{A,D}(a)$, and $\tilde{\gamma} \equiv \frac{\frac{\partial V_{A,D}}{\partial A}}{\frac{\partial V_{A,D}}{\partial D}}$, noting that $\tilde{\gamma}$ is proportional to $\gamma$ as defined in the main empirical model shown in Equation \ref{eq_ADutility}.

In all cases below, we focus only on the scenarios where $A=\{1,2\}$ and $D=\{1,2\}$ and evaluate preferences based on changes from initial values of $A=1$ and $D=1$. This simplifies the construction of the partial derivatives that define researchers' preferences. The relative value of an increase amount is: $\frac{\partial V_{1,1}}{\partial A} = V_{2,1} - V_{1,1}$, and the relative value of an increase in duration is: $\frac{\partial V_{1,1}}{\partial D} = V_{1,2} - V_{1,1}$. 

\subsubsection*{Discount rate}
To study the effect of researchers' discount rates on their grant preferences, we assume that the production and consumption functions, where grant funding $a$ is the input, can be jointly given by $a_t^{1/2}$. In this case, there is no heterogeneity in researchers' production or consumption functions, nor in their guaranteed funding flows ($\beta=0$). Focusing first on changes in grant amount for a $D=1$ grant, researchers' utility, optimal input allocations, and marginal valuations are given by:
\begin{equation*}
\begin{aligned}
V_{1,1} & = \max_{a} \; a^{1/2} \\
a^*(V_{1,1}) & = a^*(V_{2,1}) = 1 \\
\frac{\partial V_{1,1}}{\partial A} & = 2^{1/2} - 1 .
\end{aligned}
\end{equation*}
There are no costs to spending the grant funding, so all funds are used (as will be the case in all 1-year grant scenarios). Next, focusing on changes in grant duration for a $A=1$ grant, researchers' utility, optimal input allocations, and marginal valuations are given by:
\begin{equation*}
\begin{aligned}
V_{1,2} & = \max_{a} \; a^{1/2} + \delta (1-a)^{1/2} \\
a^*(V_{1,2}) & = \frac{1}{\delta^2+1} \\
\frac{\partial V_{1,1}}{\partial D} & = (\frac{1}{\delta^2+1})^{1/2} + \delta (1-(\frac{1}{\delta^2+1}))^{1/2} - 1 .
\end{aligned}
\end{equation*}
Thus, the ratio of researchers' marginal value from increases in grant amount versus duration is given by:
\begin{equation*}
\tilde{\gamma}(\delta) = \frac{(2^{1/2} - 1^{1/2})}{((\frac{1}{\delta^2+1})^{1/2} + \delta (1-(\frac{1}{\delta^2+1}))^{1/2} - 1^{1/2} )} ,
\end{equation*}
where $\partial \tilde{\gamma} / \partial \delta < 0 $. Thus, we should expect researchers that discount future periods at a larger rate (smaller $\delta$) to have a stronger preference for grant amount (larger $\gamma$). 

\subsubsection*{Capital-intensity, risk-aversion, and direct utility}
Here, we focus on the role of capital-intensity (i.e., convexity of $f$ w.r.t. $a$), risk-aversion (i.e., convexity of $u$ w.r.t. $f(a)$), and direct utility of grant funding (i.e., convexity of $u$ w.r.t. $a$). We assume that the production and consumption functions, where grant funding $a$ is the input, can be jointly given by $a_t^\kappa$. All three of these features are captured by $\kappa$: a larger value indicates a more convex production function w.r.t. funding, larger risk tolerance, and larger direct utility of grant funding. In this case, there is no discounting of the second period ($\delta=0$) and no heterogeneity in researchers' guaranteed funding flow ($\beta=0$). Focusing first on changes in grant amount for a $D=1$ grant, researchers' utility, optimal input allocations, and marginal valuations are given by:
\begin{equation*}
\begin{aligned}
V_{1,1} & = \max_{a} \; a^\kappa \\
a^*(V_{1,1}) & = a^*(V_{2,1}) = 1\\
\frac{\partial V_{1,1}}{\partial A} & = 2^\kappa - 1
\end{aligned}
\end{equation*}
There are no costs to spending the grant funding, so all funds are used. Next, focusing on changes in grant duration for a $A=1$ grant, researchers' utility, optimal input allocations, and marginal valuations are given by:
\begin{equation*}
\begin{aligned}
V_{1,2} & = \max_{a} \; a^\kappa + (1-a)^\kappa \\
a^*(V_{1,2}) & = \frac{1}{2} \\
\frac{\partial V_{1,1}}{\partial D} & = (\frac{1}{2})^\kappa + (\frac{1}{2})^\kappa - 1 .
\end{aligned}
\end{equation*}
Thus, the ratio of researchers' marginal value from increases in grant amount versus duration is given by:
\begin{equation*}
\tilde{\gamma}(\kappa) = \frac{2^\kappa - 1}{(\frac{1}{2})^\kappa + (\frac{1}{2})^\kappa - 1} ,
\end{equation*}
where $\partial \tilde{\gamma} / \partial \kappa > 0 $. Thus, we should expect researchers that are more capital-intensive, more risk-loving, and receive more direct utility from funding to have a stronger preference for grant amount (larger $\gamma$). 

\subsubsection*{Fundraising productivity}
Here, we focus on the role of researchers' ability to obtain research funding (beyond the focal grant), which we capture with $\beta$. We assume that the production and consumption functions, where grant funding $a$ is the input, can be jointly given by $(\beta+a_t)^{1/2}$. In this case, there is no discounting of the second period ($\delta=0$) and no heterogeneity in researchers' production or consumption functions. Focusing first on changes in grant amount for a $D=1$ grant, researchers' utility, optimal input allocations, and marginal valuations are given by:
\begin{equation*}
\begin{aligned}
V_{1,1} & = \max_{a} \; (\beta + a)^{1/2} \\
a^*(V_{1,1}) & = a^*(V_{2,1}) = 1\\
\frac{\partial V_{1,1}}{\partial A} & = (\beta+2)^{1/2} - (\beta+1)^{1/2}
\end{aligned}
\end{equation*}
There are no costs to spending the grant funding, so all funds are used. Next, focusing on changes in grant duration for a $A=1$ grant, researchers' utility, optimal input allocations, and marginal valuations are given by:
\begin{equation*}
\begin{aligned}
V_{1,2} & = \max_{a} \; (\beta + a)^{1/2} + (1-(\beta + a))^{1/2} \\
a^*(V_{1,2}) & = \frac{1}{2} - \beta \\
\frac{\partial V_{1,1}}{\partial D} & = (\beta + (\frac{1}{2} - \beta))^{1/2} + (1-(\beta + (\frac{1}{2} - \beta)))^{1/2} - (\beta + 1)^{1/2} .
\end{aligned}
\end{equation*}
Thus, the ratio of researchers' marginal value from increases in grant amount versus duration is given by:
\begin{equation*}
\tilde{\gamma}(\beta) = \frac{(\beta+2)^{1/2} - (\beta+1)^{1/2}}{(\beta + (\frac{1}{2} - \beta))^{1/2} + (1-(\beta + (\frac{1}{2} - \beta)))^{1/2} - (\beta + 1)^{1/2}} ,
\end{equation*}
where $\partial \tilde{\gamma} / \partial \beta > 0 $. Thus, we should expect researchers who are more productive at securing funding for their research every period to have a stronger preferences for grant amount (larger $\gamma$).

\subsection{Estimating heterogeneity in grant design preferences}\label{sec_app_prefs_results}
To summarize the predictions from the preceding subsection, we expect the strongest preferences for funding amount (relative to duration) to be amongst researchers who: (1) discount the future at a higher rate; (2) are more funding-intensive in their production functions; (3) receive more direct utility from grant funding; (4) are less risk-averse; and (5) are more productive at fundraising for their research. Below, Tables \ref{tab_indifhet_age_capintense_shrsoftmoney}-\ref{tab_indifhet_riskaversion_fundrprod} estimate $\gamma$ parameters for subsamples split according to proxies for these five dimensions.

To proxy for discount rates, we use researchers age. To proxy for funding-intensity, we take the standardized sum of researchers expected funding amounts plus their time spent fundraising under the assumption that these metrics provide a signal of how important funding is to a researchers' science. To proxy for researchers' direct utility from funding, we use the share of their salaries that is sourced via ``soft money'' (i.e., paid for directly by grant funding). To proxy for risk aversion, we take the standardized sum of the three measures of risk in the survey, relating to both researchers' own perceptions of the riskiness of their science, their beliefs about their peers' perceptions of the riskiness of their science, and their degree of risk-taking in their personal lives. To proxy for fundraising productivity, we divide researchers' expected funding in the next five years (excluding any guarantees) by the number of hours they expect to spend on fundraising in that same period to arrive at a \$-grant per hour metric.

Tables \ref{tab_indifhet_age_capintense_shrsoftmoney}-\ref{tab_indifhet_riskaversion_fundrprod} report estimates of $\gamma$ after splitting the sample along these five dimensions. In all cases, we estimate different $\gamma$ values that agree with our theoretical predictions, and in most cases the differences are statistically significant. 


Table \ref{tab_indifhet_other} provides additional sample splits to explore this heterogeneity further.

\begin{table}[ht] \centering \small
\caption{Heterogeneity in the amount-duration tradeoff}\label{tab_indifhet_age_capintense_shrsoftmoney}
\-\\
{
\def\sym#1{\ifmmode^{#1}\else\(^{#1}\)\fi}
\begin{tabular}{l*{9}{c}}
\hline\hline
& \multicolumn{2}{c}{Age} & &\multicolumn{2}{c}{Capital intensity} & &\multicolumn{2}{c}{Share soft money} \\ \cline{2-3} \cline{5-6} \cline{8-9}
            &\multicolumn{1}{c}{(1)}         &\multicolumn{1}{c}{(2)}         &            &\multicolumn{1}{c}{(3)}         &\multicolumn{1}{c}{(4)}         &            &\multicolumn{1}{c}{(5)}         &\multicolumn{1}{c}{(6)}         \\
\hline
            &                     &                     &            &                     &                     &            &                     &                     \\
$\gamma$    &       0.755\sym{***}&       0.802\sym{***}&            &       0.750\sym{***}&       0.785\sym{***}&            &       0.749\sym{***}&       0.799\sym{***}\\
            &    (0.0110)         &    (0.0148)         &            &    (0.0121)         &    (0.0119)         &            &    (0.0112)         &    (0.0112)         \\
[1em]
\hline
\\ m.r.s.   &      61,093         &      46,985         &            &      62,915         &      51,850         &            &      62,958         &      47,893         \\
\hspace{3mm} &     [7.8\%]         &     [6.0\%]         &            &     [8.0\%]         &     [6.6\%]         &            &     [8.1\%]         &     [6.1\%]         \\
[1em] \hline Sub-sample&         low         &        high         &            &         low         &        high         &            &         low         &        high         \\
$ N$ obs.   &       1,476         &       1,114         &            &       1,393         &       1,390         &            &       1,799         &       1,306         \\
\hline\hline
\end{tabular}
}

\begin{quote} \footnotesize
\emph{Note}: Larger $\gamma$ values reflect a stronger taste for grant size compared to duration; (m.r.s.) marginal rate of substitution (\$ per year) is reported at the means and is shown as a percentage of average grant size in brackets; robust standard errors in parentheses; $^{*}$ \(p<0.1\), $^{**}$ \(p<0.05\), $^{***}$ \(p<0.01\). Sample splits are described in-text, with ``low'' and ``high'' referring to the lower and upper terciles of a given feature, respectively. 
\end{quote}
\end{table}

\begin{table}[ht] \centering \small
\caption{Heterogeneity in the amount-duration tradeoff (cont'd)}\label{tab_indifhet_riskaversion_fundrprod}
\-\\
{
\def\sym#1{\ifmmode^{#1}\else\(^{#1}\)\fi}
\begin{tabular}{l*{7}{c}}
\hline\hline
& \multicolumn{2}{c}{Risk aversion} & &\multicolumn{2}{c}{Fundraising productivity} \\ \cline{2-3} \cline{5-6}
            &\multicolumn{1}{c}{(1)}         &\multicolumn{1}{c}{(2)}         &            &\multicolumn{1}{c}{(3)}         &\multicolumn{1}{c}{(4)}         &            \\
\hline
            &                     &                     &            &                     &                     &            \\
$\gamma$    &       0.786\sym{***}&       0.749\sym{***}&            &       0.721\sym{***}&       0.803\sym{***}&            \\
            &    (0.0122)         &    (0.0118)         &            &    (0.0168)         &    (0.0134)         &            \\
[1em]
\hline
\\ m.r.s.   &      51,794         &      63,012         &            &      72,286         &      46,668         &            \\
\hspace{3mm} &     [6.6\%]         &     [8.1\%]         &            &     [9.2\%]         &     [6.0\%]         &            \\
[1em] \hline Sub-sample&         low         &        high         &            &         low         &        high         &            \\
$ N$ obs.   &       1,395         &       1,387         &            &         881         &         881         &            \\
\hline\hline
\end{tabular}
}

\begin{quote} \footnotesize
\emph{Note}: Larger $\gamma$ values reflect a stronger taste for grant size compared to duration; (m.r.s.) marginal rate of substitution (\$ per year) is reported at the means and is shown as a percentage of average grant size in brackets; robust standard errors in parentheses; $^{*}$ \(p<0.1\), $^{**}$ \(p<0.05\), $^{***}$ \(p<0.01\). Sample splits are described in-text, with ``low'' and ``high'' referring to the lower and upper terciles of a given feature, respectively. 
\end{quote}
\end{table}


\begin{table}[ht] \centering \small
\caption{Heterogeneity in the size-duration tradeoff, by other features}\label{tab_indifhet_other}
\-\\
{
\def\sym#1{\ifmmode^{#1}\else\(^{#1}\)\fi}
\begin{tabular}{l*{12}{c}}
\hline\hline
& \multicolumn{5}{c}{Broad field of study} & & \multicolumn{3}{c}{Tenure status} \\ \cline{2-6} \cline{8-10} 
            &\multicolumn{1}{c}{(1)}         &\multicolumn{1}{c}{(2)}         &\multicolumn{1}{c}{(3)}         &\multicolumn{1}{c}{(4)}         &\multicolumn{1}{c}{(5)}         &            &\multicolumn{1}{c}{(6)}         &\multicolumn{1}{c}{(7)}         &\multicolumn{1}{c}{(8)}         \\
\hline
            &                     &                     &                     &                     &                     &            &                     &                     &                     \\
$\gamma$    &       0.760\sym{***}&       0.740\sym{***}&       0.779\sym{***}&       0.782\sym{***}&       0.769\sym{***}&            &       0.757\sym{***}&       0.761\sym{***}&       0.773\sym{***}\\
            &    (0.0170)         &    (0.0171)         &    (0.0127)         &    (0.0171)         &    (0.0158)         &            &    (0.0167)         &    (0.0145)         &   (0.00908)         \\
[1em]
\hline
\\ m.r.s.   &      59,651         &      66,065         &      53,643         &      52,794         &      56,779         &            &      60,677         &      59,316         &      55,575         \\
\hspace{3mm} &     [7.6\%]         &     [8.4\%]         &     [6.9\%]         &     [6.8\%]         &     [7.3\%]         &            &     [7.8\%]         &     [7.6\%]         &     [7.1\%]         \\
[1em] \hline Sub-sample&        Eng.         &        Hum.         &        Med.         &        Nat.         &        Soc.         &            &         N/A         &  Pre-tenure         &     Tenured         \\
$ N$ obs.   &         703         &         786         &       1,165         &         655         &         866         &            &         821         &         903         &       2,451         \\
\hline\hline
\end{tabular}
}

\begin{quote} \footnotesize
\emph{Note}: Larger $\gamma$ values reflect a stronger taste for grant size compared to duration; (m.r.s.) marginal rate of substitution (\$ per year) is reported at the means and is shown as a percentage of average grant size in brackets; robust standard errors in parentheses; $^{*}$ \(p<0.1\), $^{**}$ \(p<0.05\), $^{***}$ \(p<0.01\).
\end{quote}
\end{table}

\subsection{Alternative estimation of grant preferences}\label{sec_app_prefs_noncrs}
The empirical model of grant preferences in the main paper assumes constant returns to scale (CRS; $\gamma+(1-\gamma)=1$). We can relax this assumption by making use of a third thought experiment in the data. Although this third experiment was designed for another research question, it provides the variation necessary to estimate a more flexible model that has a separate utility parameter for both grant size and duration.

In this other experiment, we asked respondents to report their willingness to trade off their annual salary in exchange for increases in their guaranteed research budget, effectively, a grant. This was done twice, by asking respondents how low of a salary they would be willing to take in exchange for having access to an additional \$250,000 or \$1,000,000 for 5 years.

To begin, first consider a new indirect utility function that depends on researchers' annual salary $S$, their current guaranteed research budget over the coming 5 years $B$, and a new grant of amount $A$ and duration $D$:
\begin{align}\label{eq_ADnoncrs1}
\begin{split}
v(S_{ig}, B_{i}, A_{ig}, D_{ig}) & = \alpha  S_{ig}^\sigma (B_{i} + A_{ig})^\rho D_{ig}^\delta \,\,.
\end{split}
\end{align}
In order to use the variation from the thought experiments, we assume we can decompose Equation \ref{eq_ADnoncrs1} into two separate functions that describe researchers' preferences over the new grant (evaluating just $A$ and $D$) and their preferences over an increase in their 5-year guaranteed budget and a corresponding change in their salary:
\begin{align}\label{eq_ADnoncrs2}
\begin{split}
v_1(A_{ig}, D_{ig}) & = \alpha_{i1} A_{ig}^\rho D_{ig}^\delta \\
v_2(S{ig}, B_{i}, A_{ig}) & = \alpha_{i2} S_{ig}^\sigma (B_{i} + A_{ig})^\rho \,\,.
\end{split}
\end{align}
To summarize, we have researchers' responses to three different scenarios: (1) being indifferent between two grants that vary in their size and duration, call these grants $g=\{1,2\}$; (2) being indifferent between some decrease in their salary and an increase in their research budget by \$250,000, call this increase grant $g=3$; and (3) being indifferent between some decrease in their salary and an increase in their research budget by \$1,000,000, call this increase grant $g=4$. Thus, we have three equations:
\begin{align}\label{eq_ADnoncrs_indif}
\begin{split}
v_1(A_{i1}, D_{i1}) & = v_1(A_{i2}, D_{i2}) \\
v_2(S_{i0}, B_{i} ) & = v_2(S_{i3}, B_{i} , A_{i3}) \\
v_2(S_{i0}, B_{i} ) & = v_2(S_{i4}, B_{i} , A_{i4}) \,\,,
\end{split}
\end{align}
where $S_{i0}$ indicates their current salary and $B_i$ is their current 5-year research budget guarantee. The thought experiments involve providing ($A_{i1},D_{i1},D_{i2},A_{i3},A_{i4}$) to the respondents, who in turn report the values ($A_{i2},S_{i3},S_{i4}$) such that indifference holds. We use these three equations to generate the three moment conditions necessary to identify the key parameters $\sigma$, $\rho$, and $\delta$.

Table \ref{tab_noncrsgamma} reports the results from GMM estimation of the key parameters. The ratio of these alternative estimates of the elasticities of grant size ($\sigma$) and duration ($\delta$), $\frac{0.75}{0.22}$=3.41, are very close to our estimate of this same ratio under the CRS assumption we use in the main results, $\frac{0.77}{(1-0.77)}$=3.26. We take this as evidence that the CRS assumption is not distorting our results on the relative preferences researchers have for grant designs.

\begin{table}[htbp] \centering \small
\caption{Non-CRS based grant preference estimates}\label{tab_noncrsgamma}
\-\\
\def\sym#1{\ifmmode^{#1}\else\(^{#1}\)\fi}
\begin{tabular}{l*{1}{c}}
\hline\hline
            &\multicolumn{1}{c}{(1)}         \\
\hline
            &                     \\
$\sigma$ (salary)&       1.016\sym{***}\\
            &     (0.001)         \\
            &                     \\
$\rho$ (grant amount)&       0.075\sym{***}\\
            &     (0.004)         \\
            &                     \\
$\delta$ (grant duration)&       0.022\sym{***}\\
            &     (0.001)         \\
[0.5em]
\hline
$ N$ obs.   &       4,003         \\
\hline\hline
\end{tabular}

\-\\ \-\\
\begin{quote} \footnotesize
\emph{Note}: GMM estimates of parameters in Eq. \ref{eq_ADnoncrs2} based on the three moments generated by the indifference points shown in Eq. \ref{eq_ADnoncrs_indif}; bootstrapped standard errors in parentheses; $^{*}$ \(p<0.1\), $^{**}$ \(p<0.05\), $^{***}$ \(p<0.01\).
\end{quote}
\end{table}

\clearpage
\section{Funders' revealed preferences over grant designs}\label{sec_app_funderprefs}
\setcounter{figure}{0}
\renewcommand{\thefigure}{D\arabic{figure}}
\setcounter{table}{0}
\renewcommand{\thetable}{D\arabic{table}}
\setcounter{equation}{0}
\renewcommand{\theequation}{D\arabic{equation}}


For simplicity, assume the grant market is served by a single supplier, one ``aggregate funder'' that reflects the aggregated behavior of all individual funders. Researchers are the consumers. The aggregate funder's budget is exogenously determined and the aggregate demand for grants of all types is inelastic. No matter which set of grants the funder chooses to offer, researchers will pursue, and be awarded, all of them. This assumption seems relatively plausible given the fact that published success rates across funders are on the scale of 10-20\% (see \href{https://www.science.org/content/article/new-system-scientists-never-have-write-grant-application-again}{here} for more). This implies that the funder is ignoring researchers' preferences, which will lead us to over-estimate their preferences (i.e., our estimate of $\gamma_{\text{fund}}$ will be larger than the true value to the extent funders are in fact incorporating researchers' preferences into their portfolio decisions).

We assume the funder's maximization problem is given by:
\begin{equation}\label{eq_ADutil_fund}
\max_{\mathbf{N}} \sum_{a=\underline{A}}^{\overline{A}} \sum_{d=\underline{D}}^{\overline{D}}\pi(N_{ad}\,, a^{\gamma_{\text{fund}}}d^{1-\gamma_{\text{fund}}}) \,,
\end{equation}
where $\mathbf{N}=(N_{\underline{A},\underline{D}}, ... , N_{\overline{A},\overline{D}})$ and $\pi$ is a value function that describes the funder's net costs and benefits of awarding $N_{ad}$ grants of amount $a$ and duration $d$. Furthermore, assume that $\pi$ is well-behaved and increasing and concave in its arguments: $\pi'>0, \pi''<0$. $\pi$ would be concave if funders prefer strategies that involve diversification across different grant designs, for example, to learn about the value of projects with small investments before making larger investments. Implicitly, $\pi$ incorporates the funder's budget constraint.

Since the funder's valuation of their grant portfolio is separable in the groupings of amount and duration per $a$ and $d$, then, in equilibrium, they will choose an allocation, $\mathbf{N}^*$, that equalizes the marginal returns across all groups (i.e., $\frac{\partial \pi}{\partial N_{ad}}$ is equal for all $ad$ groups).

Here, just as changes in prices set by a monopolist firm are informative about changes in the firm's marginal costs, the observed distribution of grant types ($\mathbf{N}^*$; as shown in Figure \ref{fig_dimension_heatmap}) will be informative about the funder's preferences. Ideally, we would have some theory or data to be more specific about the shape of $\pi$ in order to estimate $\frac{\partial \pi}{\partial a}$ and $\frac{\partial \pi}{\partial d}$, but we do not. We can still speak to the \emph{relative} value the funder appears to place on grant size and duration per $\gamma_{\text{fund}}$.

Note that the funder's choice of $N^*_{ad}$ can be written as $\phi(a^{\gamma_{\text{fund}}}d^{1-\gamma_{\text{fund}}})$, where the shape of $\phi$ is governed by $\pi$. Then, the relationship between $N^*_{ad}$ and $(a,d)$ will be informative of $\gamma_{\text{fund}}$ since:
\begin{equation}
\frac{\partial N^*_{ad} / \partial a}{\partial N^*_{ad} / \partial d} = \frac{\partial \phi(a^{\gamma_{\text{fund}}}d^{1-\gamma_{\text{fund}}}) / \partial a}{\partial \phi(a^{\gamma_{\text{fund}}}d^{1-\gamma_{\text{fund}}}) / \partial d} = \frac{\gamma_{\text{fund}} d}{(1-\gamma_{\text{fund}}) a} \,.
\end{equation}
Thus, by regressing $N^*_{ad}$ on $a$ and $d$, the ratio of those coefficients (and the sample means) can be used to infer $\gamma_{\text{fund}}$.

\begin{figure}[htbp] \centering
\caption{Grant design preferences, researchers versus funders}\label{fig_indif_resvfun}
\includegraphics[width=0.8\textwidth, trim=5mm 5mm 0mm 10mm, clip]{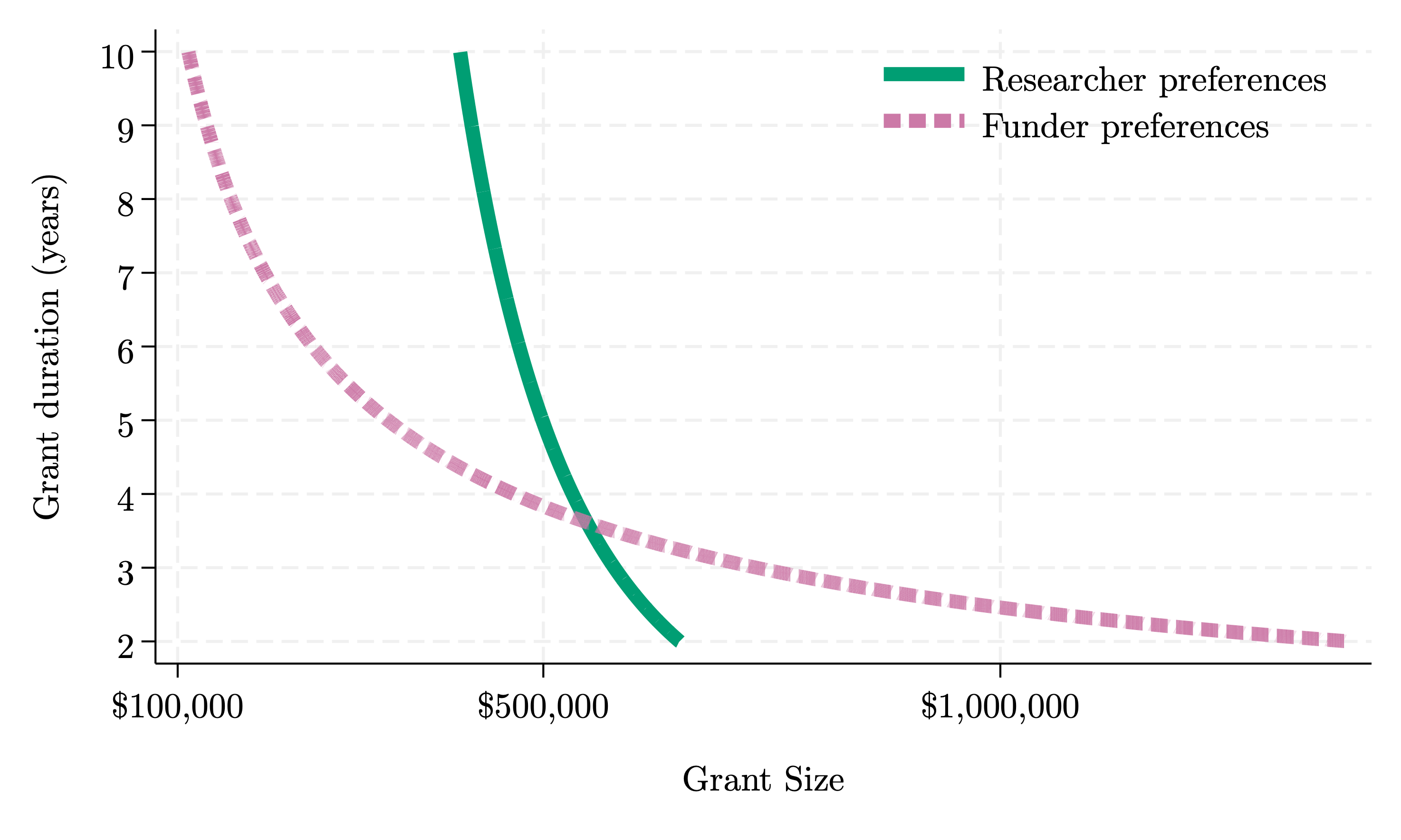}
\begin{quote} \footnotesize
\emph{Note}: Shows the preferences of researchers per the survey-based thought experiment ($\gamma$), and of funders per observed distribution of research grants in the Dimensions data ($\gamma^{\text{fund}}$); the intersection of the curves is chosen to be the average grant size and duration.
\end{quote}
\end{figure}

Using \$100,000 bins for $ a$ and 1-year bins for $ d$, a regression of the number of grants in each bin on the grant size and duration indicates that there are --150.3 (s.e.=34.7) fewer grants for every \$100,000 increase and --355.8 (s.e.=72.4) fewer grants for every 1-year increase. These coefficients imply a $\gamma_{\text{fund}}$ of 0.39.
Figure \ref{fig_indif_resvfun} plots the indifference curve implied by this estimate, contrasted with researchers' preferences. 

This approach is likely identifying an upper bound of $\gamma^{\text{fund}}$ because of our assumption that funders do not respond to researchers' demand. Thus, if they are in fact responding to demand already, and we know from the prior analyses researchers' $\gamma$ is approaching 0.8, then our estimate of funders' preferences is biased upward. Still, funders are clearly much more willing to trade off size for duration. Intuitively, the time constraints facing funders must be less important than their cash constraints. But, it is notable how different funders' and researchers' preferences appear to be. In a sense, it appears funders are already pulling researchers towards a ``longer-smaller'' portfolio of grants than they would prefer.

\end{document}